\documentclass[a4paper,twoside,11pt]{article} % to use for all kind of papers but AACA

\usepackage[english]{babel} %
\usepackage{lmodern}
\usepackage[T1]{fontenc}
\usepackage[latin9]{inputenc}	% Così le "è" vengono trasformate in \`e etc. etc.
\newlength{\defaultparindent}
\usepackage{latexsym}
\usepackage{amsmath}
\usepackage{amsfonts}
\usepackage{amsthm}
\usepackage{amssymb}
\usepackage{bbold}
\usepackage{multirow}
\usepackage{hyperref}
\hypersetup{colorlinks=true,citecolor=green,linkcolor= blue}

\usepackage{todonotes} % more appealing than \marginpar ??
\setuptodonotes{color=white, linecolor=orange} % global options for package todonotes
\usepackage{soul} % to get overstriked text
\usepackage{tikz-cd}% to draw commutative diagrams
\usepackage{enumerate}

\usepackage{algorithm} % to write algorithms, suggested by Renato on May 11, 2017
\usepackage{algpseudocode}

\usepackage[arXiv]{optional} % Possible values of the option fixing the format: x, std, arXiv, JMP, JOPA, AACA, TCS and also note control: margin_notes, final_notes
\def\mynote{\todo} % \marginpar or \todo
\opt{AACA}{% only for submission to AACA -> \cal font is not defined....
\def\cal{\mathcal}
}

\opt{arXiv,std,JMP,JOPA,TCS}{% only for arXiv, std, JMP, JOPA & TCS we need the definitions included in mbh_defs_TeX (that contains original versions)

\newtheorem{MS_theorem}{Theorem}
\newtheorem{MS_lemma}{Lemma}
\newtheorem{MS_Proposition}{Proposition}
\newtheorem{MS_Corollary}[MS_Proposition]{Corollary}

\def\myconjugate#1{\overline{#1}} % this way conjugate definitions can be easily changed (\bar did not work fine on multiple chars)
\def\ie{i.e.\ }
\def\eg{e.g.\ }
\def\myisom{\cong} % this way isomorphism symbol definition can be easily changed; possiblities: \cong \simeq
\newcommand{\R}{\ensuremath{\mathbb{R}}} % good; MB XII 2005; needs \usepackage{bbold}
\newcommand{\Identity}{\ensuremath{\mathbb{1}}} % good; MB XII 2005; needs \usepackage{bbold}
\newcommand*\sepline{%
 \begin{center}
  \rule[1ex]{.3\textwidth}{.1pt}
 \end{center}}

\def\my_span#1{\mbox{Span}\left(#1\right)} % changed from 'span' since it interfered with \multicolumn{} MB X 2009
\def\bino#1#2{#1 \choose #2} % standard definition
\def\dotinformula{\;\; \mathrm{.}} % defines space + a full stop (in \rm font) to be placed at the end of a formula
\def\OO#1{\ensuremath{\mbox{O}\!\left(#1\right)}}
\def\SO#1{\ensuremath{\mbox{SO}\!\left(#1\right)}}
\def\GL#1{\ensuremath{\mbox{GL}\!\left(#1\right)}}

\def\O1#1{\ensuremath{\mbox{O}^{#1}(1)}}
\newcommand{\comm}[2]{\ensuremath{\left[ #1, #2 \right]}}
\newcommand{\anticomm}[2]{\ensuremath{\left\{ #1, #2 \right\}}} % or \left[...\right]^+
\newcommand{\myClg}[3]{\ensuremath{{{\cal C}\ell} {\left( #3 \right)}}}	% this is a possible def of myCl that uses only the scalar product g

\DeclareMathOperator{\rank}{rank}
\def\inv#1{\ensuremath{\mbox{inv}\left(#1\right)}}
}

\opt{JMP}{% only for submission to JMP -> double line spacing
}

\def\h_eigen{\eta}
\def\g_eigen{\theta}
\def\mygen{e} % this way the definition of the generators of the algebra can be easily changed
\def\myprimidemp{\mathbb{p}} % this way the definition of the primitive idempotents can be easily changed (beware: not all fonts in math mode have lower characters, e.g. \cal)
\def\myprimidempset{\mathbb{P}} % this way the definition of the primitive idempotents can be easily changed (beware: not all fonts in math mode have lower characters, e.g. \cal)
\def\myseparation{\sepline} % \sepstars or \sepline

\def\SAT{\ensuremath{\mbox{SAT}}}
\def\bigO#1{\ensuremath{\mathcal{O}\left(#1\right)}}
\def\mylitrl{\ensuremath{\rho}}
\def\mysetS{\mathcal{I}} % this way the definition of the set S can be easily changed
\def\mySpinorS{{\mathbb{S}}} % this way the definition of the MLI of spinor spaces S can be easily changed
\def\myFockB{{\cal F}} % this way the definition of the Fock basis can be easily changed
\def\mysetM{\ensuremath{{{\cal M}_n}}} % this way the definition of the set of $2^n$ MTNS of SAT or of Witt basis can be easily changed
\def\mysnqG{\ensuremath{{{\cal N}_n}}} % this way the definition of the semi-neutral quadric Grassmannian can be easily changed
\def\myBooleanS{{\cal{S}}} % this way the definition of the a Boolean SAT problem S can be easily changed
\def\myBooleanT{\mathrm{T}} % this way the definition of the a Boolean T may be easily changed
\def\myBooleanF{\mathrm{F}} % this way the definition of the a Boolean F may be easily changed
\def\myssSum{+_{\!s}} % this way we choose easily the simple spinors sum operator \eg \oplus, \boxplus, \heartsuit, \dotplus, \plushat, etc.
\DeclareMathOperator{\mysbs}{sbs} % definition of sbs in analogy to \sup for clauses z_j

\begin{document}

\opt{x,std,arXiv,JMP,JOPA,TCS}{% in all cases - but AACA
\title{{\bf A Satisfiability algorithm based on Simple Spinors of the Clifford algebra of $\R^{n,n}$} %\\(temporary title)
	}

\author{\\
	\bf{Marco Budinich}%
%
%\footnote{on leave of absence from: University of Trieste, Trieste, Italy}%
%
\\
%	ICTP and INFN, Trieste, Italy\\
	University of Trieste, Italy\\ % - %\\
	\texttt{marco.budinich@units.it}\\
%	\texttt{http://www.ts.infn.it/\~{ }mbh/MBHgeneral.html}\\
%
%\\	Preliminary - restricted circulation (FYEO)
%
%
%	Submitted on March 17, 2016
%	Submitted to: {\em Journal of Mathematical Physics} on September 29, 2017
%		keywords of this submission: Spinor, Satisfiability problem, Clifford algebra
%	Submitted to: {\em Communications in Mathematical Physics} on March 14, 2016.\\
%	Submitted to: {\em Journal of Physics A: Mathematical and Theoretical} \\on October 20, 2017
%	Submitted to: {\em SIAM Journal on Discrete Mathematics} on March 30, 2018
%\\	Re-resubmitted to: {\em Advances in Applied Clifford Algebras} on June 12, 2017
%	Published in: {\em Advances in Applied Clifford Algebras}, 2015\\
%	{\small DOI:10.1007/s00006-015-0547-8}
%	Resubmitted to: {\em Journal of Mathematical Physics} on June 27, 2016\\
%	{\tiny First submission: March 30, 2016}
%	To appear in: {\em Journal of Mathematical Physics}\\
%	{\tiny Submitted on March 30 and June 27, 2016; accepted July 12, 2016}
%	{\em Journal of Mathematical Physics} {\bf 57} (2016) DOI: 10.1063/1.4959531\\
%\\	March 31, 2020, submitted
%\\	Submitted to: {\em Advances in Applied Clifford Algebras} on March 4, 2021
%\\	Submitted to: {\em Journal of Mathematical Physics} on July 6, 2021
%	Submitted to: {\em Theoretical Computer Science} on December 29, 2023
%	Submitted, December 29, 2023
	}
\date{ \today }
%\date{April 21, 2017}
%\date{ } % to hide the date this line must be present (believe it or not...)
\maketitle
}

\begin{abstract}
We refine the formulation of the Boolean satisfiability problem with $n$ Boolean variables in Clifford algebra $\myClg{}{}{\R^{n,n}}$ \cite{Budinich_2019} and exploit this continuous setting to outline a new unsatisfiability test. This algorithm is not combinatorial and can prove unsatisfiability in polynomial time.
\end{abstract}

%{\bf Keywords:} {Clifford algebra; orthogonal group; involutions.}

\opt{x,std,arXiv,JMP,JOPA,TCS}{% in all cases - but AACA
}

\opt{AACA}{% only for AACA
\maketitle
}

\section{Introduction}
\label{sec_Introduction}
Clifford algebra is a remarkably powerful tool initially developed to deal with automorphisms of quadratic spaces \cite{Porteous_1995} that got its major achievements with spinors of mathematical physics. Since then it has been successfully applied to very many different fields, including combinatorial problems \cite{BudinichP_2006, Budinich_2017}.%
\opt{margin_notes}{\mynote{mbh.note: solita frasetta su Cartan ?? to more than a cantury ago... Un sentiero poco battuto è quello...}}%

On the other side the Boolean SATisfiability problem is the progenitor of many combinatorial problems and, suprisingly, fits smoothly in the Boolean algebra embedded in $\myClg{}{}{\R^{n,n}}$, the Clifford algebra of $\R^{n,n}$, the realm of simple spinors \cite{Cartan_1937}. This is not the first encounter of mathematical physics with \SAT{}: also statistical mechanics applied to \SAT{} brought interesting results \cite{Knuth_2015}.

The first contribution of this work is a solid foundation of Boolean algebra within idempotents of \myClg{}{}{\R^{n,n}} and we use this framework to examine different formulations of \SAT{} from a unified standpoint. Subsequently we focus on the continuous formulation of \SAT{} in the group \OO{n}, with its equivalent spinorial and vectorial representations. The second, and major, contribution is to exploit this formulation to outline an algorithm for an unsatisfiability test. The core of the algorithm relies on linear combinations of simple spinors induced by clauses.

More in detail in sections~\ref{prop_SAT_basics} and \ref{sec_neutral_R_nn} we succintly recall the Boolean Satisfiability problem and the Clifford algebra \myClg{}{}{\R^{n,n}}. Section~\ref{sec_Boole_in_Cl} provides a sturdy formulation of Boolean algebra with idempotents of \myClg{}{}{\R^{n,n}} tailored to our needs while in section~\ref{sec_SAT_in_Cl} we apply these results to give a neat encoding of \SAT{} in \myClg{}{}{\R^{n,n}} together with a first unsatisfiability condition. To prepare the \SAT{} formulation in a more specific setting in section~\ref{sec_O(n)_MTNS} we introduce, with a simple formalism, the isomorphism between the set of all totally null subspaces of maximal dimension $n$ of $\R^{n,n}$ and the continuous group \OO{n}. In central section~\ref{sec_SAT_in_O(n)} we extend this isomorphism to simple spinors of $\myClg{}{}{\R^{n,n}}$ and we can transform a \SAT{} problem into the problem of forming a cover for $\OO{n}$ that gives a continuous formulation of \SAT{} \cite{Budinich_2019}. In final, long, section~\ref{sec_Simple_Spinors_SAT}, subdivided in six parts, we apply this continuous formulation to outline an unsatisfiability test that, exploiting the linear space properties of simple spinors of \myClg{}{}{\R^{n,n}}, ultimately induces a generalization of the resolution algorithm. This test is not combinatorial and a simple analysis indicates that it is polynomial.

For the convenience of the reader we tried to make this paper as elementary and self-contained as possible.

\section{The Boolean Satisfiability problem}
\label{prop_SAT_basics}
The Boolean Satisfiability Problem \cite[Section~7.2.2.2]{Knuth_2015} seeks an assignment of $n$ Boolean variables $\mylitrl_i \in \{ \myBooleanT, \myBooleanF \}$ (true, false), that makes $\myBooleanT$, \emph{satisfies}, a given Boolean formula $\myBooleanS$ put in Conjunctive Normal Form (CNF) \eg
%$$
%(\mylitrl_1 \lor \lnot \mylitrl_2) \land (\mylitrl_2 \lor \mylitrl_3) \land ( \lnot \mylitrl_1 \lor \lnot \mylitrl_3) \land ( \lnot \mylitrl_1 \lor \lnot \mylitrl_2 \lor \mylitrl_3) \land (\mylitrl_1 \lor \mylitrl_2 \lor \lnot \mylitrl_3)
%$$
\begin{equation}
\label{formula_SAT_std}
\myBooleanS \equiv (\mylitrl_1 \lor \myconjugate{\mylitrl}_2) \land (\mylitrl_2 \lor \mylitrl_3) \land ( \myconjugate{\mylitrl}_1 \lor \myconjugate{\mylitrl}_3) \land ( \myconjugate{\mylitrl}_1 \lor \myconjugate{\mylitrl}_2 \lor \mylitrl_3) \land (\mylitrl_1 \lor \mylitrl_2 \lor \myconjugate{\mylitrl}_3)
\end{equation}
as a logical AND ($\land$) of $m$ \emph{clauses} ${\cal C}_j$, the expressions in parenthesis, each clause being composed by the logical OR ($\lor$) of $k$ or less \emph{literals} (a variable or its complement)
\opt{margin_notes}{\mynote{mbh.note: for literal vs Boolean variable see Knuth \cite[Section~7.2.2.2, p. 2]{Knuth_2015} ... ``A literal is either a variable or the complement of a variable.''}}%
possibly preceded by logical NOT ($\lnot\mylitrl_i$, $\myconjugate{\mylitrl}_i$ for short). In (\ref{formula_SAT_std}) $n = 3, m = 5$ and $k = 3$. To underline the difference with algebraic equality $=$ in what follows we use $\equiv$ to represent logical equivalence, namely that for all possible values taken by the Boolean variables the two expressions are equal. A \emph{solution} is either an assignment of $\mylitrl_i$ that gives $\myBooleanS \equiv \myBooleanT$ or a proof that such an assignment does not exist and $\myBooleanS \equiv \myBooleanF$ in all cases.

\SAT{} was the first combinatorial problem proven to be NP-complete \cite{Cook_1971}; in particular while the case of $k = 3$, 3\SAT, can be solved only in a time that grows exponentially with $n$, $2\SAT$ and $1\SAT$ problems can be solved in polynomial time, that is \emph{fast}.%

Using the distributive properties of the logical operators $\lor, \land$ any given $k\SAT$ $\myBooleanS$ expands in a logical OR of up to $k^m$ terms each term being a $1\SAT$ problem namely a logical AND of $m$ Boolean variables. Since $\mylitrl_i \land \myconjugate{\mylitrl}_i \equiv \myBooleanF$ the presence of a literal together with its logical complement is a necessary and sufficient condition for making a $1\SAT$ formula $\myBooleanF$, namely \emph{unsatisfiable}, and thus these terms can be omitted and so $k^m$ is just an upper bound to the number of terms. Conversely a satisfiable $1\SAT$ formula has only one assignment of its Boolean variables that makes it $\myBooleanT$ and that can be read scanning the formula; thus in the sequel we will freely use $1\SAT$ formulas for assignments.

The final expanded expression can be further simplified and reordered exploiting the commutativity of the logical operators $\lor, \land$ and the properties $\mylitrl_i \land \mylitrl_i \equiv \mylitrl_i \lor \mylitrl_i \equiv \mylitrl_i$. All ``surviving'' terms of this expansion, the Disjunctive Normal Form (DNF), are $1\SAT$ terms, each of them representing an assignment that satisfies the problem. On the contrary if the DNF is empty, as happens for (\ref{formula_SAT_std}), this is a proof that there are no assignments that make the formula $\myBooleanT$: the problem is unsatisfiable.

Expansion to DNF is a dreadful algorithm for solving \SAT{}: first of all the method is an overkill since it produces all possible solutions whereas one would be enough; in second place this brute force approach gives a running time proportional to the number of expansion terms \bigO{(k^{\frac{m}{n}})^n} whereas modern $\SAT$ solvers run in \bigO{1.307^n} \cite{PaturiPudlakSaksZane_2005}. Nevertheless DNF will play a central role in the formulation of \SAT{} in Clifford algebra.
%
%We will write a \SAT{} problem $\myBooleanS$ in a more concise form as
%\begin{equation}
%\label{formula_SAT_std}
%\myBooleanS \equiv (\mylitrl_i \lor \mylitrl_j \lor \cdots \lor \mylitrl_h) (\mylitrl_l \lor \mylitrl_o \lor \cdots \lor \mylitrl_p) \cdots (\mylitrl_t \lor \mylitrl_u \lor \cdots \lor \mylitrl_z) \quad \mylitrl_i \in \{\mylitrl_i, \myconjugate{\mylitrl}_i \}
%\end{equation}
%where $\mylitrl_i$ is a \emph{Boolean variable} that stands for Boolean variable $\mylitrl_i$ or its complement, $\myconjugate{x}_i$ for short, and the ordinary product stands for logical AND.

\section{$\R^{n,n}$ and its Clifford algebra}
\label{sec_neutral_R_nn}
We review some properties of $\R^{n,n}$ and of its Clifford algebra $\myClg{}{}{\R^{n,n}}$ that are at the heart of the following results.

$\myClg{}{}{\R^{n,n}}$ is isomorphic to the algebra of real matrices $\R(2^n)$ \cite{Porteous_1995} and this algebra is more easily manipulated exploiting the properties of its Extended Fock basis (EFB, see \cite{Budinich_2016} and references therein) with which any algebra element is a linear superposition of simple spinors. The $2 n$ generators of the algebra $\mygen_{i}$ form an orthonormal basis of the linear space $\R^{n,n}$
\begin{equation}
\label{formula_generators}
%\mygen_i \mygen_j + \mygen_j \mygen_i := \anticomm{\mygen_i}{\mygen_j} = 2 \delta_{i j} (-1)^{i+1} \qquad i,j = 1,2, \ldots, 2 n
\mygen_i \mygen_j + \mygen_j \mygen_i := \anticomm{\mygen_i}{\mygen_j} = 2 \left\{ \begin{array}{l l}
- \delta_{i j} & \mbox{for} \; i \le n \\
\delta_{i j} & \mbox{for} \; i > n
\end{array} \right.
\qquad i,j = 1,2, \ldots, 2 n
\end{equation}
and we define the Witt, or null, basis of $\R^{n,n}$:
\begin{equation}
\label{formula_Witt_basis}
\left\{ \begin{array}{l l l}
p_{i} & = & \frac{1}{2} \left( \mygen_{i} + \mygen_{i + n} \right) \\
q_{i} & = & \frac{1}{2} \left( \mygen_{i} - \mygen_{i + n} \right)
\end{array} \right.
%\left\{ \begin{array}{l l l}
%p_{i} & = & \frac{1}{2} \left( \mygen_{2i-1} + \mygen_{2i} \right) \\
%q_{i} & = & \frac{1}{2} \left( \mygen_{2i-1} - \mygen_{2i} \right)
%\end{array} \right.
%\Rightarrow
%\left\{\begin{array}{l l l}
%\mygen_{2i-1} & = & p_{i} + q_{i} \\
%\mygen_{2i} & = & p_{i} - q_{i}
%\end{array} \right.
\quad i = 1,2, \ldots, n
\end{equation}
that, with $\mygen_{i} \mygen_{j} = - \mygen_{j} \mygen_{i}$ for $i \ne j$, gives
\begin{equation}
\label{formula_Witt_basis_properties}
\anticomm{p_{i}}{p_{j}} = \anticomm{q_{i}}{q_{j}} = 0
\qquad
\anticomm{p_{i}}{q_{j}} = \delta_{i j}
\end{equation}
showing that all $p_i, q_i$ are mutually orthogonal, also to themselves, that implies $p_i^2 = q_i^2 = 0$ and are thus null vectors.
\opt{margin_notes}{\mynote{mbh.note: should I mention here $\R^{2 n}_{hb}$ ? See book 780}}%
Defining
\begin{equation}
\label{formula_Witt_decomposition}
\left\{ \begin{array}{l}
P = \my_span{p_1, p_2, \ldots, p_n} \\
Q = \my_span{q_1, q_2, \ldots, q_n}
\end{array} \right.
\end{equation}
$P$ and $Q$ are two totally null subspaces of maximum dimension $n$ and form a Witt decomposition \cite{Porteous_1995} of $\R^{n,n}$ since $P \cap Q = \{ 0 \}$ and $P \oplus Q = \R^{n,n}$.

The $2^{2 n}$ simple spinors forming EFB are given by all possible sequences
\begin{equation}
\label{formula_EFB_def}
\psi = \psi_1 \psi_2 \cdots \psi_i \cdots \psi_n \qquad \psi_i \in \{ q_i p_i, p_i q_i, p_i, q_i \} \qquad i = 1, 2, \ldots, n
\end{equation}
where each $\psi_i$ takes one of its $4$ possible values \cite{Budinich_2016} and
\opt{margin_notes}{\mynote{mbh.note: should I use old version with $\mygen_{2 i - 1} \mygen_{2 i} = \comm{q_i}{p_i}$ ?}}%
each $\psi_i$ is uniquely identified by two ``bits'' $h_i, g_i = \pm1$: $h_i = 1$ if the leftmost vector of $\psi_i$ is $q_i$, $-1$ otherwise; $g_i = 1$ if $\psi_i$ is even, $-1$ if odd. The $h$ and $g$ \emph{signatures} of $\psi$ are respectively the vectors $(h_1, h_2, \ldots, h_n)$ and $(g_1, g_2, \ldots, g_n)$.

Since $\mygen_{i} \mygen_{i + n} = q_i p_i - p_i q_i := \comm{q_i}{p_i}$ in EFB the identity $\Identity$ and the volume element $\omega$ (scalar and pseudoscalar) assume similar expressions \cite{Budinich_2016}:
\begin{equation}
\label{formula_identity_omega}
\begin{array}{l l l}
\Identity & := & \anticomm{q_1}{p_1} \anticomm{q_2}{p_2} \cdots \anticomm{q_n}{p_n} \\
\omega & := & \mygen_1 \mygen_2 \cdots \mygen_{2 n} = (-1)^\frac{n(n-1)}{2} \comm{q_1}{p_1} \comm{q_2}{p_2} \cdots \comm{q_n}{p_n}
\end{array}
\end{equation}
%With (\ref{formula_Witt_basis_properties}) and $q_i p_i + p_i q_i = 1$ we easily obtain
%$$
%q_i p_i \, q_i p_i = q_i p_i \quad p_i q_i \, p_i q_i = p_i q_i \quad q_i p_i \, p_i q_i = p_i q_i \, q_i p_i = 0 \quad q_i p_i \, q_j p_j = q_j p_j \, q_i p_i
%$$
%that shows that $q_i p_i$ and $p_i q_i$ are part of a family of orthogonal, commuting, idempotents. Since $\myClg{}{}{\R^{n,n}}$ is a simple algebra, the unit element of the algebra is the sum of $2^n$ primitive (indecomposable) idempotents $\myprimidemp_i$
and since $\myClg{}{}{\R^{n,n}}$ is a simple algebra, the algebra identity is also the sum of its $2^n$ primitive (indecomposable) idempotents $\myprimidemp_i$ we gather in set $\myprimidempset$
\begin{equation}
\label{formula_identity_def}
\Identity = \sum_{i = 1}^{2^n} \myprimidemp_i %= \prod_{j = 1}^{n} \anticomm{q_j}{p_j}
\qquad \myprimidemp_i \in \myprimidempset \dotinformula
\end{equation}
Comparing the two expressions of $\Identity$ we observe that the full expansion of the anticommutators of (\ref{formula_identity_omega}) contains $2^n$ terms each term being one of the primitive idempotents \emph{and} a simple spinor. $\myprimidempset$ is thus a proper subset of EFB (\ref{formula_EFB_def}) and its elements are
\begin{equation}
\label{formula_primitive_idempotents_def}
\myprimidemp = \psi_1 \psi_2 \cdots \psi_i \cdots \psi_n \qquad \psi_i \in \{ q_i p_i, p_i q_i \} \qquad i = 1, 2, \ldots, n \dotinformula
\end{equation}
We recall the standard properties of primitive idempotents
\begin{equation}
\label{formula_primitive_idempotents}
\myprimidemp_i^2 = \myprimidemp_i \quad (\Identity - \myprimidemp_i)^2 = \Identity - \myprimidemp_i \quad \myprimidemp_i (\Identity - \myprimidemp_i) = 0 \quad \myprimidemp_i \myprimidemp_j = \delta_{i j} \myprimidemp_i
\end{equation}
and define the set
\begin{equation}
\label{formula_S_def}
\mysetS := \left\{ \sum_{i = 1}^{2^n} \delta_i \myprimidemp_i : \delta_i \in \left\{0, 1 \right\}, \myprimidemp_i \in \myprimidempset \right\}
\end{equation}
in one to one correspondence with the power set of $\myprimidempset$. $\mysetS$ is closed under Clifford product but not under addition and is thus not even a subspace. With (\ref{formula_primitive_idempotents}) we easily prove
\begin{MS_Proposition}
\label{prop_S_properties}
For any $s \in \mysetS$ then $s^2 = s$.
\end{MS_Proposition}
\noindent $\mysetS$ is thus the set of the idempotents, in general not primitive; a simple consequence is that for any $s \in \mysetS$ also $(\Identity - s) \in \mysetS$.
%$\mysetS$ properties are easier to grasp observing that $\delta_i \in \left\{0, 1 \right\}$, the only idempotents of $\R$.

\bigskip

Any EFB element of (\ref{formula_EFB_def}) is a simple spinor, uniquely identified in EFB by its $h$ signature, while the minimal left ideal, or spinor space $\mySpinorS$, to which it belongs is identified by its $h \circ g = (h_1 g_1, h_2 g_2, \ldots, h_n g_n)$ signature \cite{Budinich_2016}. The algebra, as a linear space, is the direct sum of these $2^n$ spinor spaces that, in isomorphic matrix algebra $\R(2^n)$, are usually associated to linear spaces of matrix columns.

For each of these $2^n$ spinor spaces $\mySpinorS$ its $2^n$ simple spinors (\ref{formula_EFB_def}) \cite{Budinich_2016} form a Fock basis $\myFockB$ and any spinor $\psi \in \mySpinorS$ is a linear combination of the simple spinors $\psi_\lambda \in \myFockB$ \cite{BudinichP_1989, Budinich_2016}, namely
\begin{equation}
\label{formula_Fock_basis_expansion}
\psi = \sum_\lambda \alpha_\lambda \psi_\lambda \qquad \alpha_\lambda \in \R, \quad \psi_\lambda \in \myFockB \dotinformula
\end{equation}
We illustrate this with the simplest example in $\R^{1,1}$, the familiar Minkowski plane of physics, here $\myClg{}{}{\R^{1,1}} \myisom \R(2)$ and the EFB (\ref{formula_EFB_def}) is formed by just 4 elements: $\{ qp_{+ +}, pq_{- -}, p_{- +}, q_{+ -} \}$ with the subscripts indicating respectively $h$ and $h \circ g$ signatures that give the binary form of the integer matrix indexes; its EFB matrix is
\begin{equation*}
\label{formula_EFB_1_1}
\bordermatrix{& + & - \cr
+ & q p & q \cr
- & p & p q \cr }
\end{equation*}
and, as anticipated, we can write the generic element $\mu \in \myClg{}{}{\R^{1,1}}$ in EFB
$$
\mu = \xi_{+ +} qp_{+ +} + \xi_{- -} pq_{- -} + \xi_{- +} p_{- +} + \xi_{+ -} q_{+ -} \qquad \xi \in \R \dotinformula
$$
The two columns are two minimal left ideals namely two (equivalent) spinor spaces $\mySpinorS_{+}$ and $\mySpinorS_{-}$. The two elements of each column are the simple spinors of Fock basis $\myFockB$ while $qp$ and $pq$ are the primitive idempotents and $qp + pq = \Identity$.

In turn simple spinors of a Fock basis $\myFockB$ are in one to one correspondence with $\R^{n,n}$ null subspaces of maximal dimension $n$. For any $\psi \in \myFockB$ we define its associated maximal null subspace $M(\psi)$ as
\begin{equation}
\label{formula_MTNS_M_psi}
M(\psi) = \my_span{x_1, x_2, \ldots, x_n} \quad x_i = \left\{ \begin{array}{l l}
p_i & \mbox{iff} \; \psi_i = p_i, p_i q_i \\
q_i & \mbox{iff} \; \psi_i = q_i, q_i p_i
\end{array} \right.
\quad i = 1,2, \ldots, n
\end{equation}
\noindent and $x_i$ is determined by the $h$ signature of $\psi$ in EFB \cite{BudinichP_1989, Budinich_2016}. For example in $\myClg{}{}{\R^{3,3}}$ given the simple spinor $\psi = p_1 q_1 \, q_2 p_2 \, q_3 p_3$
$$
\psi = p_1 q_1 \, q_2 p_2 \, q_3 p_3 \quad \implies \quad M(\psi) = \my_span{p_1, q_2, q_3}
$$
and with (\ref{formula_Witt_basis_properties}) we see that for any $v \in M(\psi)$ then $v \psi = 0$.

We gather these $2^n$ maximal null subspaces of $\R^{n,n}$ in set \mysetM{} each of its elements being the span of the $n$ null vectors obtained choosing one null vector from each couple $(p_i, q_i)$ (\ref{formula_Witt_basis}). The set \mysetM{} is the same for all the $2^n$ different possible spinor spaces being identified by the $h$ signatures of $\psi$ in EFB and so it can be defined also starting from primitive idempotents (\ref{formula_primitive_idempotents_def}) and in summary we can give three equivalent definitions for \mysetM{}
\begin{equation}
\label{formula_Mn}
\mysetM = \left\{ \begin{array}{l l}
\{ M(\psi) : \psi \in \myFockB \} \\
\{ \my_span{x_1, x_2, \ldots, x_n} : x_i \in \{ p_i, q_i \} \} \\
\{ M(\myprimidemp) : \myprimidemp \in \myprimidempset \} \dotinformula
\end{array} \right.
\end{equation}

\section{The Boolean algebra of \myClg{}{}{\R^{n,n}}}
\label{sec_Boole_in_Cl}
We exploit the known fact that in any associative, unital, algebra every family of commuting, orthogonal, idempotents generates a Boolean algebra to prove that the $2^{2^n}$ idempotents of $\mysetS$ (\ref{formula_S_def}) form a Boolean algebra.

A finite Boolean algebra is a set equipped with the inner operations of logical AND, OR and NOT that satisfy well known properties
but we will use an axiomatic definition \cite{Huntington_1933, Givant_Halmos_2009} that needs only a binary and a unary inner operations satisfying three simple axioms to prove:
\begin{MS_Proposition}
\label{prop_Boolean_algebra}
The set $\mysetS$ equipped with the two inner operations
\begin{equation}
\label{formula_S_properties}
\begin{array}{llll}
%\mathrm{binary} & \to & \mysetS \times \mysetS \to \mysetS; s_1, s_2 \to s_1 s_2 \quad \mathrm{(Clifford \; product)} \\
%\mathrm{unary} & \to & \mysetS \to \mysetS; s \to \Identity - s \\
\mysetS \times \mysetS \to \mysetS & s_1, s_2 & \to & s_1 s_2 %\quad \mathrm{(Clifford \; product)}
\\
\mysetS \to \mysetS & s & \to & \Identity - s \\
\end{array}
\end{equation}
is a finite Boolean algebra.
\end{MS_Proposition}
\begin{proof}
We already observed that $\mysetS$ is closed under operations (\ref{formula_S_properties}) that moreover satisfy Boolean algebra axiomatic definition \cite{Huntington_1933}: the binary operation is associative since Clifford product is and commutative because all $\mysetS$ elements commute. The third (Huntington's) axiom requires that for any $s_1, s_2 \in \mysetS$
$$
(\Identity - (\Identity - s_1) s_2) (\Identity - (\Identity - s_1) (\Identity - s_2)) = s_1
$$
that is easily verified.
\end{proof}
\noindent We remark that $\mysetS$ is not a subalgebra of \myClg{}{}{\R^{n,n}} since it is not closed under addition. Any finite Boolean algebra is isomorphic to the power set of its Boolean atoms \cite{Givant_Halmos_2009}. In this case $\mysetS$ elements are in one to one correspondence with the power set of $\myprimidempset$ and we thus identify the Boolean atoms with the $2^n$ primitive idempotents (\ref{formula_primitive_idempotents_def}).

With simple manipulations we get all Boolean expressions in \myClg{}{}{\R^{n,n}}: in the unary operation of (\ref{formula_S_properties}) we recognize the logical NOT and associating the logical AND to Clifford product from $s (\Identity - s) = 0 \in \mysetS$ we deduce that $0$ stands for $\myBooleanF$ and consequently that $\Identity$ stands for $\myBooleanT$. For the logical OR we use De Morgan's relations
$$
\mylitrl_1 \lor \mylitrl_2 \equiv \myconjugate{\myconjugate{\mylitrl}_1 \land \myconjugate{\mylitrl}_2} \quad \to \quad \Identity - (\Identity - s_1) (\Identity - s_2) = s_1 + s_2 - s_1 s_2
$$
and we can easily verify that $\mysetS$ is closed also under this binary operation.

We formulate Boolean expressions in \myClg{}{}{\R^{n,n}} associating literals $\mylitrl_i$ to idempotents and we gather associations in this table where $p_i$ and $q_i$ are vectors of the Witt basis (\ref{formula_Witt_basis})
\begin{equation}
\label{formula_Boolean_substitutions}
\begin{array}{lll}
\myBooleanF & \to & 0 \\
\myBooleanT & \to & \Identity \\
\mylitrl_i & \to & q_i p_i \\
%L_1 \land L_2 \to l_1 l_2 \\
%\lnot \mylitrl_i \equiv
\myconjugate{\mylitrl}_i & \to & \Identity - q_i p_i = p_i q_i \\
\mylitrl_i \land \mylitrl_j & \to & q_i p_i \; q_j p_j \\
\mylitrl_i \lor \mylitrl_j & \to & q_i p_i + q_j p_j - q_i p_i \; q_j p_j \dotinformula
\end{array}
\end{equation}
For example given some simple Boolean expressions with (\ref{formula_Witt_basis_properties}) we easily verify
$$
%\begin{array}{rclrclrccclrcl}
%\mylitrl_i \land \mylitrl_i & \equiv & \mylitrl_i & \myconjugate{\mylitrl}_i \land \myconjugate{\mylitrl}_i & \equiv & \myconjugate{\mylitrl}_i & \mylitrl_i \land \myconjugate{\mylitrl}_i & \equiv & \myconjugate{\mylitrl}_i \land \mylitrl_i & \equiv & \myBooleanF & \mylitrl_i \land \mylitrl_j & \equiv & \mylitrl_j \land \mylitrl_i \\
%q_i p_i \; q_i p_i & = & q_i p_i & p_i q_i \; p_i q_i & = & p_i q_i & q_i p_i \; p_i q_i & = & p_i q_i \; q_i p_i & = & 0 & q_i p_i \; q_j p_j & = & q_j p_j \; q_i p_i \\
%\end{array}
\begin{array}{lll}
\mylitrl_i \land \mylitrl_i \equiv \mylitrl_i & \to & q_i p_i \; q_i p_i = q_i p_i \\
\myconjugate{\mylitrl}_i \land \myconjugate{\mylitrl}_i \equiv \myconjugate{\mylitrl}_i & \to & p_i q_i \; p_i q_i = p_i q_i \\
\mylitrl_i \land \myconjugate{\mylitrl}_i \equiv \myconjugate{\mylitrl}_i \land \mylitrl_i \equiv \myBooleanF & \to & q_i p_i \; p_i q_i = p_i q_i \; q_i p_i = 0 \\
\mylitrl_i \land \mylitrl_j \equiv \mylitrl_j \land \mylitrl_i & \to & q_i p_i \; q_j p_j = q_j p_j \; q_i p_i
\end{array}
$$
and from now on we will use $\mylitrl_i$ and $\myconjugate{\mylitrl}_i$ also in $\myClg{}{}{\R^{n,n}}$ meaning respectively $q_i p_i$ and $p_i q_i$ and Clifford product will stand for logical AND $\land$%
\opt{margin_notes}{\mynote{mbh.note: here there is a commented repetition of $1\SAT$ of section 2}}%
%
%
%
%
%A $1\SAT$ problem is just a logical AND of $m$ Boolean variables. For both assignments of $\mylitrl_i$, $\mylitrl_i \land \myconjugate{\mylitrl}_i \equiv \myBooleanF$ and thus the presence of a Boolean variable together with its logical complement is a necessary and sufficient condition for making a $1\SAT$ formula unsatisfiable. We can interpret a satisfiable $1\SAT$ formula as an assignment of variables since there is only one assignment that makes it $\myBooleanT$ and that can be read scanning the formula; in the sequel we will freely use $1\SAT$ formulas for assignments and we will switch between the two forms as and when it suits us. We also easily see that $1\SAT$ formulas are idempotents.%
%
{} and in full generality we can prove \cite{Budinich_2017}
\begin{MS_Proposition}
\label{prop_logical_formulas_in_S}
Any Boolean expression $\myBooleanS$ with $n$ Boolean variables is represented in $\myClg{}{}{\R^{n,n}}$ by $S \in \mysetS$ obtained with substitutions (\ref{formula_Boolean_substitutions}) moreover $\myconjugate{\myBooleanS}$ is represented by $\Identity - S$ both being idempotents of $\myClg{}{}{\R^{n,n}}$. Given another Boolean expression ${\cal Q}$ the logical equivalence $\myBooleanS \equiv {\cal Q}$ holds if and only if $S = Q$ for their respective idempotents in $\myClg{}{}{\R^{n,n}}$.
\end{MS_Proposition}
\noindent In summary with substitutions (\ref{formula_Boolean_substitutions}) we can safely encode any Boolean expression, and thus \SAT{} problems, in Clifford algebra.

\section{\SAT{} in Clifford algebra \myClg{}{}{\R^{n,n}}}
\label{sec_SAT_in_Cl}

The straightest way of encoding a \SAT{} problem in CNF (\ref{formula_SAT_std}) in Clifford algebra is exploiting De Morgan relations to rewrite its clauses as
$$
{\cal C}_j \equiv (\mylitrl_{j_1} \lor \mylitrl_{j_2} \lor \cdots \lor \mylitrl_{j_k}) \equiv \myconjugate{\myconjugate{\mylitrl}_{j_1} \myconjugate{\mylitrl}_{j_2} \cdots \myconjugate{\mylitrl}}_{j_k}
$$
and thus the expression of a clause in Clifford algebra is
\begin{equation}
\label{formula_clause}
{\cal C}_j \to \Identity - \myconjugate{\mylitrl}_{j_1} \myconjugate{\mylitrl}_{j_2} \cdots \myconjugate{\mylitrl}_{j_k} := \Identity - z_j
\end{equation}
and the expression of a \SAT{} problem in CNF with $m$ clauses is
\begin{equation}
\label{formula_SAT_EFB_2}
S = \prod_{j = 1}^m (\Identity - z_j)
\end{equation}
and from Proposition~\ref{prop_logical_formulas_in_S} easily descends
\begin{MS_Proposition}
\label{prop_SAT_in_Cl_2}
Given a \SAT{} problem $\myBooleanS$ then $\myBooleanS \equiv \myBooleanF$ if and only if, for the corresponding algebraic expression in $\myClg{}{}{\R^{n,n}}$ (\ref{formula_SAT_EFB_2}) $S = 0$
\end{MS_Proposition}
\noindent that transforms a Boolean problem in an algebraic one. To master the implications of (\ref{formula_SAT_EFB_2}) we need the full expression of a $1\SAT$ formula \eg $\mylitrl_1 \myconjugate{\mylitrl}_2$ namely $q_1 p_1 \; p_2 q_2$: with (\ref{formula_identity_omega}), (\ref{formula_identity_def}) and (\ref{formula_primitive_idempotents_def})
\begin{equation}
\label{formula_literal_projection}
q_1 p_1 \; p_2 q_2 = q_1 p_1 \; p_2 q_2 \; \Identity = q_1 p_1 \; p_2 q_2 \prod_{j = 3}^{n} \anticomm{q_j}{p_j}
\end{equation}
since $q_1 p_1 \anticomm{q_1}{p_1} = q_1 p_1$ and $p_2 q_2 \anticomm{q_2}{p_2} = p_2 q_2$ and the full expansion of this expression is the sum of $2^{n - 2}$ primitive idempotents $\myprimidemp$ (\ref{formula_primitive_idempotents_def}) and thus $q_1 p_1 \; p_2 q_2$ is an idempotent of $\mysetS$. From the Boolean standpoint this can be interpreted as the property that given the $1\SAT$ formula $\mylitrl_1 \myconjugate{\mylitrl}_2$ the other, unspecified, $n-2$ literals $\mylitrl_3, \ldots, \mylitrl_n$ can take all possible $2^{n - 2}$ values or, more technically, that $\mylitrl_1 \myconjugate{\mylitrl}_2$ has a \emph{full} DNF made of $2^{n - 2}$ Boolean atoms.

More in general any $1\SAT$ formula with $m$ literals is a sum of $2^{n - m}$ primitive idempotents, namely Boolean atoms. With (\ref{formula_Boolean_substitutions}) we can rewrite (\ref{formula_primitive_idempotents_def}) as
$$
\myprimidemp = \psi_1 \psi_2 \cdots \psi_i \cdots \psi_n \qquad \psi_i \in \{ \mylitrl_i, \myconjugate{\mylitrl}_i \} \qquad i = 1, 2, \ldots, n
$$
showing that the $2^n$ primitive idempotents $\myprimidemp$ are just the possible $2^n$ $1\SAT$ formulas with $n$ literals, the Boolean atoms, for example:
\begin{equation}
\label{formula_atoms_primitive}
\mylitrl_1 \myconjugate{\mylitrl}_2 \mylitrl_3 \cdots \mylitrl_n \to q_1 p_1 \; p_2 q_2 \; q_3 p_3 \cdots q_n p_n \in \myprimidempset \dotinformula
\end{equation}
%
% here there is the substitution of an assignment in S
%Given an assignment of its $n$ Boolean variables, \eg $\mylitrl_1 \myconjugate{\mylitrl}_2 \cdots \mylitrl_n$, $\myBooleanS \equiv \myBooleanF$ if and only if $\mylitrl_1 \myconjugate{\mylitrl}_2 \cdots \mylitrl_n \land \myBooleanS \equiv \myBooleanF$, becoming $\mylitrl_1 \myconjugate{\mylitrl}_2 \cdots \mylitrl_n \, S= 0$ in Clifford algebra. We interpret these formulas as the \emph{substitution} of the only assignment satisfying $\mylitrl_1 \myconjugate{\mylitrl}_2 \cdots \mylitrl_n$ into $S$. By (\ref{formula_SAT_EFB_2}) $\mylitrl_1 \myconjugate{\mylitrl}_2 \cdots \mylitrl_n \, S = 0$ if and only if there exists a clause $z_j$ such that $\mylitrl_1 \myconjugate{\mylitrl}_2 \cdots \mylitrl_n (\Identity - z_j) = 0$ namely $\mylitrl_1 \myconjugate{\mylitrl}_2 \cdots \mylitrl_n = \mylitrl_1 \myconjugate{\mylitrl}_2 \cdots \mylitrl_n z_j$ \cite{Budinich_2017}.
%
By Proposition~\ref{prop_logical_formulas_in_S} $S \in \mysetS$ (\ref{formula_S_def}) and is thus the sum of primitive idempotents (\ref{formula_primitive_idempotents_def}) that now we know represent Boolean atoms and ultimately (\ref{formula_SAT_EFB_2}) gives the full DNF expansion of the \SAT{} problem $S$ each term being one assignment that makes the problem $\myBooleanT$ while if the expansion is empty the problem is unsatisfiable and thus expansion of the CNF $S$ of (\ref{formula_SAT_EFB_2}) reproduces faithfully the Boolean expansion to DNF outlined in section~\ref{prop_SAT_basics}.

From the computational side Proposition~\ref{prop_SAT_in_Cl_2} is not a big deal since the expansion of (\ref{formula_SAT_EFB_2}) corresponds to the DNF expansion that in section~\ref{prop_SAT_basics} we named a ``dreadful'' algorithm. But porting \SAT{} to Clifford algebra offers other advantages since we can exploit algebra properties. For example the unsatisfiability condition $S = 0$ makes $S$ a scalar whereas if satisfiable $S$ is not a scalar. Exploiting scalar properties in Clifford algebra we proved \cite{Budinich_2017}
\begin{MS_theorem}
\label{theorem_SAT_unsat_thm}
A given nonempty \SAT{} problem in $\myClg{}{}{\R^{n,n}}$ (\ref{formula_SAT_EFB_2}) is unsatisfiable ($S = 0$) if and only if, for all generators (\ref{formula_generators}) of $\myClg{}{}{\R^{n,n}}$
\begin{equation}
\label{formula_SAT_symmetry}
\mygen_i \; S \; \mygen_i^{-1} = S \qquad \forall \; 1 \le i \le 2 n \dotinformula
\end{equation}
\end{MS_theorem}

\noindent This result gives an unsatisfiability test based on the symmetry properties of its CNF expression $S$ (\ref{formula_SAT_EFB_2}). We remark that as far as computational performances are concerned an efficient unsatisfiability test would bring along also an efficient solution algorithm. Suppose the test (\ref{formula_SAT_symmetry}) fails and thus that $S$ is satisfiable, to get an actual solution we choose a Boolean variable, \eg $\mylitrl_i$, and replace it with $\myBooleanT$ and apply again the test to the derived problem $S_i$. If the test on $S_i$ fails as well this means that $\mylitrl_i \equiv \myBooleanT$ otherwise, necessarily, $\mylitrl_i \equiv \myBooleanF$ and repeating this procedure $n$ times for all literals we obtain an assignment that satisfies $S$. The algorithmic properties of unsatisfiability test (\ref{formula_SAT_symmetry}) have been preliminarly explored in \cite{Budinich_2017}.

\bigskip

$\myClg{}{}{\R^{n,n}}$ epitomizes the geometry of linear space $\R^{n,n}$ and thus \SAT{} encoding (\ref{formula_SAT_EFB_2}) brings along also a geometric interpretation that is at the root of further encodings of \SAT{} in Clifford algebra.

$S$ (\ref{formula_SAT_EFB_2}) is ultimately a sum of primitive idempotents (\ref{formula_primitive_idempotents_def}) that are in one to one correspondence with the null maximal subspaces of \mysetM{} (\ref{formula_Mn}).

It follows that $S$, and more in general any $\mysetS$ element, induces a subset of \mysetM{}, the empty subset if $S = 0$. More precisely the elements of this subset are all and only those maximal totally null subspaces (\ref{formula_Mn}) corresponding to the Boolean atoms making $S$. For any $s \in \mysetS$ (\ref{formula_S_def}) let $I_s$ such that
\begin{equation}
\label{formula_J_s_def}
s = \sum_{i \in I_s} \myprimidemp_i \qquad I_s \subseteq \{ 1, 2, \ldots 2^n \}
\end{equation}
and so all $s \in \mysetS$ induce a subset of \mysetM{}
\begin{equation}
\label{formula_cal_T_s_def}
{\cal T}_s' := \{M(\myprimidemp_i) : i \in I_s \} \subseteq \mysetM \quad \implies \quad {\cal T}_{\Identity - s}' = \mysetM \setminus {\cal T}_s' \dotinformula
\end{equation}
Applying this definition to clauses idempotents (\ref{formula_clause}) Proposition~\ref{prop_SAT_in_Cl_2} becomes:
\begin{MS_Proposition}
\label{prop_SAT_in_M_n}
Given a \SAT{} problem $S$ in $\myClg{}{}{\R^{n,n}}$ (\ref{formula_SAT_EFB_2}) then the problem is unsatisfiable ($S = 0$) if and only if
\begin{equation}
\label{formula_SAT_in_M_n}
\cup_{j = 1}^m {\cal T}_{z_j}' = \mysetM \dotinformula
\end{equation}
\end{MS_Proposition}
\begin{proof}
For any $s_1, s_2 \in \mysetS$ from $\mysetS$ definition (\ref{formula_S_def}) we easily get
$$
{\cal T}_{s_1 s_2}' = {\cal T}_{s_1}' \cap {\cal T}_{s_2}'
$$
and in this setting Proposition~\ref{prop_SAT_in_Cl_2} $S$ states that $S = 0$ if and only if
\begin{equation}
\label{formula_SAT_in_T}
{\cal T}_S' = \cap_{j = 1}^m {\cal T}_{\Identity - z_j}' = \emptyset \dotinformula
\end{equation}
The thesis follows by (\ref{formula_cal_T_s_def}) and by elementary set properties.
\end{proof}
\noindent The \SAT{} problem has now the form of a problem of subsets of \mysetM{} that provides also an interpretation of (\ref{formula_SAT_EFB_2}). Since $z_j$ is the unique assignment of the $k$ literals of ${\cal C}_j$ that give ${\cal C}_j \equiv \myBooleanF$, and thus $S = 0$, then if the union of all these cases (\ref{formula_SAT_in_M_n}) covers \mysetM{} the problem is unsatisfiable. This in turn implies that any $M(\myprimidemp) \notin \cup_{j = 1}^m {\cal T}_{z_j}'$ is a solution of $S$.

Given $z_j$ with $k$ literals we define $M(z_j)$ as the $k$ dimensional null subspace (\ref{formula_MTNS_M_psi}) induced by the literals of $z_j$, and adopting the lighter notation ${\cal T}_{j}' $ for ${\cal T}_{z_j}'$, it is easy to see that
\begin{equation}
\label{formula_cal_T_j_def}
{\cal T}_j' := \{M(\myprimidemp_i) : M(z_j) \subseteq M(\myprimidemp_i) \} \subseteq \mysetM \dotinformula
\end{equation}

To proceed further we review the isomorphism between the set of all totally null subspaces of maximal dimension of $\R^{n,n}$ and the group $\OO{n}$.

\section{The orthogonal group $\OO{n}$ and the set \mysnqG{}}
\label{sec_O(n)_MTNS}
Let \mysnqG{} be the set of all totally null subspaces of maximal dimension $n$ of $\R^{n,n}$, a quadric Grassmannian for Ian Porteous \cite[Chapter~14]{Porteous_1995}. \mysnqG{} is isomorphic to subgroup $\OO{n}$ of $\OO{n,n}$ and $\OO{n}$ acts transitively on \mysnqG{}.

We review these relations: seeing the linear space $\R^{n,n}$ as $\R^n \times \R^n$ we can write its generic element as $(x,y)$ and $(x,y)^2 = - x^2 + y^2$. Any $n$ dimensional subspace of $\R^{n,n}$ may be represented as the image of an injective map $\R^n \to \R^n \times \R^n; x \to (s(x), t(x))$ for $s, t \in \GL{n}$. This subspace is made by all pairs $(s(x), t(x))$ and we denote it with $(s, t) \in \GL{n} \times \GL{n}$. By same mechanism, with $s = \Identity$ and $t \in \OO{n}$, for any $x \in \R^n$ $(x,t(x)) \in \R^{n,n}$ is a null vector since $(x,t(x))^2 = - x^2 + t(x)^2 = 0$ and it belongs to the $n$ dimensional null subspace $(\Identity, t)$. Given that from now on $s \equiv \Identity$ it is natural to identify $\R^n$ as $\R^n \times \{0\}$, the timelike subspace of $\R^{n,n}$, and we will do so unless differently specified.

Isometries (orthogonal transformations) $t \in \OO{n}$ establish the quoted isomorphism since any subspace $(\Identity, t) \subset \R^{n,n}$ is in \mysnqG{} and conversely any element of \mysnqG{} can be written as $(\Identity, t)$ \cite[Corollary~14.13]{Porteous_1995} and thus
\begin{equation}
\label{formula_Nn}
\mysnqG = \{ (\Identity, t) : t \in \OO{n} \}
\end{equation}
and the isomorphism between \mysnqG{} and $\OO{n}$ is realized by map
\opt{margin_notes}{\mynote{mbh.note: is this the Cayley chart ?}}%
\begin{equation}
\label{formula_bijection_Nn_O(n)}
\mysnqG \to \OO{n}; (\Identity, t) \to t \dotinformula
\end{equation}

For example, assuming that the map $(\Identity, \Identity): \R^n \to \R^n \times \R^n$ is such that $\mygen_i \to (\mygen_i, \mygen_{i + n})$, then two generic null vectors of $P$ and $Q$ (\ref{formula_Witt_decomposition}) are respectively $(x, x)$ and $(y, -y)$ and in this notation $P$ and $Q$ are thus
\opt{margin_notes}{\mynote{mbh.ref: see pp. 55, 56}}%
\begin{equation}
\label{formula_P_Q_def}
\left\{ \begin{array}{l}
P = (\Identity, \Identity) \\
Q = (\Identity, -\Identity) \dotinformula
\end{array} \right.
\end{equation}
The action of $\OO{n}$ is transitive on \mysnqG{} since for any $t, u \in \OO{n}$, $(\Identity,ut) \in \mysnqG$ and the action of $\OO{n}$ is trivially transitive on $\OO{n}$.

We examine isomorphism (\ref{formula_bijection_Nn_O(n)}) when restricted to subset $\mysetM \subset \mysnqG$ (\ref{formula_Mn}) taking $P = (\Identity, \Identity)$ as our ``reference'' element of \mysetM{}.
\opt{margin_notes}{\mynote{mbh.note: in physics language this corresponds to the choice of the vacuum spinor}}%
Let $\lambda_i \in \OO{n}$ be the hyperplane reflection inverting spacelike vector $\mygen_{i + n}$, namely
\begin{equation*}
\label{formula_t_i_def}
\lambda_i(\mygen_{j}) =
\left\{ \begin{array}{l l l}
-\mygen_{j} & \quad \mbox{for} \quad j = i + n\\
\mygen_{j} & \quad \mbox{otherwise}
\end{array} \right.
\quad i = 1,2, \ldots, n \quad j = n+1, n+2, \ldots, 2n
\end{equation*}
its action on the Witt basis (\ref{formula_Witt_basis}) exchanges the null vectors $p_i$ and $q_i$. Starting from $P$ (\ref{formula_Witt_decomposition}) we can get any other \mysetM{} element inverting a subset of the $n$ spacelike vectors $\mygen_{i + n}$. Each isometry $\lambda_i$ is represented, in the vectorial representation of $\OO{n}$, by a diagonal matrix $\lambda \in \R(n)$ with $\pm 1$ on the diagonal and these matrices form the group
$$
\OO{1} \times \OO{1} \cdots \times \OO{1} = \stackrel{n}{\times} \OO{1} := \O1{n}
$$
immediate to get since $\OO{1} = \{ \pm 1 \}$.%
\opt{margin_notes}{\mynote{mbh.ref: remember that the tensor product is associative \cite[Proposition~11.5]{Porteous_1995}; $\O1{n}$ is not normal since in general $t \lambda t^{-1} \notin \O1{n}$}}%
{} \O1{n} is a discrete, abelian, subgroup of involutions of $\OO{n}$, namely linear maps $t$ such that $t^2 = \Identity$. It is thus clear that since $P = (\Identity, \Identity)$ for any $M(\myprimidemp) \in \mysetM$ there exists a unique $\lambda \in \O1{n}$ such that
$$
M(\myprimidemp) = (\Identity, \lambda)
$$
and thus we proved constructively
\begin{MS_Proposition}
\label{prop_bijection_restricted}
Isomorphism (\ref{formula_bijection_Nn_O(n)}) restricted to $\mysetM \subset \mysnqG$ has for image subgroup \O1{n} of $\OO{n}$
\begin{equation}
\label{formula_bijection_Mn_O^n(1)}
\mysetM = \{ (\Identity, \lambda) : \lambda \in \O1{n} \} \qquad \implies \qquad \mysetM \to \O1{n}; (\Identity, \lambda) \to \lambda \dotinformula
\end{equation}
\end{MS_Proposition}
\noindent Given reference $P$ let $\psi_\Identity = p_1 q_1 \; p_2 q_2 \; p_3 q_3 \cdots p_n q_n$ be the \emph{reference} simple spinor, such that $M(\psi_\Identity) = P = (\Identity, \Identity)$, the vacuum spinor of physics, we resume concisely the action of $\lambda \in \O1{n}$ on spinors and vectors with (see \eg \cite{BudinichP_1989, Budinich_2016} for more extensive treatments)
\begin{equation}
\label{formula_lambda_action}
M(\lambda(\psi_\Identity)) := M(\psi_\lambda) = (\Identity, \lambda)
\end{equation}
and by the action of $\lambda$ we get respectively from $\psi_\Identity$ all spinors of the Fock basis $\myFockB$ and from null subspace $P$ all \mysetM{} elements.

Isomorphism (\ref{formula_bijection_Mn_O^n(1)}) adds a fourth definition of \mysetM{} (\ref{formula_Mn}) with which we can port \SAT{} within group $\O1{n}$ and, later, $\OO{n}$. We redefine ${\cal T}_s'$ (\ref{formula_cal_T_s_def}) as a subset of \O1{n}
\begin{equation}
\label{formula_cal_T_s_def2}
{\cal T}_s' := \{ \lambda \in \O1n : (\Identity, \lambda) = M(\myprimidemp_i), i \in I_s \} \subset \O1{n}
\end{equation}
and with this definition we can transform Proposition~\ref{prop_SAT_in_M_n} to
\begin{MS_Proposition}
\label{prop_SAT_in_On1}
Given a \SAT{} problem $S$ in $\myClg{}{}{\R^{n,n}}$ (\ref{formula_SAT_EFB_2}) then the problem is unsatisfiable ($S = 0$) if and only if
\begin{equation}
\label{formula_SAT_in_On1}
\cup_{j = 1}^m {\cal T}_{j}' = \O1{n}
\end{equation}
\end{MS_Proposition}
\noindent that gives the first formulation of \SAT{} problems in group language. From the computational point of view of there are no improvements since $\O1{n}$ is a discrete group and checking if subsets ${\cal T}_{j}'$ form a cover essentially requires testing all $2^n$ group elements, just the same as testing all $2^n$ Boolean atoms to see if any solves \SAT{}.

We resume all this in a commutative diagram in which numbers refer to formulas
$$
\begin{tikzcd}[column sep=small]
& \myprimidemp \in \myprimidempset \arrow[dl, Leftrightarrow, "(\ref{formula_Mn})" '] \arrow[dr, Leftrightarrow, "(\ref{formula_lambda_action})"] & \\
M(\myprimidemp)% = (\Identity, \lambda)
\in \mysetM \arrow[Leftrightarrow, "(\ref{formula_bijection_Mn_O^n(1)})"]{rr} & & \lambda \in \O1{n}
\end{tikzcd}
$$
that authorizes us, from now on, to freely switch between:
\begin{itemize}
\item the $2^n$ primitive idempotents $\myprimidemp$, or Boolean assignments (atoms) of $n$ literals,
\item the $2^n$ simple spinors of the Fock basis $\psi_\lambda \in \myFockB$,
\item the $2^n$ totally null subspaces of maximal dimension $M(\psi_\lambda) \in \mysetM$,
\item the $2^n$ elements of the discrete, abelian group $\lambda \in \O1{n}$.
%\item
\end{itemize}

\section{\SAT{} in orthogonal group $\OO{n}$}
\label{sec_SAT_in_O(n)}
Before going on we remind some basics properties of spinor space $\mySpinorS$, a minimal left ideal of $\myClg{}{}{\R^{n,n}}$, and the particular case of \emph{simple spinors} that we will indicate with $\mySpinorS_s \subset \mySpinorS$. Simple spinors \cite{Cartan_1937} are an elusive subject that rarely surfaces in recent literature, a noteworthy exception being \cite{BudinichP_1989}. In a nutshell: we saw that to $\psi_\lambda \in \myFockB$ are associated $M(\psi_\lambda) = (\Identity, \lambda) \in \mysetM{}$ and more in general simple spinors are those spinors that are associated to a null subspace of maximal dimension $n$, namely
\begin{equation}
\label{formula_simple_spinors}
\psi_t \in \mySpinorS \quad \mbox{such that} \quad M(\psi_t) = (\Identity, t) \in \mysnqG{} \qquad t \in \OO{n}
\end{equation}
this relation between simple spinors and \OO{n} being bijective \cite{BudinichP_1989}.

In the last step we show that, when problem $S$ is unsatisfiable, subsets induced by clauses not only form a cover of \O1{n} (\ref{formula_SAT_in_On1}) but also of its parent group $\OO{n}$ that opens new computational perspectives and we summarize here the needed results of \cite{Budinich_2019} to which the reader is addressed for a more exhaustive treatment. We start extending the definition of isometries induced by a clause (\ref{formula_cal_T_j_def}) with (\ref{formula_cal_T_s_def2}) to
\begin{equation}
\label{formula_cal_T_j_def2}
{\cal T}_j := \{t \in \OO{n} : M(z_j) \subseteq (\Identity, t)\} \subset \OO{n}
\end{equation}
this being an obvious generalization of (\ref{formula_cal_T_j_def}), moreover $\O1n \subset \OO{n}$ implies ${\cal T}_j' \subseteq {\cal T}_j$ and we can give \cite{Budinich_2019} different equivalent definitions for ${\cal T}_j$ \eg
$$
{\cal T}_j =
\left\{ \begin{array}{l}
\{ t \in \OO{n} : M(z_j) \subseteq (\Identity, t) \} \subset \OO{n} \\
\{ \psi_t \in \my_span{\psi_\lambda} : \lambda \in {\cal T}_j' \subset \myFockB \mbox{ and } \psi_t \in \mySpinorS_s \mbox{ with } M(\psi_t) = (\Identity, t) \} \subset \mySpinorS_s \\
\{ t \in \OO{n} : (\Identity, t) = M(\psi_t) \mbox{ for } \psi_t \mbox{ as above} \}
\end{array} \right.
$$
and moreover for any ${\cal T}_j$ then \cite[Lemma 1]{Budinich_2019}
\begin{equation}
\label{formula_cal_T_j_cal_T_j'}
{\cal T}_j \cap \O1n = {\cal T}_j' \dotinformula
\end{equation}

Given ${\cal T}_j$ definition as the span of the subset of the Fock basis given by ${\cal T}_j'$ and being spinor space $\mySpinorS$ a linear space, we can define the set%
\opt{margin_notes}{\mynote{mbh.note: not introduced $\myssSum$ here to maintain compatibility with \cite{Budinich_2019}}}%
\begin{equation}
\label{formula_cal_T_j+k_def}
{\cal T}_{j} + {\cal T}_{k} := \left\{ \psi = \alpha \psi_j + \beta \psi_k : \quad
\begin{array}{l}
\alpha, \beta \in \R \\
\psi_j \in {\cal T}_{j}, \quad \psi_k \in {\cal T}_{k}, \\
\mbox{such that} \quad \psi \in \mySpinorS_s
\end{array}
\right\}
\end{equation}
%that is commutative
%
%and (most probably not) associative.%
%
%and we will focus our attention on the elements of ${\cal T}_{j} + {\cal T}_{k}$ that are not in ${\cal T}_{j} \cup {\cal T}_{k}$.
%
with which \cite{Budinich_2019}

\begin{MS_theorem}
\label{theorem_SAT_in_O(n)}
A given \SAT{} problem $S$ in $\myClg{}{}{\R^{n,n}}$ (\ref{formula_SAT_EFB_2}) with $n$ literals is unsatisfiable if and only if the isometries induced by its $m$ clauses (\ref{formula_cal_T_j_def2}), (\ref{formula_cal_T_j+k_def}) form a cover for $\OO{n}$:
\begin{equation}
\label{formula_SAT_in_O(n)}
\sum_{j = 1}^m {\cal T}_j = \OO{n} \dotinformula
\end{equation}
\end{MS_theorem}

\noindent Here $\sum_{j = 1}^m {\cal T}_j$ does not imply an addition between sets ${\cal T}$'s but only stands for the set of spinors linear combinations of spinors taken from different ${\cal T}$'s.

We remark that this theorem is not a straightforward generalization of Proposition~\ref{prop_SAT_in_On1} and that definition (\ref{formula_cal_T_j+k_def}) is pivotal: replacing $\sum$ with $\cup$ in (\ref{formula_SAT_in_O(n)}) the result does not hold: %and it is easy to show unsatisfiable problems for which $\cup_{j = 1}^m {\cal T}_j \ne \OO{n}$ \cite{Budinich_2019}.
for example in the only unsatisfiable 2\SAT{} problem with $n = 2$:
\begin{equation}
\label{formula_2SAT_n2}
z_1 = \mylitrl_1 \mylitrl_2 \quad z_2 = \mylitrl_1 \myconjugate{\mylitrl}_2 \quad z_3 = \myconjugate{\mylitrl}_1 \mylitrl_2 \quad z_4 = \myconjugate{\mylitrl}_1 \myconjugate{\mylitrl}_2
\end{equation}
the corresponding $2^2$ diagonal matrices of $\R(2)$ satisfy Proposition~\ref{prop_SAT_in_On1} but do not form a cover of $\OO{2}$ (riddle solved in Section~\ref{subsec_8_three_cases_Proposition_13}).%
\opt{margin_notes}{\mynote{mbh.note: see \cite[p. 11]{Budinich_2019} \& log pp. 774, 784}}%

With these (and following) results we can port the commutative diagram of Section~\ref{sec_O(n)_MTNS} to the continuous case and to simple spinors
\opt{margin_notes}{\mynote{mbh.note: should I add that some of these results appear to be new ? (or at page 10)}}%
$$
\begin{tikzcd}[column sep=small]
& \psi_t \in \mySpinorS_s \arrow[dl, Leftrightarrow, "(\ref{formula_spinorial_representation})(\ref{formula_simple_spinors_v1k})
" '] \arrow[dr, Leftrightarrow, "(\ref{formula_simple_spinors})"] & \\
M(\psi_t) %= (\Identity, t)
\in \mysnqG{} \arrow[Leftrightarrow, "(\ref{formula_bijection_Nn_O(n)})"]{rr} & & t \in \OO{n}
\end{tikzcd}
$$
that authorizes us, from now on, to \emph{freely switch} between:
\begin{itemize}
\item simple spinors $\psi_t \in \mySpinorS_s$,
\item isometries $t \in \OO{n}$,
\item totally null subspaces of maximal dimension $(\Identity, t) \in \mysnqG{}$.
%\item
\end{itemize}

%\newpage
\section{An unsatisfiability test made with simple spinors: the idea}
\label{sec_Simple_Spinors_SAT}
To apply the new formulation to the computational side of \SAT{} we focus our attention to the continuous setting of Theorem~\ref{theorem_SAT_in_O(n)} where we need to prove that with $\sum_{j = 1}^m {\cal T}_j$ we can cover \OO{n} since we already remarked \cite{Budinich_2019} that an unsatisfiability test exploiting Proposition~\ref{prop_SAT_in_On1} offers no real advantages.

A known characteristic of \SAT{} problems is that while checking if a single assignment is in $\cup_{j = 1}^m {\cal T}_j'$ is polynomial (easy), to give a proof that all $2^n$ assignments are in this set, providing an unsatisfiability certificate, can require ${\cal O}(2^n)$ tests (hard).

In the continuous setting of Theorem~\ref{theorem_SAT_in_O(n)} we can also presume that checking if a single $t \in \OO{n}$ is in $\sum_{j = 1}^m {\cal T}_j$ is easy. But in \OO{n} things are quite different and we show that, for some $t$, just two of these tests can provide a certificate of unsatisfiability.

With isomorphisms (\ref{formula_simple_spinors}) to any $t \in \OO{n}$ corresponds the simple spinor $\psi_t \in \mySpinorS_s$ such that $M(\psi_t) = (\Identity, t)$ and this spinor can be expanded in Fock basis $\myFockB$ (\ref{formula_Fock_basis_expansion}) and we define its \emph{support}
\begin{equation}
\label{formula_support_spinor}
\sup \psi_t := \{ \psi_\lambda \in \myFockB : \alpha_\lambda \ne 0 \; \mbox{in (\ref{formula_Fock_basis_expansion})} \} \subseteq \myFockB
\end{equation}
so that $\psi_t \in \my_span{\sup \psi_t}$ and clearly $0 < | \sup \psi_t | \le 2^n$. So any $t \in \OO{n}$ induces $\sup \psi_t$, namely a set of $\lambda \in \O1{n}$ that in turn can be seen as a set of Boolean assignments (\ref{formula_bijection_Mn_O^n(1)}). Applying this to our case, given any $t \in {\cal T}_j$, since ${\cal T}_j \cap \O1n = {\cal T}_j'$ \cite[Lemma 1]{Budinich_2019} then
$$
\sup \psi_t \subseteq {\cal T}_j' \quad \implies \quad {\cal T}_j \subseteq \my_span{{\cal T}_j'}
$$
provided we identify $\psi_\lambda \in \myFockB$ with corresponding $\lambda \in \O1{n}$ (more precisely $M(\psi_\lambda) = (\Identity, \lambda)$) and where in second relation we put $\subseteq$ since not all linear combinations of $\psi_\lambda \in {\cal T}_j'$ are simple spinors. In other words $\sup \psi_t$ are the Boolean assignments induced by $t$ that make the problem unsatisfiable. In following Corollary~\ref{coro_sum_support} we show that this generalizes to any $t \in \sum_{j} {\cal T}_j$ of Theorem~\ref{theorem_SAT_in_O(n)} and that, if $t \in \sum_{j} {\cal T}_j$, all Boolean assignments of $\sup \psi_t$ make the problem at hand unsatisfiable.

In the next step we show that there exist simple spinors such that $| \sup \psi_t | = 2^{n-1}$ and thus, if one of them is in $\sum_{j = 1}^m {\cal T}_j$, this excludes $2^{n-1}$ \SAT{} assignments in one shot. With another $t' \in \OO{n}$ we can exclude the complementary $2^{n-1}$ terms (corresponding respectively to cases of $\det t, t' = \pm1$) and so we can conclude that if $t, t' \in \sum_{j = 1}^m {\cal T}_j$ then $\cup_{j = 1}^m {\cal T}_j'$ covers the entire Fock basis $\myFockB$ and thus the problem at hand is unsatisfiable by Proposition~\ref{prop_SAT_in_On1}.

The advantage of the continuous formulation is now manifest: in the discrete formulation a single $\lambda \in \cup_{j = 1}^m {\cal T}_j'$ excludes just \emph{one} assignment whereas in the continuous case $t \in \sum_{j = 1}^m {\cal T}_j$ can exclude \emph{up to} $2^{n-1}$.

Our \SAT{} problem is defined in the Clifford algebra of $\R^{n,n}$ and precisely in the spinorial representation of $\OO{n}$ and since $\myClg{}{}{\R^{n,n}} \myisom \R(2^n)$, \SAT{} is a problem in the algebra of real matrices of dimension $2^n \times 2^n$
\opt{margin_notes}{\mynote{mbh.note: doubly check spinorial vs vectorial representations of \OO{n}, see log pp. 736, 759, M\_532}}%
and, from the computational side, the situation looks problematic. The turning point is that the spinorial representation of group $\OO{n}$ is equivalent to its vectorial representation corresponding to the much more manageable and familiar algebra of real $n \times n$ matrices $\R(n)$.

In a nutshell to any $t \in \OO{n}$ corresponds an orthogonal matrix $T \in \R(n)$ of the vectorial representation and the action of $T$ on $u \in \R^n$ is given by $T u$. In the spinorial representation of $t$ in $\myClg{}{}{\R^{n,n}} \myisom \R(2^n)$ the action of $t$ on $u$ is given by
\begin{equation}
\label{formula_spinorial_representation}
(-1)^k v_1 v_2 \cdots v_k \; u \; (v_1 v_2 \cdots v_k)^{-1}
\end{equation}
for some $v_1, v_2, \ldots v_k \in \{0\} \times \R^n$%
\opt{margin_notes}{\mynote{mbh.note: here implicitly $v_i$ are spacelike vectors and thus $v_i^2 > 0$ and invertible}}%
, linearly independent and with $k \le n$, this being nothing else than the Cartan theorem for Euclidean spaces in disguise \cite[Theorem~5.15]{Porteous_1995}: vectors $v_1, v_2, \ldots v_k$ give the directions of the $k \le n$ hyperplane reflections in which any isometry of $\OO{n}$ can be decomposed.

Through equivalence between the representations of $\OO{n}$ and exploiting commutative diagram of Section~\ref{sec_SAT_in_O(n)}, our results can be equally formulated either in $\R(2^n)$ or in $\R(n)$ (and in particular Theorem~\ref{theorem_SAT_in_O(n)}) but we do not insist on this.

Summarizing there are two main ingredients in this recipe for \SAT{}: the first is that a single $t \in \sum_{j = 1}^m {\cal T}_j$ can rule out $2^{n-1}$ assignments while the second is the equivalence between the spinorial and vectorial representations of $\OO{n}$ respectively in matrix algebras $\R(2^n)$ and $\R(n)$. Moreover in this formulation there is no combinatorics since $\myFockB$ is a proper basis of the linear space of spinors $\mySpinorS$ and expansion (\ref{formula_Fock_basis_expansion}) is \emph{unique} and if an element of the Fock basis $\psi_\lambda \notin \sum_{j = 1}^m {\cal T}_j$ necessarily it is a solution of the \SAT{} problem.

\subsection{The theory}
\label{subsec_8_Simple_Spinors_SAT_theory}
We now proceed proving formally various results we will need in last part and we do it within the frame of $\myClg{}{}{\R^{n,n}} \myisom \R(2^n)$ that offers a more structured theoretical setting since it contains vectors, bivectors, spinors and the neatly defined Boolean algebra we exploited to formulate \SAT{} problems.

We warn the reader that subsequent steps assume good familiarity with Clifford algebras and simple spinors but she/he can find all details in quoted references, excluding few new results and adaptations to \SAT{}. However, through equivalence between the spinorial and vectorial representations of $\OO{n}$ and exploiting commutative diagram of Section~\ref{sec_SAT_in_O(n)}, \emph{all} following results could also be equally formulated in the familiar matrix algebra $\R(n)$, without resorting to spinors.%
\opt{margin_notes}{\mynote{mbh.note: commented here there is a reference to an old Appendix supposed to contain this formulation}}%
%
%and some of them are presented in this form in the Appendix at page~\pageref{Appendix}.
%%
%\opt{margin_notes}{\mynote{mbh.note: Appendix ??}}%
%%

We start by some very general simple spinors properties \cite{Budinich_2012}:
\begin{MS_Proposition}
\opt{margin_notes}{\mynote{mbh.note: log pp. 775, 776; this result generalizes \cite{Budinich_2012} (M\_598) Propositions~6 \& 7}}%
\label{prop_all_SSpinors}
All simple spinors $\psi$ of $\myClg{}{}{\R^{n,n}}$ can be written as
\begin{equation}
\label{formula_simple_spinors_v1k}
\psi = v_1 v_2 \cdots v_k \psi_\Identity
\end{equation}
for some $v_1, v_2, \ldots v_k \in \{0\} \times \R^n$, linearly independent and with $k \le n$.
\end{MS_Proposition}
This is again a consequence of the Cartan theorem for Euclidean spaces \cite[Theorem~5.15]{Porteous_1995} : vectors $v_1, v_2, \ldots v_k$ give the $\{0\} \times \R^n$ directions of the $k \le n$ reflections in which any isometry of $\OO{n}$ can be decomposed (\ref{formula_spinorial_representation}). We just hint that expanding each couple
\opt{margin_notes}{\mynote{mbh.note: first $\wedge$ appearance}}%
$v_i v_{i+1} = v_i \cdot v_{i+1} + v_i \wedge v_{i+1}$ we can expand $\psi$ and moreover expanding each $v_i \wedge v_{i+1}$ in bivector basis $\mygen_{k} \mygen_{l}$
\opt{margin_notes}{\mynote{mbh.note: from \cite[p. 2129]{BudinichP_1989} The Lie algebra spin(g) of the group Spin(g) can be identified with the subspace [V, V] of Cl(g) spanned by all the commutators [u,v] where $u, v \in V$, see log. pp. 351-355.}}%
$$
v_i \wedge v_{i+1} = \left( \sum_k \alpha_{i, k} \mygen_{k} \right) \wedge \left( \sum_l \alpha_{i+1, l} \mygen_{l} \right) = \sum_{k,l; \, l > k} (\alpha_{i, k} \alpha_{i+1, l} - \alpha_{i, l} \alpha_{i+1, k}) \mygen_{k} \mygen_{l}
$$
we finally arrive at $\psi$ expansion (\ref{formula_Fock_basis_expansion}) since all $\mygen_{i_1} \mygen_{i_2} \cdots \mygen_{i_r} \psi_\Identity \in \myFockB$. This shows how the expansion of a simple spinor in the Fock basis is deeply intertwined with bivectors of $\myClg{}{}{\R^{n,n}}$ and the corresponding Lie algebra. In the vectorial representation of \OO{n} this explains how any $t \in \SO{n}$ can be decomposed in $g \le {\bino{n}{2}}$ \SO{2} rotations acting in subspaces $\my_span{\mygen_i, \mygen_j}$ giving the Givens expansion of $t$ \cite{Hoffman_Raffenetti_Ruedenberg_1972} but we do not insist on this.

Given any two null subspaces of $\R^{n,n}$ $(\Identity, t_1), (\Identity, t_2) \in \mysnqG{}$ they have necessarily an intersection of dimension $r$ with $0 \le r \le n$, their \emph{incidence}, that is given by all vectors $u \in \R^n$ such that $(u, t_1 u) = (u, t_2 u)$, namely $t_1 u = t_2 u$. We give a pivotal property of simple spinors \cite[Proposition~5]{BudinichP_1989} here slightly adapted to our needs:
\begin{MS_Proposition}
\label{prop_5_BudinichP_1989}
Given any two linearly independent simple spinors $\psi, \phi \in \mySpinorS_s$ then their linear combinations $\alpha \psi + \beta \phi$ ($\alpha, \beta \in \R$, $\alpha^2 + \beta^2 = 1$) are simple if and only if the incidence of their associated $n$ dimensional null subspaces is $n-2$ namely
$$
\dim \left( M(\psi) \cap M(\phi) \right) = n-2
$$
and then $M(\psi) \cap M(\alpha \psi + \beta \phi) = M(\phi) \cap M(\alpha \psi + \beta \phi) = M(\psi) \cap M(\phi)$.
\end{MS_Proposition}

\noindent We put this result in a more usable form:

\begin{MS_Proposition}
\opt{margin_notes}{\mynote{mbh.note: log pp. 778, 780.2}}%
\label{prop_subspace_SSpinors}
Given any simple spinor $\psi$ of $\myClg{}{}{\R^{n,n}}$ and given any two, linearly independent, $v_1, v_2 \in \{0\} \times \R^n$ then
$$
v_1 v_2 \psi = (v_1 \cdot v_2 + v_1 \wedge v_2) \psi
$$
is a simple spinor and if $M(\psi) = (\Identity, t)$ then $M(v_1 v_2 \psi) = (\Identity, t t_\theta)$ where $t_\theta \in \SO{2}$ acts in $\my_span{v_1, v_2}$; moreover their common null subspace is
$$
M(\psi) \cap M(v_1 v_2 \psi) = \{ (u, t u) \in \R^{n,n} : u \in \my_span{v_1, v_2}^\perp \subset \{0\} \times \R^n \} \dotinformula
$$
\end{MS_Proposition}
\begin{proof}
By standard properties of Clifford product $v_1 \wedge v_2$ identifies the two dimensional subspace $\my_span{v_1, v_2}$ and any $u \in \my_span{v_1, v_2}^\perp$ commutes with $v_1 v_2$. By hypothesis $M(\psi) = (\Identity, t) = \{(x, t x) \in \R^{n,n}: (x, t x) \psi = 0 \}$ and moreover $M(v_1 v_2 \psi) = v_1 v_2 M(\psi) (v_1 v_2)^{-1}$ \cite{BudinichP_1989} this being the spinorial representation of $t_\theta \in \SO{2}$ acting in $\my_span{v_1, v_2}$ (in matrix formalism $T_\theta x = (\Identity - 2 v_1 v_1^T) (\Identity - 2 v_2 v_2^T) x$). It follows that $(x, t x) \psi = 0 = v_1 v_2 (x, t x) (v_1 v_2)^{-1} v_1 v_2 \psi$ so that for any $u \in \my_span{v_1, v_2}^\perp$ then $(u, tu) \in M(\psi) \cap M(v_1 v_2 \psi)$; the incidence of the two null subspaces is $n-2$ and, by Proposition~\ref{prop_5_BudinichP_1989}, $v_1 v_2 \psi$ is a simple spinor.
\opt{margin_notes}{\mynote{mbh.note: this proof is reasonable but not crystal-clear... it assumes the map $\R^n \to \R^n \times \R^n$ after formula (\ref{formula_bijection_Nn_O(n)})}}%
\end{proof}
%
%
%\begin{MS_Corollary}
%\label{coro_incidence}
%Given any simple spinors $\psi$ of $\myClg{}{}{\R^{n,n}}$ and given any two, linearly independent, $v_1, v_2 \in \{0\} \times \R^n$ as in previous Proposition then for the incidence of $M(\psi)$ and $M(v_1 v_2 \psi)$
%$$
%M(\psi) \cap M(v_1 v_2 \psi) = \{ (x,x) \in \R^{n,n} : x \in \my_span{v_1, v_2}^\perp \subset \{0\} \times \R^n \}
%$$
%\end{MS_Corollary}
%\begin{proof}
%Let $M(\psi) = (\Identity, t)$, by previous result $M(\psi) \cap M(v_1 v_2 \psi) = \{ x \in\{0\} \times \R^n: t x = t (\Identity - 2 v_1 v_1^T) (\Identity - 2 v_2 v_2^T) x \}$ that is $x = (\Identity - 2 v_1 v_1^T) (\Identity - 2 v_2 v_2^T) x$ namely all and only vectors left unaffected by reflections by $v_1$ and $v_2$.
%\end{proof}

All simple spinors are Weyl \cite[Proposition 3]{BudinichP_1989}, namely eigenvctors of the volume element $\omega$ (\ref{formula_identity_omega}) of $\myClg{}{}{\R^{n,n}}$ of eigenvalue $\pm 1$: their \emph{helicity} and with (\ref{formula_simple_spinors}) given $\psi \in \mySpinorS_s$ and $t \in \OO{n}$ such that $M(\psi) = (\Identity, t)$ the helicity of $\psi$ is equal to $\det t = \pm1$.

\begin{MS_lemma}
\label{lemma_same_support}
Given any simple $\psi \in \mySpinorS_s$ and its expansion in the Fock basis (\ref{formula_Fock_basis_expansion}) of $|\sup \psi| = r$ and given any generator $\mygen_{i}$ of $\R^{n,n}$, then $\mygen_{i} \psi$: is a simple spinor of opposite helicity, $|\sup \mygen_{i} \psi| = r$ and $\sup \psi \cap \sup \mygen_{i} \psi = \emptyset$.%
\opt{margin_notes}{\mynote{mbh.note: both $\mygen_{i}$ and $\mygen_{i+n}$ give the same $\sup \mygen_{i} \psi$ as it's easy to check with (\ref{formula_Witt_basis})}}%
\end{MS_lemma}
\begin{proof}
Given $\psi \in \mySpinorS_s$ of given helicity and its Fock basis expansion (\ref{formula_Fock_basis_expansion}), being Fock basis elements simple spinors themselves, it is easy to see that all $\psi_\lambda \in \sup \psi$ must have the same helicity of $\psi$. For any generator $\mygen_{i}$, $\mygen_{i} \psi$ is a simple spinor of opposite helicity \cite{BudinichP_1989} and thus all its Fock basis elements also have opposite helicity thus proving that $\sup \psi \cap \sup \mygen_{i} \psi = \emptyset$%
\opt{margin_notes}{\mynote{mbh.note: beware! does not extend to $(\mygen_{i} + \mygen_{j})\psi$, see 'floating leaflets' p. 121}}%
. To prove that the size of the support of the two spinors is identical we remark that for any $\psi_\lambda$ of (\ref{formula_Fock_basis_expansion}) there exist one and only one $\mygen_{i} \psi_\lambda \in \myFockB$ (that, in the language of \cite{Budinich_2016}, has opposite $h$ and $g-$signatures with respect to $\psi_\lambda$ and thus same $h \circ g-$signature being in the same spinor space).
\opt{margin_notes}{\mynote{mbh.note: commented here there is previous corollary to Lemma \ref{lemma_same_support}}}%
\end{proof}
%
%\begin{MS_Corollary}
%\label{coro_Fock_partition}
%Given a simple spinors $\psi \in \mySpinorS_s$ such that $| \sup \psi | = 2^{n-1}$ and given any $\mygen_{i} \in \{0\} \times \R^n$, then $\sup \psi$ and $\sup \mygen_{i} \psi$ form a partition of the Fock basis, namely $\sup \psi \cap \sup \mygen_{i} \psi = \emptyset$ and $\sup \psi \cup \sup \mygen_{i} \psi = \myFockB$.
%\end{MS_Corollary}
%\begin{proof}
%That $\psi \in \mySpinorS_s$ always exists is guaranteed by Proposition~\ref{prop_full_sup_SSpinors}, the rest follows directly from previous Lemma.
%\end{proof}

\begin{MS_Proposition}
\opt{margin_notes}{\mynote{mbh.note: log pp. 780.2', 790, commented there is a part on Fock basis covering. Apparently could be in conflict with Proposition~9 of M\_577 but probably that proposition refers to subspaces such that any element is simple.}}%
\label{prop_full_sup_SSpinors}
For any $n > 1$ in $\myClg{}{}{\R^{n,n}}$ there exist infinite simple spinors $\psi \in \mySpinorS_s$ (\ref{formula_simple_spinors}) such that $| \sup \psi | = 2^{n-1}$.
%, moreover for any of these spinors, given any $\mygen_{i} \in \{0\} \times \R^n$, then $\mygen_{i} \psi$ has: minus $\psi$ helicity, $| \sup \mygen_{i} \psi | = 2^{n-1}$ and support exactly complementary to that of $\psi$, in other words the supports of $\psi$ and $\mygen_{i} \psi$ form a partition of Fock basis $\myFockB$.
\end{MS_Proposition}
\begin{proof}
We proceed by induction on $n$ starting from $n = 2$, in this case a Fock basis of spinor space is given by \eg \cite{Budinich_2016}
$$
\myFockB = \{ q_1 \, q_2, \; q_1 \, p_2 q_2, \; p_1 q_1 \, q_2, \; p_1 q_1 \, p_2 q_2 \}
$$
and thus $\mySpinorS = \my_span{q_1 \, q_2, \; q_1 \, p_2 q_2, \; p_1 q_1 \, q_2, \; p_1 q_1 \, p_2 q_2}$. By Proposition~\ref{prop_5_BudinichP_1989} any spinor of the linear subspace $\my_span{q_1 \, q_2, \; p_1 q_1 \, p_2 q_2}$ is simple:
\opt{margin_notes}{\mynote{mbh.note: see 'floating leaflets' pp. 122 ff \& log pp. 759, 777}}%
\eg $\psi = \cos \frac{\theta}{2} \; q_1 \; q_2 + \sin \frac{\theta}{2} \; p_1 q_1 \; p_2 q_2$ is simple with $M(\psi) = (\Identity, \left(\begin{array}{r r} \cos \theta & - \sin \theta \\ \sin \theta & \cos \theta \end{array}\right))$ moreover $\mygen_{2} \psi = (p_2 + q_2) \psi = \sin \frac{\theta}{2} \; p_1 q_1 \; q_2 - \cos \frac{\theta}{2} \; q_1 \; p_2 q_2$ is simple with $M(\mygen_{2} \psi) = (\Identity, \left(\begin{array}{r r} \cos \theta & \sin \theta \\ \sin \theta & - \cos \theta \end{array}\right))$ and so the proposition is true for $n = 2$. We remark that the helicities of $\psi$ and $\mygen_{2} \psi$ are opposite and that the determinant of $M(\psi)$ and $M(\mygen_{2} \psi)$ are respectively $1$ and $-1$. For the induction step let the proposition be true for $n-1$ and let $\phi$ be a simple spinor of $\myClg{}{}{\R^{n-1,n-1}}$ of $| \sup \phi | = 2^{n-2}$ and we move to $\myClg{}{}{\R^{n,n}}$: here $\varphi = \phi p_n q_n$ is a simple spinor of support $2^{n-2}$ with $M(\varphi) = M(\phi) \oplus \R\{p_n\}$ how it is simple to check. Let $v = \cos \frac{\theta}{2} \mygen_{n} + \sin \frac{\theta}{2} \mygen_{n-1}$ with $\mygen_{n-1} \phi$ with support of size $2^{n-2}$ in $\myClg{}{}{\R^{n-1,n-1}}$ and opposite helicity to $\phi$. Spinor
$$
\begin{array}{l l l}
\psi = \mygen_{n} v \varphi & = & \mygen_{n} v \phi p_n q_n = (\mygen_{n} \cdot v + \mygen_{n} \wedge v) \phi p_n q_n = (\cos \frac{\theta}{2} + \sin \frac{\theta}{2} \mygen_{n} \mygen_{n-1}) \phi p_n q_n = \\
& = & \cos \frac{\theta}{2} \phi p_n q_n - \sin \frac{\theta}{2} \mygen_{n-1} \phi q_n
\end{array}
$$
where in last equality we exploited the fact that $\mygen_{n}$ has no effect on spinor $\phi$ of $\myClg{}{}{\R^{n-1,n-1}}$ and, without loss of generality, we have assumed that $\mygen_{n}$ commutes with $\phi$. The two final terms have equal helicities and by induction hypothesis have support in Fock basis of $\myClg{}{}{\R^{n,n}}$ of size $2^{n-1}$. The spinor
$$
\mygen_{n} \psi = \mygen_{n} (\cos \frac{\theta}{2} \phi p_n q_n - \sin \frac{\theta}{2} \mygen_{n-1} \phi q_n) = \cos \frac{\theta}{2} \phi q_n + \sin \frac{\theta}{2} \mygen_{n-1} \phi p_n q_n
$$
is of opposite helicity and covers the missing half of the Fock basis of $\myClg{}{}{\R^{n,n}}$ thus completing the proof.
\end{proof}

Real coefficients $\cos \frac{\theta}{2}, \sin \frac{\theta}{2}$ are used to prove that there are infinite spinors of maximal support but for our purposes we will need just one of these spinors and thus, to ease notation, from now on we will tacitly assume $\theta = \frac{\pi}{2}$ that will give equal real coefficients so that \eg $\psi = \frac{1}{\sqrt 2} (\Identity + \mygen_{n} \mygen_{n-1}) \phi p_n q_n = \frac{1}{\sqrt 2} \mygen_{n} (\mygen_{n} + \mygen_{n-1}) \phi p_n q_n$. In more general case $v_1 v_2 \psi$ the coefficient is chosen to give $v_1^2 = v_2^2 = 1$, \eg factor $\frac{1}{\sqrt 2}$ normalizes $\mygen_{n} + \mygen_{n - 1}$.

Given any $\varphi \in \mySpinorS_s$ such that $| \sup \varphi | = 2^{n-1}$ by Lemma~\ref{lemma_same_support} we have $\sup \varphi \oplus \sup \mygen_{i} \varphi = \myFockB$. We already defined before (\ref{formula_lambda_action}) the \emph{reference} spinor $\psi_\Identity = p_1 q_1 \; p_2 q_2 \; p_3 q_3 \cdots p_n q_n$; clearly $\psi_\Identity \in \myFockB$ and $M(\psi_\Identity) = \my_span{p_{1}, p_{2}, \ldots, p_{n}} = P = (\Identity, \Identity)$ and with $\psi_\Identity$ we build explicitly one of these spinors:

\begin{MS_Corollary}
\label{coro_phi_full_sup}
For any $n > 1$ given $\myClg{}{}{\R^{n,n}}$ the spinor
\begin{equation}
\label{formula_phi_full_sup}
\varphi := 2^\frac{1 - n}{2} \; \prod_{i=1}^{n-1} (\Identity + \mygen_{i} \mygen_{i+1}) \psi_\Identity
\end{equation}
is simple with $| \sup \varphi | = 2^{n-1}$,
%, in other words the supports of $\varphi$ and $\mygen_{i} \varphi$ together form a partition of Fock basis $\myFockB$.
moreover for any $ \mygen_{i}$ then $\sup \varphi \oplus \sup \mygen_{i} \varphi = \myFockB$.%
\opt{margin_notes}{\mynote{mbh.note: see extensions 'floating leaflets' p. 121}}%
\end{MS_Corollary}
\begin{proof}
That $\varphi$ is simple is manifest being the result of $n - 1$ applications of Proposition~\ref{prop_subspace_SSpinors} to $\psi_\Identity$. For $| \sup \varphi |$ instead of proceeding by induction like in Proposition~\ref{prop_full_sup_SSpinors} we give a simpler proof observing that the product contains $n-1$ terms in parenthesis each of them being the sum of identity and bivector $\mygen_{i} \mygen_{i+1}$ so that the full expansion of $\varphi$ contains all the $2^{n-1}$ subsets of the product obtained choosing one of the two terms in each parentheses and observing that it is impossible to get complete cancellations between bivector terms.
\opt{margin_notes}{\mynote{mbh.note: commented here there are: the proof by induction, the definition of succession $\varphi_k$ and discussion of $\psi_j + \psi_k \in {\cal T}_{j} + {\cal T}_{k}$ being also in ${\cal T}_{j} \cup {\cal T}_{k}$ or not}}%
\end{proof}

From now on we apply general spinors properties directly to \SAT{} problems and clauses and before plunging into this we refresh the notation. Each literal $\mylitrl_i$ induces sets ${\cal T}'$ (\ref{formula_cal_T_s_def2}) and ${\cal T}$ (\ref{formula_cal_T_j_def2}) that, by commutative diagram of Section~\ref{sec_SAT_in_O(n)}, can equivalently be seen as a subset of: simple spinors $\mySpinorS_s$, \OO{n} or $\mysnqG$ (\ref{formula_Nn}) depending on what better fits the case under scrutiny. Moreover any $\mylitrl_i$ uniquely identifies generator $\mygen_{i}$ that for any $\psi \in {\cal T}$ with $M(\mylitrl_i) \subseteq M(\psi)$ is such that $M(\myconjugate{\mylitrl}_i) \subseteq M(\mygen_{i} \psi)$ and we will switch back and forth between literal $\mylitrl_i$ and associate generator $\mygen_i$ bearing in mind that they are completely different mathematical objects: $\mylitrl_i$ is an idempotent of $\mysetS$ (\ref{formula_Boolean_substitutions}) while $\mygen_i$ is a base vector of $\R^{n,n}$, a generator of \myClg{}{}{\R^{n,n}} whose action induces an hyperplane reflection. Exploiting this one to one correspondence we juxtapose to $\sup z_j$, the literals of clause $z_j$ so that \eg $\sup \mylitrl_i \myconjugate{\mylitrl}_j \mylitrl_k = \{ \mylitrl_i, \mylitrl_j, \mylitrl_k \}$, also $\mysbs z_j$ the subspace of $\{0\} \times \R^n$ induced by the set of generators associated to literals so that $\mysbs \mylitrl_i \myconjugate{\mylitrl}_j \mylitrl_k = \{ \mygen_i, \mygen_j, \mygen_k \}$, the subspace associated to $M(z_j)$. We start with a technical result:

\begin{MS_lemma}
\label{lemma_eit_in_Tj}
Given any $t \in {\cal T}_j$ (\ref{formula_cal_T_j_def2}) and any $v \in \{0\} \times \R^n$ then $t_v := - v t v^{-1} \in {\cal T}_j$ if and only if $v \in \my_span{\mysbs z_j}^\perp$.
\end{MS_lemma}
\begin{proof}
For any $t \in {\cal T}_j$ $M(z_j) \subseteq (\Identity, t)$ its action on $u$ is given by (\ref{formula_spinorial_representation}) and the action of $t_v$ is $(-1)^{k+1} v v_1 v_2 \cdots v_k \; u \; (v v_1 v_2 \cdots v_k)^{-1}$. Let $v \in \my_span{\mysbs z_j}^\perp$ then $v$ anticommutes with all $\mygen_{i} \in \mysbs z_j$ and thus $M(z_j) \subseteq (\Identity, t_v)$, namely $t_v \in {\cal T}_j$. Conversely given $t_v \in {\cal T}_j$ necessarily $M(z_j) \subseteq (\Identity, t_v)$ and since we know that also $M(z_j) \subseteq (\Identity, t)$ it follows $v$ anticommutes with all $\mygen_{i} \in \mysbs z_j$ and thus, necessarily, $v \in \my_span{\mysbs z_j}^\perp$.
\opt{margin_notes}{\mynote{mbh.note: commented here there is a weaker version with $v = \mygen_{i}$}}%
\end{proof}

%\begin{MS_lemma}
%%
%\opt{margin_notes}{\mynote{mbh.note: what happens for $\mygen_{i} + \mygen_{j} \in \my_span{\mysbs z_j}$ ?}}%
%%
%\label{lemma_eit_in_Tj}
%Given any $t \in {\cal T}_j$ (\ref{formula_cal_T_j_def2}) and any generator $\mygen_{i}$ then $\mygen_{i} t \in {\cal T}_j$ if and only if $\mygen_{i} \notin \mysbs z_j$.
%\end{MS_lemma}
%\begin{proof}
%For any $t \in {\cal T}_j$ $M(z_j) \subseteq (\Identity, t)$ and for any $\mygen_{i} \notin \mysbs z_j$, $M(z_j) \subseteq (\Identity, \mygen_{i} t)$ and thus $\mygen_{i} t \in {\cal T}_j$. Conversely given $\mygen_{i} t \in {\cal T}_j$ necessarily $M(z_j) \subseteq (\Identity, \mygen_{i} t)$ and since we know that also $M(z_j) \subseteq (\Identity, t)$ it follows $\mygen_{i} \notin \mysbs z_j$.
%\end{proof}

\begin{MS_Proposition}
\opt{margin_notes}{\mynote{mbh.note: log pp. 795, 806, 806'. Do we need associativity: not really ?? Do we need to report here the discussion of ${\cal T}_{j}$ not being a subspace of \cite[p. 11]{Budinich_2019}. Commented all parts on $\psi \notin {\cal T}_{j} \cup {\cal T}_{k}$}}%
\label{prop_Tau_j_k}
Given two clauses $z_j, z_k$ with induced sets ${\cal T}_{j}, {\cal T}_{k}$ then $\psi = \frac{1}{\sqrt 2} (\psi_j + \psi_k) \in {\cal T}_{j} + {\cal T}_{k}$ if and only if the incidence of $M(\psi_j), M(\psi_k)$ is $n-2$ and thus if $\psi = v_1 v_2 \psi_j$ like in Proposition~\ref{prop_subspace_SSpinors} and thus $v_1 \wedge v_2 \psi_j \in {\cal T}_{k}$ and in particular only in the three following cases depending on the number of common opposite literals of $z_j$ and $z_k$:
\begin{itemize}
\item no common, opposite literals (but possibly equal ones): if $v_1, v_2 \in \my_span{\mysbs z_j \cap \mysbs z_k}^\perp$,
% and in this case in general $\psi \in {\cal T}_{j} \cup {\cal T}_{k}$. If moreover $v_1 \in \my_span{\mysbs z_j \setminus (\mysbs z_j \cap \mysbs z_k)}$ and $v_2 \in \mbox{Span}(\mysbs z_k \setminus (\mysbs z_j \cap \mysbs z_k))$ then $\psi \notin {\cal T}_{j} \cup {\cal T}_{k}$, % not using \my_span here since it doesn't allow breaking of the long formula
\item one common, opposite literal $\mylitrl_i$: if $v_1 = \mygen_{i}$ and $v_2 \in \mbox{Span}((\mysbs z_j \cap \mysbs z_k) \setminus \{\mygen_{i}\})^\perp$%
\opt{margin_notes}{\mynote{mbh.note: $\{\mygen_{i}\}$ is necessary because \eg $v_2 = \mygen_{i} + \mygen_{r}$ is a must, see 'floating leaflets' pp. 150, 154}}%
,
\item two common, opposite literals $\mylitrl_{i}, \mylitrl_{l}$: if $v_1 = \mygen_{i}$ and $v_2 = \frac{1}{\sqrt 2} (\mygen_{i} + \mygen_{l})$.
%\item
\end{itemize}
\end{MS_Proposition}
\noindent For the purposes of this Proposition cases with more than two opposite literals are excluded because incidence $n-2$ would be impossible.
\begin{proof}
%We start observing that ${\cal T}_{j} \cap {\cal T}_{k} \ne \emptyset$ if and only if $z_j$ and $z_k$ have no common opposite literals since for any $\psi \in {\cal T}_{j} \cap {\cal T}_{k}$ then both $M(z_j), M(z_k) \subseteq M(\psi)$ and this may happen only if clauses are compatible \cite{Budinich_2019}%
%%
%\opt{margin_notes}{\mynote{mbh.note: compatible clauses are not defined in this paper}}%
%%
%, namely only if they have no common opposite literals (they still can have common equal literals).
For no common opposite literals%
%, namely ${\cal T}_{j} \cap {\cal T}_{k} \ne \emptyset$
: given $\psi = \frac{1}{\sqrt 2} (\psi_j + \psi_k) \in {\cal T}_{j} + {\cal T}_{k}$
\opt{margin_notes}{\mynote{mbh.note: log p. 780.1}}%
by Proposition~\ref{prop_subspace_SSpinors} $\psi = v_1 v_2 \psi_j$ and the incidence part is given by all $(u, t u) \in \R^{n,n}$ with $u \in \my_span{v_1, v_2}^\perp \subset\{0\} \times \R^n$ it follows that any $v_1, v_2 \in \my_span{\mysbs z_j \cap \mysbs z_k}^\perp$ can do since $\sup z_j \cap \sup z_k$ can contain only common equal literals, excluded by Proposition~\ref{prop_subspace_SSpinors}.

For one common, opposite literal $\mylitrl_{i}$ necessarily it is not in the incidence part and by Proposition~\ref{prop_subspace_SSpinors} there must exist another $v_2$, linearly independent from $v_1 = \mygen_{i}$, such that $v_1 \wedge v_2 \psi_j \in {\cal T}_{k}$. Any common literal of $\sup z_j \cap \sup z_k \setminus \{ \mylitrl_i \}$ is necessarily equal and thus in incidence part $M(\psi_j) \cap M(\psi_k)$ and by Proposition~\ref{prop_subspace_SSpinors} follows $v_2 \in \my_span{(\mysbs z_j \cap \mysbs z_k) \setminus \{\mygen_{i}\}}^\perp$%
%; in this case, because of opposite $\mylitrl_{i}$, for $\alpha \beta \ne 0$ $\psi \notin {\cal T}_{j} \cup {\cal T}_{k}$
.

For two common opposite literals, namely \eg $\mylitrl_{i}, \mylitrl_{l}$ then, being incidence of $M(\psi_j)$ and $M(\psi_k)$ $n-2$ and being necessarily $\mygen_{i}, \mygen_{l}$ not in the incidence part, by Proposition~\ref{prop_subspace_SSpinors}, necessarily $v_1 \wedge v_2 \psi_j = \mygen_{i} \mygen_{l} \psi_j \in {\cal T}_{k}$%
%and also in this case for $\alpha \beta \ne 0$ $\psi \notin {\cal T}_{j} \cup {\cal T}_{k}$
.
\opt{margin_notes}{\mynote{mbh.note: Commented here there is a longer proof that used a weaker version of Lemma~\ref{lemma_eit_in_Tj}.}}%

To prove the converse we remark that in all cases $v_1 \wedge v_2 \psi_j \in {\cal T}_{k}$.
%
%\opt{margin_notes}{\mynote{mbh.note: commented there is the additional condition to have $\psi \notin {\cal T}_{j} \cup {\cal T}_{k}$ also contained in old Lemma~\ref{lemma_sum_psi_in_union}}}%
%
%and since in all cases at least one of these vectors is a generator in $\mysbs z_j$ it follows by Lemma~\ref{lemma_eit_in_Tj} $v_1 v_2 \psi_j \notin {\cal T}_{j}$ and $v_1 v_2 \psi_j \in {\cal T}_{k}$ so that $\psi_j + \psi_k \in {\cal T}_{j} + {\cal T}_{k} \setminus ({\cal T}_{j} \cup {\cal T}_{k})$.
\end{proof}

\begin{MS_Corollary}
\label{coro_Tau_j_k}
Given two clauses $z_j, z_k$ with their induced sets ${\cal T}_{j}, {\cal T}_{k}$ then ${\cal T}_{j} + {\cal T}_{k} \ne \emptyset$ (\ref{formula_cal_T_j+k_def}) if and only if the clauses have $0, 1$ or $2$ common opposite literals.
\opt{margin_notes}{\mynote{mbh.note: could also add that for $1$ or $2$ common opposite the sum $\notin {\cal T}_{j} \cup {\cal T}_{k}$, see also above}}%
\end{MS_Corollary}

\noindent Some other technical Lemmas:
\begin{MS_lemma}
\opt{margin_notes}{\mynote{mbh.note: log p. 783}}%
\label{lemma_sum_intersection}
Given a \SAT{} problem let $J$ be any non empty subset of the $m$ clauses, then
$$
\left( \sum_{j \in J} {\cal T}_j \right) \cap \O1n = \cup_{j \in J} {\cal T}_j'
$$
\end{MS_lemma}
\begin{proof}
By (\ref{formula_cal_T_j_cal_T_j'}) $ {\cal T}_j \cap \O1n = {\cal T}_j'$ so we just need to prove that $({\cal T}_{j} + {\cal T}_{k}) \cap \O1n = {\cal T}_j' \cup {\cal T}_k'$ but this follows trivially from the definition of ${\cal T}_{j} + {\cal T}_{k}$ (\ref{formula_cal_T_j+k_def}) and from the fact that $\myFockB$ is a proper basis of $\mySpinorS$ and thus all elements of ${\cal T}_{j} + {\cal T}_{k}$ are necessarily in $\my_span{{\cal T}_j' \cup {\cal T}_k'}$ and the relation is correctly $=$ (and not only $\subseteq$) since all $\psi_\lambda \in {\cal T}_j'$ are in ${\cal T}_j$%
\opt{margin_notes}{\mynote{mbh.note: provided we accept $\psi_j$ as a member of ${\cal T}_{j} + {\cal T}_{k}$, see 'floating leaflets' p. 191}}%
.
\end{proof}

\begin{MS_Corollary}
\label{coro_sum_support}
Given $\psi \in \mySpinorS_s$ such that $\psi \in \sum_{j \in J} {\cal T}_j$, where $J$ is any non empty subset of the $m$ clauses, then
$$
\sup \psi \subseteq \cup_{j \in J} {\cal T}_j' \dotinformula
$$
\end{MS_Corollary}

\noindent Thus $\psi$ excludes all Boolean assignments of $\sup \psi$ since any $\psi_\lambda \in \sup \psi$ is necessarily in at least one $ {\cal T}_j'$ and thus is an assignment that renders $\myBooleanF$ the problem at hand and thus also $\varphi$ (\ref{formula_phi_full_sup}) excludes $2^{n-1}$ assignments.

With spinor sum (\ref{formula_cal_T_j+k_def}) we reformulate Theorem~\ref{theorem_SAT_in_O(n)} in $\mySpinorS_s$ exploiting the one to one correspondence between simple spinors and \OO{n}
\opt{margin_notes}{\mynote{mbh.note: log pp. 787, 795}}%
and we put it in a form more amenable to an actual algorithm testing unsatisfiability:
\begin{MS_theorem}
\label{theorem_SAT_in_O(n)_4}
A given \SAT{} problem is unsatisfiable if and only if given one $\varphi \in \mySpinorS_s$ with $| \sup \varphi | = 2^{n-1}$ like (\ref{formula_phi_full_sup}) and any $\mygen_{i}$ then the isometries induced by a non empty subset $J$ of its $m$ clauses (\ref{formula_cal_T_j_def2}) are such that:%
\opt{margin_notes}{\mynote{mbh.note: probably $\mygen_{i} \varphi$ is necessary only in 'low $n$' cases , see 'floating leaflets' pp. 186, 191}}%
\begin{equation}
\label{formula_SAT_in_O(n)_4}
\varphi, \mygen_{i} \varphi \in \sum_{j \in J} {\cal T}_j \dotinformula
\end{equation}
\end{MS_theorem}
\begin{proof}
Supposing that (\ref{formula_SAT_in_O(n)_4}) holds this implies that the Fock basis expansion of $\varphi$, with $| \sup \varphi | = 2^{n-1}$, is in $\sum_{j \in J} {\cal T}_j$ and by Corollary~\ref{coro_sum_support} $\cup_{j \in J} {\cal T}_j'$ covers the first half of the Fock basis. Repeating the same procedure with $\mygen_{i} \varphi$ by Lemma~\ref{lemma_same_support} we cover the second half of the Fock basis and thus $\cup_{j \in J} {\cal T}_j'$ contains the full Fock basis and the problem is unsatisfiable by Proposition~\ref{prop_SAT_in_On1}; the converse follows immediately from same Proposition.
\end{proof}

An unsatisfiability test exploiting Theorem~\ref{theorem_SAT_in_O(n)_4} in essence requires to check wether $\varphi \in \mySpinorS_s$ (\ref{formula_phi_full_sup}) of maximal support is in $\sum_{j \in J} {\cal T}_j$ and, in the affirmative, repeat the same test on $\mygen_{i} \varphi$
%\begin{itemize}
%\item build $\varphi \in \mySpinorS_s$ of maximal support (we did it in Corollary~\ref{coro_phi_full_sup}),
%\item verify whether $\varphi \stackrel{?}{\in} \sum_{j = 1}^m {\cal T}_j$ (we will do it in Proposition~\ref{prop_time_test}),
%\item perform the same test on $\mygen_{i} \varphi$.
%\end{itemize}
and if both tests succeed we have a certificate that the \SAT{} problem at hand is unsatisfiable.
\opt{margin_notes}{\mynote{mbh.note: commented here are considerations on succession $\varphi_k$}}%
%
%The remarkable fact is that we will not need to check the $2^{n-1}$ Fock basis components of $\varphi$ (\ref{formula_phi_full_sup}) but we can exploit the recursive construction of the succession $\varphi_k$ (\ref{formula_phi_succession}) breaking the initial problem into $n$ subproblems each of them corresponding to the application of Propositions~\ref{prop_subspace_SSpinors} and \ref{prop_Tau_j_k} to $\varphi_k$, in turn corresponding to the Givens expansion of $\varphi$.
%
The crucial pending question remains computational complexity of (\ref{formula_SAT_in_O(n)_4}) that we address in next sections.

\subsection{Preliminaries for an actual 3\SAT{} algorithm}
\label{subsec_8_Simple_Spinors_SAT_algorithm_new}
We recall some standard properties of $\R^n$, its isometry group $\OO{n}$ and ${\cal T}_j$ definition: any $t \in \OO{n}$ has only $3$ eigenvalues: $\pm 1$ and couples of complex conjugates and all eigenvectors corresponding to different eigenvalues are reciprocally orthogonal. Moreover any $t \in \OO{n}$ having only $\pm 1$ eigenvalues is an \emph{involution}: $t^2 = \Identity$ and in $\R(n)$ $T = T^T$, a prominent, but not exhaustive, example being the $2^n$ involutions $\lambda \in \O1{n}$ of Section~\ref{sec_O(n)_MTNS}.

It follows that any $t \in \OO{n}$ defines univocally three reciprocally orthogonal subspaces of $\R^n$ corresponding respectively to eigenvectors of $\pm 1$ eigenvalues and to their orthogonal complement and at least one of these subspaces have non null dimensions (these subspaces are in close connection with the Wall parametrization of $t \in \OO{n}$ \cite[Chapter 11]{Taylor_1992}).

Given $t \in {\cal T}_j$ (\ref{formula_cal_T_j_def2}) for any $\mygen_{i} \in \mysbs z_j$ it follows $t \mygen_{i} = \pm \mygen_{i}$ and thus for the corresponding simple spinor $\psi_t$ that either $p_i$ or $q_i$ are in $M(\psi_t)$ (\ref{formula_P_Q_def}) and thus all $\psi_\lambda \in \sup \psi_t$ share corresponding terms $p_i q_i$ or $q_i$ the proof being that $p_i$ or $q_i$ is in $M(\psi_t)$ and in all $M(\psi_\lambda)$ of $\sup \psi_t$ (\ref{formula_Fock_basis_expansion}).%
\opt{margin_notes}{\mynote{mbh.note: this is actually formalized in a subsequent Lemma (-> lemma\_pi\_qi) not included in present version, moreover here there is commented the extension to $\inv \psi$}}%
%
%
%We thus define also for simple spinor $\psi$, $\inv \psi$ as the subset of terms $p_i q_i$ or $q_i$ in $\psi_\lambda \in \sup \psi$ corresponding to $p_i$ or $q_i$ in $M(\psi)$ and in all $M(\psi_\lambda)$ of its support.

${\cal T}_j$ definition (\ref{formula_cal_T_j_def2}) obviously generalizes to arbitrary involutory parts, \eg for any given $z_x$ we write in spinor space
\opt{margin_notes}{\mynote{mbh.note: here we introduce ${\cal T}_x$, see 'floating leaflets' pp. 193, 198, old vrs paper p. 24}}%
$$
{\cal T}_{x} := \{\psi \in \mySpinorS_s : M(z_x) \subseteq M(\psi)\} \subseteq \mySpinorS_s
$$
and also remark that any $z_x$ partitions the space $\{0\} \times \R^n$ into $\mysbs z_x$ and its orthogonal complement. The directions of $\mysbs z_x$ are ``frozen'' in the sense that all $\psi \in {\cal T}_x$ have necessarily equal corresponding Fock basis terms, $p_i q_i$ or $q_i$, to give $M(z_x) \subseteq M(\psi)$. The other directions, $(\mysbs z_x)^\perp$, are ``free'' and we define ${\cal T}_{x}$ (and ${\cal T}_{x}'$) \emph{coverage}%
\opt{margin_notes}{\mynote{mbh.note: first appearance of coverage defined also, implicitly, after (\ref{formula_literal_projection}) at page~\pageref{formula_literal_projection} and in Proposition~\ref{prop_full_sup_SSpinors} at page~\pageref{prop_full_sup_SSpinors}: collect all together ? see note -> 22}}%
{} as the number of Fock basis elements in this set and for $|\mysbs z_x| = r$ the coverage is $2^{n-r}$ that coincides with DNF expansion of $z_x$ (\ref{formula_literal_projection}).
%We conclude recalling that there exist $2^r {\bino{n}{r}} = \bigO{n^r}$ different involutory parts $z_x$ of size $r$.

We now summarize the classical \emph{resolution} algorithm for \SAT{} that will turn out to be a particular case of the spinor sum and we will exploit its proven correctness to prove the correctness of our algorithm. Given a problem containing two clauses with one common opposite literal, \eg the first two clauses of $\myBooleanS$ (\ref{formula_SAT_std}): $(\mylitrl_1 \lor \myconjugate{\mylitrl}_2) \land (\mylitrl_2 \lor \mylitrl_3)$, the ``resolvent'' clause is the union of the two clauses with the common opposite literal $\mylitrl_2$ removed \ie $(\mylitrl_1 \lor \mylitrl_3)$. Resolution defines a binary operation in the set of clauses
$$
\diamond: {\cal C} \times {\cal C} \to {\cal C}
\qquad \qquad {\cal C}_i \diamond_k {\cal C}_j = {\cal C}_i \lor {\cal C}_j \setminus \{ \mylitrl_{k}, \myconjugate{\mylitrl}_{k} \}
$$
and we will occasionally write $\diamond_k$ to underline that $\mylitrl_k$ is the common opposite literal removed and thus in our example $(\mylitrl_1 \lor \myconjugate{\mylitrl}_2) \diamond_2 (\mylitrl_2 \lor \mylitrl_3) = (\mylitrl_1 \lor \mylitrl_3)$. It is easy to verify that $\myBooleanS$ is satsfiable if and only if such is the problem augmented by the resolvent clause $\myBooleanS \land (\mylitrl_1 \lor \mylitrl_3)$%
\opt{margin_notes}{\mynote{mbh.note: for Knuth resolution \& c. see 'floating leaflets' p. 109}}%
. To represent clause ${\cal C}_j$ in Section~\ref{sec_SAT_in_Cl} we introduced equivalent notation $z_j \equiv \myconjugate{{\cal C}}_j$ that gives the unique assignment of ${\cal C}_j$ literals that makes ${\cal C}_j \equiv \myBooleanF$ and is more suited to prove unsatisfiability in \myClg{}{}{\R^{n,n}}. With this notation previous example of the resolvent clause reads
\begin{equation}
\label{formula_resolvent}
\myconjugate{\mylitrl}_1 \mylitrl_2 \diamond_2 \myconjugate{\mylitrl}_2 \myconjugate{\mylitrl}_3 = \myconjugate{\mylitrl}_1 \myconjugate{\mylitrl}_3 \dotinformula
\end{equation}

\begin{MS_Proposition}
\label{prop_resolution_algorithm}
A given \SAT{} problem is unsatisfiable if and only if, by repeated $\diamond$ application to the set of clauses, we find the empty clause $\epsilon$, the clause with no literals \cite[Section~7.2.2.2, p. 54]{Knuth_2015}; \eg we find $\mylitrl_1 \diamond \myconjugate{\mylitrl}_1 = \epsilon$.%
\opt{margin_notes}{\mynote{mbh.note: first appearance of empty clause $\epsilon$ but a similar empty concept is used after (\ref{formula_atoms_primitive}) at page~\pageref{formula_atoms_primitive}}}%
\end{MS_Proposition}

For example in the only unsatisfiable 2\SAT{} problem with $n = 2$ (\ref{formula_2SAT_n2}) we easily get \eg $(z_1 \diamond z_3) \diamond (z_2 \diamond z_4) = \epsilon$. This result applies to both 2 and 3\SAT{} and in the first case the $\diamond$ operation in general produces a 2\SAT{} clause and since the set of all 2 clauses is \bigO{n^2}, the running time of the resolution algorithm for 2\SAT{} is polynomial.

%\pagebreak % to keep footnote in same page

For 3\SAT{} the $\diamond$ operation in general produces a
4\SAT{} clause%
\footnote{more precisely given clauses $z_i$ and $z_j$ with $k_i$ and $k_j$ literals then $z_i \diamond z_j$ has $k_l$ literals with $\max(k_i, k_j) - 1 \le k_l \le k_i + k_j - 2$.}%
{} but repeated application of $\diamond$ can produce clauses of up to $n - 1$%
\opt{margin_notes}{\mynote{mbh.note: easy proof: common opposite is never there}}%
{} literals and the set of these clauses is \bigO{2^n} and thus, when applied to 3\SAT{} problems, resolution algorithm can require exponential time.

In Section~\ref{sec_SAT_in_O(n)} we introduced the spinor sum (\ref{formula_cal_T_j+k_def}) that is the standard sum operation in the linear space of spinors restricted to simple spinors%
\opt{margin_notes}{\mynote{mbh.note: for $\myssSum$ definition see 'floating leaflets' p. 193}}%
$$
\myssSum: \mySpinorS_s \times \mySpinorS_s \to \mySpinorS_s
$$
and given \eg two spinors belonging to sets ${\cal T}_{j}, {\cal T}_{k}$, induced by clauses $z_j, z_k$, their sum (assuming factor $\frac{1}{\sqrt 2}$ is in $\myssSum$) $\psi = \psi_j \myssSum \psi_k = v_1 v_2 \psi_j \in {\cal T}_{j} + {\cal T}_{k}$ is ruled by Propositions~\ref{prop_subspace_SSpinors} and \ref{prop_Tau_j_k}. Willing to iterate the procedure adding to $\psi$ another simple spinor \eg $\phi \in {\cal T}_{r}$ to get $\phi \myssSum \psi$, quoted Propositions apply again and thus necessarily
$$
\phi \myssSum \psi = u_1 u_2 \psi = (u_1 \cdot u_2 + u_1 \wedge u_2) (v_1 \cdot v_2 + v_1 \wedge v_2) \psi_j = \phi \myssSum (\psi_j \myssSum \psi_k)
$$
where we used brackets to indicate that, differently from its parent sum, $\myssSum$ is \emph{not} associative and that in general $\phi \myssSum (\psi_j \myssSum \psi_k) \ne (\phi \myssSum \psi_j) \myssSum \psi_k$. For example let $\psi = \psi_j \myssSum \psi_k = \psi_\Identity \myssSum \mygen_{1} \mygen_{2} \psi_\Identity = \frac{1}{\sqrt 2} (\Identity + \mygen_{1} \mygen_{2}) \psi_\Identity$ that is simple and we may add $\phi = \mygen_{3} \mygen_{4} \psi$ to get $\phi \myssSum \psi = \frac{1}{\sqrt 2} (\Identity + \mygen_{3} \mygen_{4}) \psi = \mygen_{3} \mygen_{4} \psi \myssSum (\psi_\Identity \myssSum \mygen_{1} \mygen_{2} \psi_\Identity)$ where the parenthesis is strictly necessary since $\mygen_{3} \mygen_{4} \psi \myssSum \psi_\Identity$ is not a simple spinor because the incidence of the two simple spinors in the sum is $n-4$.

We now compare $\diamond$ with $\myssSum$: unsatisfiable problems made by two $1$\SAT{} clauses $\mylitrl_i, \myconjugate{\mylitrl}_i$ are immediately solved by resolution that gives $\mylitrl_i \diamond \myconjugate{\mylitrl}_i = \epsilon$ but proving this result in spinor terms requires some care. We begin assuming that also in spinor language the empty clause $\epsilon$ represents full \OO{n} coverage, even and odd, implied by a clause with no frozen part that thus can contain $\varphi, \mygen_{i} \varphi$ (\ref{formula_phi_full_sup}) of Theorem~\ref{theorem_SAT_in_O(n)_4}. Even with this specification something similar to $\mylitrl_i \diamond \myconjugate{\mylitrl}_i = \epsilon$ does not exist with simple spinors since operator $\myssSum$, acting in $\mySpinorS_s$, adds only spinors of equal helicity%
\opt{margin_notes}{\mynote{mbh.note: helicity and parity are defined just before Lemma \ref{lemma_same_support}}}%
{} producing a result of same helicity and we must thus prove full coverage separately for the two parities.

In case $n = 1$ unsatisfiability of clauses $\mylitrl_1, \myconjugate{\mylitrl}_1$ descends immediately from Theorem~\ref{theorem_SAT_in_O(n)_4} given that $\OO{1} = \{ \pm 1 \}$ and by (\ref{formula_cal_T_s_def2}) the two clauses represent the two isometries of \OO{1}, respectively even and odd (unsatisfiability comes also from Proposition~\ref{prop_SAT_in_On1} interpreting the clauses as the involutions associated to $2^1$ elements of the Fock basis $\myFockB$ of $\myClg{}{}{\R^{1,1}}$). To prove more easily unsatisfiability in case $n > 1$ we use a more general result.

\begin{MS_Proposition}
\label{prop_one_comm_different}
\opt{margin_notes}{\mynote{mbh.note: here we introduce ${\cal T}_{j \diamond k}$, see 'floating leaflets' pp. 193, 198}}%
Given two clauses $z_j, z_k$ with induced sets ${\cal T}_{j}, {\cal T}_{k}$ having exactly one common opposite literal, let it be $\mylitrl_i$, and such that there exists at least one literal that is free both in ${\cal T}_{j}$ and ${\cal T}_{k}$, let it be $\mylitrl_f$, then
$$
{\cal T}_{j \diamond k} % = \{\psi \in \mySpinorS_s : M(z_j \diamond z_k) \subseteq M(\psi)\}
\subseteq {\cal T}_{j} + {\cal T}_{k} \subseteq \mySpinorS_s
$$
where ${\cal T}_{j \diamond k} = \{\psi \in \mySpinorS_s : M(z_j \diamond_i z_k) \subseteq M(\psi)\}$ and $\mylitrl_{i}$ is a free literal.
\end{MS_Proposition}
\begin{proof}
Since the clauses have exactly one common opposite literal $\mylitrl_i$ then $z_j \diamond_i z_k$ is defined. By hypotheses and Proposition~\ref{prop_Tau_j_k} in the case of one common opposite literal we have $\psi_j \myssSum \psi_k = \frac{1}{\sqrt 2} (\Identity + \mygen_{i} \mygen_f) \psi_j \in {\cal T}_{j} + {\cal T}_{k}$ all other $n - 2$ literals associated to $\{ \mygen_{i}, \mygen_f \}^\perp$ being necessarily equal in the two addends. Since $\mylitrl_f$ is free in both clauses then $\sup (z_j \diamond_i z_k) \subseteq \{\mylitrl_1, \mylitrl_2, \ldots, \mylitrl_n \} \setminus \{\mylitrl_{i}, \mylitrl_f \}$ that implies $M(z_j \diamond_i z_k) \subseteq M(\psi_j \myssSum \psi_k)$.
\end{proof}
The condition on existence of a common free literal $\mylitrl_f$ is necessary: if in example (\ref{formula_resolvent}) we suppose $n = 3$ there is not a second generator $\mygen_f$ to build $\psi_j \myssSum \psi_k$. If there are one or more, \eg $\{ \mygen_{f}, \mygen_{g}, \ldots, \mygen_{z} \}$ any of them can do and, for any choice, it is plain that $\mylitrl_{2}$, frozen involutive for both $\psi_j \in {\cal T}_{j}$ and $\psi_k \in {\cal T}_{k}$, is in the set of free literals for $\psi_j \myssSum \psi_k$ since for both values of $\mylitrl_{2}$ the sum remains in ${\cal T}_{j} + {\cal T}_{k}$. More in general we will define $r$ generators $\{ \mygen_{i_1}, \mygen_{i_2}, \ldots, \mygen_{i_r} \}$ free for $\psi \in {\cal T}$ if for all possible $2^r$ combinations of their corresponding literals $\psi \in {\cal T}$.

While $z_j \diamond z_k$ is defined independently of $n$, to construct a spinor sum we need a free literal that is equivalent to $n > | \sup (z_j \diamond z_k) | + 1$,%
\opt{margin_notes}{\mynote{mbh.note: see low $n$ cases 'floating leaflets' p. 184; in Proposition~\ref{prop_Tau_j_k} this is implicit in \eg $v_2 \in \mbox{Span}((\mysbs z_j \cap \mysbs z_k) \setminus \{\mygen_{i}\})^\perp$}}%
{} and thus in all, non pathological, real world \SAT{} problems:

\begin{MS_Corollary}
\label{coro_resolvent_implies_ssum}
Given two clauses $z_j, z_k$ with induced sets ${\cal T}_{j}, {\cal T}_{k}$ having exactly one common opposite literal and with $n > | \sup (z_j \diamond z_k) | + 1$, then
$$
z_j \diamond z_k \quad \Rightarrow \quad \psi_j \myssSum \psi_k \qquad \mbox{with} \qquad {\cal T}_{j \diamond k} \subseteq {\cal T}_{j} + {\cal T}_{k} \dotinformula
$$
\end{MS_Corollary}

\noindent It follows that Proposition~\ref{prop_resolution_algorithm} of the resolution algorithm holds also for the clause composition algorithm induced by spinor sum. In cases of repeated applications of this Corollary clearly we get \eg
$$
{\cal T}_{r \diamond (j \diamond k)} \subseteq {\cal T}_{r} + ({\cal T}_{j} + {\cal T}_{k}) \subseteq \mySpinorS_s
$$
where parenthesis are necessary in both terms since also $\diamond$ is not associative that is no chance since we now may look at $\diamond$ as at an operation in the set of clauses induced by $\myssSum$.

With Corollary~\ref{coro_resolvent_implies_ssum} we prove unsatisfiability of clauses $z_j = \mylitrl_i, z_k = \myconjugate{\mylitrl}_i$ in case $n > 1$: given $z_j \diamond z_k = \epsilon$, equivalent to $M(z_j \diamond z_k) = \OO{n}$, we get $\OO{n} \subseteq {\cal T}_{j} + {\cal T}_{k}$. This can be understood observing that the even part of \OO{n} can be covered by $\mylitrl_i$ composed with an even isometry in the free $n-1$ coordinates together with $\myconjugate{\mylitrl}_i$ composed with an odd isometry in the free $n-1$ coordinates and conversely for the odd part of \OO{n}. With this caveats in what follows we will tolerate $\mylitrl_i \myssSum \myconjugate{\mylitrl}_i = \epsilon$ mainly to ease comparisons with $\diamond$.

\subsection{Generalized clauses}
\label{subsec_8_generalized_clauses}

We try the new arrows in our quiver on problem (\ref{formula_2SAT_n2}) and start from the case $n = 2$: here it is not possible to use Corollary~\ref{coro_resolvent_implies_ssum} since \eg $| \sup (z_1 \diamond z_2) | + 1 = 2$. In spinor terms unsatisfiability is proved summing the two cases with two common opposite literals along Proposition~\ref{prop_Tau_j_k}.
$$
\psi_1 \myssSum \psi_4 = \frac{1}{\sqrt 2} (\Identity + \mygen_{1} \mygen_{2} ) \psi_\Identity \in {\cal T}_{1} + {\cal T}_{4} \qquad \psi_2 \myssSum \psi_3 = \frac{1}{\sqrt 2} (\mygen_{1} + \mygen_{2}) \psi_\Identity = \frac{1}{\sqrt 2} (\Identity + \mygen_{1} \mygen_{2}) \mygen_{2} \psi_\Identity \in {\cal T}_{2} + {\cal T}_{3}
$$
that cover respectively the even and odd parts of \OO{2} in $\my_span{\mygen_{1}, \mygen_{2}}$ and unsatisfiability descends from Theorem~\ref{theorem_SAT_in_O(n)_4} (again unsatisfiability of (\ref{formula_2SAT_n2}) can come also from Proposition~\ref{prop_SAT_in_On1} since every clause has coverage $1$ and represents one member of the Fock basis $\myFockB$ of $\myClg{}{}{\R^{2,2}}$ made by $2^2$ terms). Similarly to previous case also in this case for $n > 2$ we can exploit a free direction $\mygen_{f}$ to sum the even part
$$
\frac{1}{\sqrt 2} (\Identity + \mygen_{1} \mygen_{2}) \psi_\Identity \myssSum \frac{1}{\sqrt 2} (\Identity + \mygen_{1} \mygen_{2}) \mygen_{2} \mygen_{f} \psi_\Identity = \frac{1}{2} (\Identity + \mygen_{1} \mygen_{2}) (\Identity + \mygen_{2} \mygen_{f}) \psi_\Identity \in ({\cal T}_{1} + {\cal T}_{4}) + ({\cal T}_{2} + {\cal T}_{3})
$$
while for the odd part
$$
\frac{1}{\sqrt 2} (\Identity + \mygen_{1} \mygen_{2}) \mygen_{f} \psi_\Identity \myssSum \frac{1}{\sqrt 2} (\Identity + \mygen_{1} \mygen_{2}) \mygen_{2} \psi_\Identity = \frac{1}{2} (\Identity + \mygen_{1} \mygen_{2}) (\mygen_{2} + \mygen_{f}) \psi_\Identity = \frac{1}{2} (\Identity + \mygen_{1} \mygen_{2}) (\Identity + \mygen_{2} \mygen_{f}) \mygen_{f} \psi_\Identity
$$
and together they cover \OO{3} in $\my_span{\mygen_{1}, \mygen_{2}, \mygen_{f}}$. Full expansion of these sums cover the even and odd parts of $\myFockB$ basis of $\myClg{}{}{\R^{3,3}}$ and thus also the possible $2^2$ combinations of literals $\mylitrl_1, \mylitrl_2$ that thus result free like $\mylitrl_f$. Full expansion of a spinor in $\myFockB$ basis gives an equivalent definition of coverage that we will use later on.

Spinor sum $\psi_1 \myssSum \psi_4 = \frac{1}{\sqrt 2} (\Identity + \mygen_{1} \mygen_{2} ) \psi_\Identity$ used to tackle (\ref{formula_2SAT_n2}) is not associated to a standard clause since it has no involutory part but only a frozen \OO{2} term in $\my_span{\mygen_{1}, \mygen_{2}}$. While a frozen involutory part represents a literal that has a fixed value, namely one half of the possible cases, a literal entering an \OO{2} term can take both values provided \emph{it changes} together with the other literal of the bivector. In other words of the $2^2$ possible combinations of the two literals, when in a frozen \OO{2} term, they may assume only two, again one half of all possible cases.
We define a \emph{generalized clause} as formed by a set of frozen involutive directions, none in this case, and a set of two dimensional subspaces containing frozen \OO{2} terms.

By the unsatisfiability proof of (\ref{formula_2SAT_n2}) for $n > 2$ we see that also summing two \OO{2} terms acting in same subspace and of opposite parity, we get the empty clause namely $(\psi_1 \myssSum \psi_4) \myssSum (\psi_2 \myssSum \psi_3) = \epsilon$. We resume these cases in a generalization of Proposition~\ref{prop_one_comm_different}:
\begin{MS_Proposition}
\label{prop_comm_different_parity}
Given two generalized clauses $z_j, z_k$ with induced sets ${\cal T}_{j}, {\cal T}_{k}$ having \emph{either} one common opposite literal \emph{or} two \OO{2} terms acting in same subspace and of opposite parity and, in both cases, at least one common free literal, then spinors sum $\psi_j \myssSum \psi_k \in {\cal T}_{j} + {\cal T}_{k} $ is always defined and, in the generalized clause they form, the frozen parts of the two addends, either the vector or the two vectors forming the bivector, are promoted to free variables in $\psi_j \myssSum \psi_k$.
\end{MS_Proposition}
\begin{proof}
In given hypotheses it is easy to verify that incidence is $n - 2$ and $\psi_j \myssSum \psi_k$ existence follows immediately by Proposition~\ref{prop_5_BudinichP_1989}. In first case the common opposite literal, being not a frozen literal of $\psi_j \myssSum \psi_k$, is necessarily free while in second case, with a common free literal, we build a cover of \OO{3} that contain all possible $2^2$ combinations of the two literals corresponding to the directions of action of \OO{2} that thus result free.
\end{proof}

A generalized clause is thus defined by a subset of the $n$ literals $\mylitrl_{i}$ and a subset of the $\bino{n}{2}$ bivectors $\mygen_{i} \mygen_{j}$%
\opt{margin_notes}{\mynote{mbh.note: this def does not cover bivector $\mygen_{i} (\mygen_{j} + \mygen_{k})$ used in Appendix}}%
, in one to one correspondence with two dimensional subspaces, and together they form the basis of vectors \emph{and} bivectors of \myClg{}{}{\R^{n,n}} containing $n + {\bino{n}{2}} = {\bino{n+1}{2}}$ elements. While there are $2^r {\bino{n}{r}} = \bigO{n^r}$ different clauses with $r$ literals, there exist $2^r {\bino{{\bino{n+1}{2}}}{r}} = \bigO{n^{2 r}}$ different generalized clauses with $r$ terms among vectors and bivectors.
\opt{margin_notes}{\mynote{mbh.note: see 'floating leaflets' p. 181}}%

\begin{MS_Proposition}
\label{prop_generalized_clause_coverage}
A generalized clauses with $r_v$ frozen directions among vectors and $r_b$ frozen \OO{2} terms has a Fock basis coverage of $2^{n - (r_v + r_b)}$.
\end{MS_Proposition}
\begin{proof}
We already know this for clauses with $r_v$ frozen literals so we need to prove the proposition just for \OO{2} terms. A frozen \OO{2} term uses two literals that cover just $2^{-1}$ of the $2^2$ possible cases and so any frozen \OO{2} term reduces possible coverage exactly as a single frozen literal.
%(old, wrong, proof of a more powerful proposition) We already proved this result in cases of clauses with only $r_v$ frozen vector directions so we need to prove the proposition just for \SO{2} terms and we proceed by induction on $r_b$; for $r_b = 1$ the proposition is true since supposing \eg $\psi$ has coverage $2^{n - 2}$ then $(\Identity + \mygen_{1} \mygen_{2}) \psi$ covers twice $\psi$ coverage \ie $2^{n - 1}$ (this is not true without hypotheses on directions free or frozen). For the induction step we assume that for $r_b$ bivectors there is a coverage of $2^{n - r_b}$ (that is to say that these are the Fock basis elements obtained by expansion of the $r_b$ \SO{2} terms) and we move to $r_b + 1$ bivectors and by hypothesis the new bivector has at least one vector that was not contained in the $r_b$ bivector set. It follows that the expansion of its \SO{2} term surely gives a doubling of the Fock basis coverage (similarly to proof of Corollary~\ref{coro_phi_full_sup})
\end{proof}

\subsection{The three cases of $\myssSum$ in Proposition~\ref{prop_Tau_j_k}}
\label{subsec_8_three_cases_Proposition_13}
We examine in detail the three cases of spinor sum $\myssSum$ in Proposition~\ref{prop_Tau_j_k} starting by the simplest one, that of two common opposite literals, let \eg
$$
z_j = \mylitrl_1 \mylitrl_2 \mylitrl_3 \quad z_k = \myconjugate{\mylitrl}_2 \myconjugate{\mylitrl}_3 \mylitrl_4 \qquad \qquad \psi_j \myssSum \psi_k = \frac{1}{\sqrt 2} (\Identity + \mygen_{2} \mygen_{3} ) \psi_\Identity
$$
where $\psi_\Identity \in {\cal T}_{j}$ and $\mygen_{2} \mygen_{3} \psi_\Identity \in {\cal T}_{k}$. In this case $\psi_j \myssSum \psi_k$ has an involutive part given by $(z_j \cup z_k) \setminus \{ \mylitrl_{2}, \myconjugate{\mylitrl}_{2}, \mylitrl_{3}, \myconjugate{\mylitrl}_{3} \} = \mylitrl_1 \mylitrl_4$ and a frozen $\SO{2}$ isometry in subspace $\my_span{\mygen_{2}, \mygen_{3}}$ and is thus a generalized clause defined by
$$
{\cal T}_{j} + {\cal T}_{k} = \{\psi \in \mySpinorS_s : \{ M(\mylitrl_{1} \mylitrl_{4}), M((\Identity + \mygen_{2} \mygen_{3} ) \psi_\Identity) \} \subseteq M(\psi)\} \subseteq \mySpinorS_s \dotinformula
$$
A complementary case of two common opposite literals is
$$
z_r = \mylitrl_1 \myconjugate{\mylitrl}_2 \mylitrl_3 \quad z_s = \mylitrl_2 \myconjugate{\mylitrl}_3 \mylitrl_4 \qquad \qquad \psi_r \myssSum \psi_s = \frac{1}{\sqrt 2} (\mygen_{2} + \mygen_{3} ) \psi_\Identity = \frac{1}{\sqrt 2} (\Identity + \mygen_{2} \mygen_{3} ) \mygen_{3} \psi_\Identity
$$
in which we recognize a case of an odd isometry in same subspace $\my_span{\mygen_{2}, \mygen_{3}}$ of the previous even example so by Proposition~\ref{prop_comm_different_parity} we get that $(\psi_j \myssSum \psi_k) \myssSum (\psi_r \myssSum \psi_s)$ gives full coverage in this subspace, $\mylitrl_{2}, \mylitrl_{3}$ become free and we get
$$
M\left((\psi_j \myssSum \psi_k) \myssSum (\psi_r \myssSum \psi_s)\right) = M(\mylitrl_{1} \mylitrl_{4})
$$
a result confirmed by resolution with $(z_j \diamond_2 z_r) \diamond_3 (z_k \diamond_2 z_s) = \mylitrl_{1} \mylitrl_{4}$.

\smallskip

The case of one common opposite literal is central, let \eg
%We now study a 3\SAT{} example of case of one common opposite literal
$$
z_j = \mylitrl_1 \mylitrl_2 \mylitrl_3 \quad z_k = \myconjugate{\mylitrl}_2 \mylitrl_3 \mylitrl_4 \qquad \qquad \psi_j \myssSum \psi_k = \frac{1}{\sqrt 2} (\Identity + \mygen_{2} \mygen_{5} ) \psi_\Identity
$$
where, as in the previous case, $\psi_\Identity \in {\cal T}_{j}$ while $\mygen_{2} \mygen_{5} \psi_\Identity \in {\cal T}_{k}$ and by Corollary~\ref{coro_resolvent_implies_ssum} we get ${\cal T}_{j \diamond k} \subseteq {\cal T}_{j} + {\cal T}_{k}$ and common opposite literal $\mylitrl_2$ becomes free. But in this case $\myssSum$ offers also other possibilities: instead of choosing a literal free for both ${\cal T}_{j}$ and ${\cal T}_{k}$, we can also choose \eg $\mylitrl_{1}$ that is frozen involutive for ${\cal T}_{j}$ and free for ${\cal T}_{k}$ and in this case we would get
$$
\psi_j \myssSum \psi_k = \frac{1}{\sqrt 2} (\Identity + \mygen_{1} \mygen_{2} ) \psi_\Identity
$$
where as above $\psi_\Identity \in {\cal T}_{j}$ and $\mygen_{1} \mygen_{2} \psi_\Identity \in {\cal T}_{k}$. In this case $\psi_j \myssSum \psi_k$ has an involutive part given by $(z_j \cup z_k) \setminus \{ \mylitrl_{1}, \myconjugate{\mylitrl}_{1}, \mylitrl_{2}, \myconjugate{\mylitrl}_{2} \} = \mylitrl_3 \mylitrl_4$ but in subspace $\my_span{\mygen_{1}, \mygen_{2}}$ spinor $\psi_j \myssSum \psi_k$ has a new frozen $\SO{2}$ isometry. In this case $\mylitrl_{2}$ is not added to the set of free literals of $\psi_j \myssSum \psi_k$ but its associated generator $\mygen_{2}$ enters in bivector $\mygen_{1} \mygen_{2}$ of frozen \SO{2} isometry in $\my_span{\mygen_{1}, \mygen_{2}}$.

We thus discovered, for one common opposite literal, that $\psi_j \myssSum \psi_k$ can generate \emph{different} generalized clauses all these cases being compliant with definition of ${\cal T}_{j} + {\cal T}_{k}$ (\ref{formula_cal_T_j+k_def}) that contains \emph{all possible} sums $\psi_j \myssSum \psi_k$ that, as this example shows, can have also different involutive parts; this is in sharp contrast with $\diamond$ where $z_j \diamond z_k$ is unique. In particular in the first case we get ${\cal T}_{j \diamond k} \subseteq {\cal T}_{j} + {\cal T}_{k}$ while in second case we get an involutive part together with an \SO{2} term and, by Proposition~\ref{prop_generalized_clause_coverage}, both cases give same coverage $2^{n - 3}$.

This could look unsettling at first but keep in mind that this is not an obstacle in building $\sum_j {\cal T}_{j}$ since we are looking for a \emph{single} spinor satisfying Theorem~\ref{theorem_SAT_in_O(n)_4} and producing more generalized clauses from a single couple of clauses is certainly more ``efficient''.

\smallskip

In the last case of no common opposite literals, let \eg
$$
z_j = \mylitrl_1 \mylitrl_2 \mylitrl_3 \quad z_k = \mylitrl_1 \myconjugate{\mylitrl}_4 \mylitrl_5 \qquad \qquad \psi_j \myssSum \psi_k = \frac{1}{\sqrt 2} (\Identity + \mygen_{3} \mygen_{4} ) \psi_\Identity
$$
where $\psi_\Identity \in {\cal T}_{j}$ and $\mygen_{3} \mygen_{4} \psi_\Identity \in {\cal T}_{k}$ and $\psi_j \myssSum \psi_k$ has an involutive part given by $(z_j \cup z_k) \setminus \{ \mylitrl_{3}, \myconjugate{\mylitrl}_{3}, \mylitrl_{4}, \myconjugate{\mylitrl}_{4} \} = \mylitrl_1 \mylitrl_2 \mylitrl_5$ and in subspace $\my_span{\mygen_{3}, \mygen_{4}}$ has a frozen \SO{2} term. But also $\psi_j \myssSum \psi_k = \frac{1}{\sqrt 2} (\Identity + \mygen_{2} \mygen_{4} ) \psi_\Identity$ with $\mygen_{2} \mygen_{4} \psi_\Identity \in {\cal T}_{k}$ and the involutive part of the sum is $\mylitrl_1 \mylitrl_3 \mylitrl_5$ and this time frozen \SO{2} term is in subspace $\my_span{\mygen_{2}, \mygen_{4}}$. ${\cal T}_{j} + {\cal T}_{k}$ contains also other generalized clauses and we can thus conclude that, of all cases of Proposition~\ref{prop_Tau_j_k}, only in that of two common opposite literals ${\cal T}_{j} + {\cal T}_{k}$ contains just one generalized clause.
\opt{margin_notes}{\mynote{mbh.note: 'floating leaflets' p. 193}}%

\subsection{The main result}
\label{subsec_8_main_result}
We stop here the exploitation of spinor sums aimed at satisfying conditions of Theorem~\ref{theorem_SAT_in_O(n)_4} to focus instead on the complexity of the derived unsatisfiability testing algorithm and we prove a less ambitious but quicker result.

Unleashing the $\myssSum$ operator to all cases of Proposition~\ref{prop_Tau_j_k} we produce much more generalized clauses than those produced by $\diamond$ and we assume that by repeated application of $\myssSum$ in the set of generalized clauses we get a set ${\cal U}$ of generalized clauses. Gven any unsatisfiable problem, by resolution and Corollary~\ref{coro_resolvent_implies_ssum} we know that the empty clause is certainly in ${\cal U}$; we have thus proved:

\begin{MS_Proposition}
\label{prop_spinor_sum_unsatisfiability}
A given \SAT{} problem is unsatisfiable if and only if, given the set ${\cal U}$ obtained by repeated $\myssSum$ application to the set of generalized clauses, we find $\epsilon \in {\cal U}$.
\end{MS_Proposition}

\noindent This result is not yet interesting since also ${\cal U}$ can contain exponentially many generalized clauses.

In the next and final step we restrain the generalized clause composition algorithm induced by spinor sum to produce only generalized clauses of coverage greater than $2^{n-4}$ that, by Proposition~\ref{prop_generalized_clause_coverage}, means only generalized clauses with less than $4$ frozen terms counting vectors and bivectors. With this limitation repeated application of the spinor sum algorithm produces a set $U \subseteq {\cal U}$ containing less than \bigO{n^8} generalized clauses.
\opt{margin_notes}{\mynote{mbh.note: or \bigO{n^9} because of $R_3$?}}%
The same holds also for resolution limited to produce at most 3\SAT{} clauses but in this case resolution algorithm is known to fail.

\begin{MS_Proposition}
\label{prop_limited_spinor_sum_unsatisfiability}
A given 3\SAT{} problem is unsatisfiable if and only if, given the set of generalized clauses $U \subseteq {\cal U}$ obtained by repeated application of $\myssSum$ on its clauses limited to produce only generalized clauses of coverage greater than $2^{n-4}$, we find $\epsilon \in U$.%
\opt{margin_notes}{\mynote{mbh.note: 'floating leaflets' p. 182.1}}%
\end{MS_Proposition}
\begin{proof}
If $\epsilon \in U$ the problem is trivially unsatisfiable and we prove the converse by induction on $n$.
%For $n = 3$ there exists only one 3\SAT{} unsatisfiable problem with $8$ clauses and it is easy to see that application of the $\myssSum$ operator quickly arrives at the empty clause (the exercise is very similar to that done for the only $n = k = 2$ unsatisfiable problem (\ref{formula_2SAT_n2})).
In subsequent Lemma~\ref{lemma_case_n5} in the Appendix we give the long and tedious proof that the proposition is true for $n = 5$, the minimal $n$ for which $\diamond$ can produce 4\SAT{} clauses. For the induction step we suppose the proposition true for $n - 1$ and proceed to $n$. Given that the problem at hand is unsatisfiable by hypothesis so are its two $n - 1$ derived problems obtained setting any of its literals respectively to $\mylitrl_i \equiv \myBooleanT$ and $\mylitrl_i \equiv \myBooleanF$. In the first case clauses containing $\mylitrl_i$ are removed while $\myconjugate{\mylitrl}_i$ is removed from clauses containing it (so called unit propagation). By induction hypothesis $\myssSum$ applied on this $n-1$ problem with $2^{n-4}$ coverage bound finds the empty clause $\epsilon$ that implies that the initial $n$ problem is reduced to $\myconjugate{\mylitrl}_i$ (apart from the case in which $\myconjugate{\mylitrl}_i$ didn't appear in clauses of the $n$ problem but in this case we get immediately unsatisfiability for the initial $n$ problem). Repeating the procedure for the reduced problem obtained setting $\mylitrl_i \equiv \myBooleanF$ we get $\mylitrl_i$ and thus the empty clause $\epsilon$ for the $n$ problem.
\opt{margin_notes}{\mynote{mbh.note: I wonder if there a neater proof. Commented here there are old proof versions and two objections to the induction step for $\myssSum$}}%
\end{proof}

Being the set of generalized clauses $U$ obtained by repeated application of $\myssSum$ polynomially bounded by \bigO{n^8} this proves that this unsatisfiability test of 3\SAT{} problems is polynomial.

The rationale is that only by $\myssSum$ we can produce all spinor sums implied by Proposition~\ref{prop_subspace_SSpinors} and thus all possible generalized clauses and not only the standard clauses produced by $\diamond$ and this allows to limit the production to generalized clauses of coverage greater than $2^{n-4}$. On the contrary there are instances in which $\diamond$ needs to produce 4\SAT{} clauses how is explicitly shown in the proof of Lemma~\ref{lemma_case_n5}.

A conclusive remark is that we provided only a proof of an upper bound for this algorithm which, instead of blindly trying all possible sums, would be probably much better if it could wisely steer the buildup of spinor sums towards the spinors of maximal coverage for Theorem~\ref{theorem_SAT_in_O(n)_4}.

\newpage
\section*{Appendix}
\label{Appendix}
Here we prove that Proposition~\ref{prop_limited_spinor_sum_unsatisfiability} is true for $n = 5$ and in particular that for all unsatisfiable problems $\myBooleanS$ quoted spinor sum algorithm finds $\epsilon \in U$.

The first step is to prove that for any 3\SAT{}, $n = 5$%
\opt{margin_notes}{\mynote{mbh.note: 'floating leaflets' p. 215}}%
, unsatisfiable problem such that, repeated application of $\diamond$ to the set of clauses produces at max 3\SAT{} clauses, $\epsilon \in U$. In all these cases since $z_i \diamond z_j$ is at max a 3\SAT{} clause, $n > | \sup (z_i \diamond z_j) | + 1$ and we are in conditions of Corollary~\ref{coro_resolvent_implies_ssum} that proves the thesis for these cases.

We are thus left with unsatisfiable 3\SAT{} problems in which repeated application of $\diamond$ produces one or more 4\SAT{} clauses that resolution uses to find the empty clause and in what follows we show that all these problems are equivalent to just one type of problem.

To prove this we need to dig a little deeper in resolution algorithm: the succession of applications of $\diamond$ to clauses of any unsatisfiable problem that ultimately produces $\epsilon$, \eg those of (\ref{formula_2SAT_n2}), defines a binary tree (more generally a directed acyclic graph) induced by $\diamond$ (\ref{formula_resolvent}) each node being labeled by a clause \cite[Section~7.2.2.2]{Knuth_2015}. Such a tree is \emph{regular} if no path from the root ($\epsilon$) to a leaf (a clause) uses the same literal twice in forming $z_i \diamond z_j$. Since there are $n$ literals any path of a regular tree arriving at the empty clause can have at most $n$ steps. Moreover any tree $T$ arriving at the empty clause can be converted in a regular tree not larger than $T$ \cite[Exercise~225]{Knuth_2015} so that we can safely assume that our trees producing $\epsilon$ are always regular.

Given two clauses $z_i$ and $z_j$ with $k_i$ and $k_j$ literals and $z_i \diamond z_j$ with $k_l$ literals we define this to be a case of \emph{clause merging} if $k_l < \min(k_i, k_j)$, namely if $k_l$ is lower than both $k_i$ and $k_j$.

\begin{MS_lemma}
\label{lemma_clause_merging}
Given clauses $z_i$ and $z_j$ with $k_i$ and $k_j$ literals then $z_i \diamond z_j$ with $k_l$ literals is a clause merging if and only if $k_i = k_j = | \sup z_i \cap \sup z_j |$ and then $k_l = k_i - 1$.
\end{MS_lemma}
\begin{proof}
It is easy to verify that in all cases $z_i \diamond z_j$ is defined then
\begin{equation}
\label{formula_resolvent_length}
k_l = k_i + k_j - | \sup z_i \cap \sup z_j | - 1 \;\; \mbox{and} \;\; 1 \le | \sup z_i \cap \sup z_j | \le \min(k_i, k_j)
\end{equation}
and supposing $k_i = k_j = | \sup z_i \cap \sup z_j |$ it follows that $k_l = k_i - 1$ and thus $k_l < \min(k_i, k_j)$ proving that we have clause merging. Conversely let $k_l < \min(k_i, k_j)$ if we suppose by absurdum $k_i \ne k_j$ namely $ \min(k_i, k_j) < \max(k_i, k_j)$ and writing (\ref{formula_resolvent_length}) as $k_l = \min(k_i, k_j) + \max(k_i, k_j) - | \sup z_i \cap \sup z_j | - 1$ we get $k_l \ge \max(k_i, k_j) - 1$ namely $k_l \ge \min(k_i, k_j)$ that falsifies our hypothesis so that necessarily $k_i = k_j$ and since by hypothesis $k_l < \min(k_i, k_j)$ by (\ref{formula_resolvent_length}) this implies $k_i = | \sup z_i \cap \sup z_j |$ and $k_l = k_i - 1$.
\end{proof}

\begin{MS_lemma}
\label{lemma_tree_structure}
In a regular tree proving unsatisfiability by resolution of an $n$, $k$\SAT{}, problem in all paths any node at $r$ steps from the root ($\epsilon$) must be labeled by a clause of $r$ or less literals.
\end{MS_lemma}
\begin{proof}
By its definition a regular tree proving unsatisfiability by resolution contains paths of up to $n$ steps; to prove the statement we proceed by induction on $r$. For $r = 1$ the proposition is true since $\mylitrl_i \diamond \myconjugate{\mylitrl}_i = \epsilon$ is the only possibility to get the empty clause; for the induction step we suppose the proposition true for $r-1$ and proceed to $r$, given clauses $z_i, z_j$ and $z_i \diamond z_j$ with respectively $k_i, k_j$ and $k_l$ literals and let $z_i$ be the clause in our path, by (\ref{formula_resolvent_length}) $\max(k_i, k_j) - 1 \le k_l \le k_i + k_j - 2$ and since by induction hypothesis $k_l \le r - 1$ it follows $\max(k_i, k_j) \le r$ and thus the thesis.
\end{proof}

In summary for our unsatisfiable 3\SAT{} problem with $n = 5$ in a regular tree obtaining $\epsilon$ any path must contain a number of steps in $[3, 5]$ that reduces to $[4, 5]$ if there are 4\SAT{} clauses. If $z_i \diamond z_j$ is a 4\SAT{} clause it must necessarily be at first step since we need at least 4 more steps to arrive at $\epsilon$ and thus $z_i$ and $z_j$ must be ``native'' clauses of the unsatisfiable problem $\myBooleanS$. Moreover at next step we must necessarily get a 3\SAT{} clause. It follows that any 4\SAT{} clause can be either composed with another 4\SAT{} clause in merging or with a clause with less than 4 literals.

We prove that in all cases of 4\SAT{} clause merging we get the same result without the need to pass through 4\SAT{} clauses and thus these clauses can be discarded. Let $z_k, z_r$ be two 4\SAT{} clauses that can be merged, by Lemma~\ref{lemma_clause_merging} $| \sup z_k \cap \sup z_r | = 4$ and thus, given $n = 5$, both clauses are produced by native 3\SAT{} clauses with identical common opposite literal; let \eg
\begin{align*}
z_i = \mylitrl_1 \mylitrl_2 \mylitrl_5 \quad
z_j = \mylitrl_3 \myconjugate{\mylitrl}_4 \myconjugate{\mylitrl}_5 \qquad \qquad
z_k := z_j \diamond_5 z_i = \mylitrl_1 \mylitrl_2 \mylitrl_3 \myconjugate{\mylitrl}_4
\\
z_s = \mylitrl_1 \mylitrl_3 \mylitrl_5 \quad
z_t= \mylitrl_2 \mylitrl_4 \myconjugate{\mylitrl}_5 \qquad \qquad
z_r := z_s \diamond_5 z_t = \mylitrl_1 \mylitrl_2 \mylitrl_3 \mylitrl_4
\end{align*}
and 4\SAT{} clause merging gives $z_k \diamond_4 z_r = \mylitrl_1 \mylitrl_2 \mylitrl_3$. But $\mylitrl_1 \mylitrl_2 \mylitrl_3$ can be obtained also with $(z_j \diamond_4 z_t) \diamond_5 z_i$ without passing through 4\SAT{} clauses. It is easy (and tedious) to check that $\mylitrl_1 \mylitrl_2 \mylitrl_3$ (or even $\mylitrl_1 \mylitrl_2$) can be obtained without resorting to 4\SAT{} clauses in all $6$ possible subdivisions of literals $\mylitrl_1, \mylitrl_2, \mylitrl_3$ and $\mylitrl_4$ among clauses $z_s$ and $z_t$.

We thus need to consider only cases in which two native clauses produce a 4\SAT{} clause that is successively composed with a clause with $3$ or less literals to produce at most a 3 clause.

Given clause $z_i \in \myBooleanS$ let $z_{(i)}$ be the subset of clauses that have the resolvent with $z_i$ namely
$$
z_{(i)} = \{ z_j \in \myBooleanS : \exists \; z_j \diamond z_i \} \subseteq \myBooleanS
$$
these clauses having one common opposite literal with $z_i$ and possibly also common equal literals. Clearly $z_i \notin z_{(i)}$ and $z_j \in z_{(i)}$ if and only if $z_i \in z_{(j)}$ being $\diamond$ commutative.

For any $n \ge 5$ given two 3\SAT{} clauses $z_i$ and $z_j$ such that $z_k := z_i \diamond z_j$ exists and is a 4\SAT{} clause by (\ref{formula_resolvent_length}) this implies that $z_i$ and $z_j$ have no common literals beyond the common opposite literal, let it be $\mylitrl_u$, and thus in $z_k$ we find all literals of $z_i$ and $z_j$ but $\mylitrl_u$. It follows that in this case any $z_r \in z_{(k)}$ \emph{induces} one of the two clauses, $z_i$ or $z_j$, that is the clause that has (at least) one common opposite literal with $z_r$.

For $n=5$ instead all 4\SAT{} clauses like $z_k$ contains 4 of the 5 literals and thus all 3\SAT{} clauses, including any $z_r := \mylitrl_x \mylitrl_y \mylitrl_z \in z_{(k)}$, have at least two literals in common with $z_k$. In particular any $z_r$ has one common opposite, let it be $\mylitrl_x$, and one common equal, let it be $\mylitrl_y$; the third literal $\mylitrl_z$ can be either a second common equal with $z_k$ or literal $\mylitrl_u$, common opposite between $z_i$ and $z_j$, that is not in $z_k$.

We now compare $z_r = \mylitrl_x \mylitrl_y \mylitrl_z \in z_{(k)}$ with its induced clause, let it be $z_j$: by definition literal $\mylitrl_x$, common opposite between $z_r$ and $z_k$, is common opposite also between $z_r$ and $z_j$ and it is different from $\mylitrl_u$, common opposite between $z_i$ and $z_j$, that is not in $z_k$. $\mylitrl_y$ is common equal between $z_r$ and $z_k$ and thus also common equal with either $z_i$ or $z_j$. Only $\mylitrl_z$ could possibly be a second common opposite literal between $z_r$ and $z_j$ and, given $z_r \in z_{(k)}$, the unique possibility would be literal $\mylitrl_u$, not in $z_k$, and, being common opposite with $z_j$, common equal with $\mylitrl_u$ in $z_i$.

We are now ready to examine one at the time all mutually exclusive possibilities that may occur in the two literals $\mylitrl_y$ and $\mylitrl_z$ between $z_r$ and $z_j$:
\begin{itemize}

\item[$\alpha$] no common (equal or opposite) literals: then, necessarily $z_r \diamond_x z_j$ is defined and is a 4\SAT{} clause moreover, since $n=5$, $z_r$ has two common literals with $z_i$ and, given $z_r \in z_{(k)}$ with induced clause $z_j$, it follows that they must be common equal literals;
\opt{margin_notes}{\mynote{mbh.note: Lemma~\ref{lemma_n4_case_alpha} shows that there can at max 2 clauses in $ z_{(k)}$ inducing same clause $z_j$}}%

\item[$\beta$] one common equal literal: it can be either $\mylitrl_u$ or the third literal of $z_j$: in both cases $z_r \diamond_x z_j$ is defined and is a 3\SAT{} clause;

\item[$\gamma$] two common equal literals: they are necessarily the two literals of $z_j$ different from $\mylitrl_x$, $z_r \diamond_x z_j$ is defined and is a 2\SAT{} clause (clause merging)%
%and also $z_r \diamond_u z_i$ is defined
;

\item[$\delta$] one common opposite literal: we already remarked that the unique possibility is that $\mylitrl_z$ is equal to $\mylitrl_u$ in $z_i$ and thus $z_r \notin z_{(j)}$, $z_r \notin z_{(i)}$ having respectively 2 and 0 common opposite literals; in this case $z_r \diamond_x z_k$ is a 4\SAT{} clause equal to $z_k$ without $\mylitrl_x$ and with $\mylitrl_u$ as in $z_i$ added and this independently of $\mylitrl_y$ being common equal with $z_i$ or $z_j$.
%If there are one common opposite and one common equal literals as above the common opposite literal must be $\mylitrl_u$ as in $z_i$ and again $z_r \notin z_{(j)}$, $z_r \notin z_{(i)}$ (case $\delta$). In both cases $\gamma$ and $\delta$ then $z_r \notin z_{(j)}$, $z_r \notin z_{(i)}$ and moreover $z_r \diamond_x z_k$ is the same 4\SAT{} clause in the two cases being equal to $z_k$ with $\mylitrl_x$ removed and same $\mylitrl_u$ added.

\end{itemize}

In following Lemma we prove that in three of these cases the 4\SAT{} clause $z_k = z_i \diamond_u z_j$ can be safely discarded since it is not needed to resolution algorithm to prove unsatisfiability of the given \SAT{} problem.

\begin{MS_lemma}
\label{lemma_removable_4clauses}
\opt{margin_notes}{\mynote{mbh.note: 'floating leaflets' p. 219}}%
Given an unsatisfiable 3\SAT{} problem \myBooleanS{} with $n = 5$ then any 4\SAT{} clause $z_k := z_i \diamond z_j$ can be safely discarded if either $z_{(k)} = \emptyset$ or all $z_r = (\mylitrl_x, \mylitrl_y, \mylitrl_z) \in z_{(k)}$ satisfy one of these conditions:
\begin{enumerate}

\item $z_r$ and its induced clause $z_j$ (or $z_i$) have one common opposite literal, $\mylitrl_x$, and one or two common equal literals (cases $\beta$ and $\gamma$);

\item $z_r$ and its induced clause $z_j$ (or $z_i$) have two common opposite literals (case $\delta$).

\end{enumerate}
\end{MS_lemma}
\begin{proof}
Given 4\SAT{} clause $z_k = z_i \diamond_u z_j$ it goes without saying that if $z_{(k)} = \emptyset$, $z_k$ can be discarded without affecting the resolution algorithm.
\opt{margin_notes}{\mynote{mbh.note: here implicit after repeated application of $\diamond$}}%

Given $z_k$ and $z_r = \mylitrl_x \mylitrl_y \mylitrl_z \in z_{(k)}$ with induced clause $z_j$ with one common opposite $\mylitrl_x$ and two common equal literals then $z_r \diamond_x z_j$ is defined and is a 2\SAT{} clause while for one common equal literal it is a 3\SAT{} clause and in both cases contains literal $\mylitrl_u$ of $z_j$. In the case of one common equal the third literal of $z_r$, $\mylitrl_z$, is not in $z_j$ and thus, since $n = 5$, necessarily is in $z_i$ and, since $z_r \in z_{(k)}$, is a common equal literal with $z_i$. It follows that in both cases $(z_r \diamond_x z_j) \diamond_u z_j$ is defined and is a 3\SAT{} clause. It is a simple, even if tedious, exercise to verify that, with these hypotheses, $z_r \diamond_x z_k = z_r \diamond_x (z_j \diamond_u z_i) = (z_r \diamond_x z_j) \diamond_u z_i$ and thus that clause $z_r \diamond_x z_k$ can be obtained avoiding 4\SAT{} clause $z_k$ but only through 3\SAT{} clauses $(z_r \diamond_x z_j) \diamond_u z_i$ and thus $z_k$ can be safely discarded.

For last case $\delta$ we already remarked that $z_r \diamond_x z_k$ is a 4\SAT{} clause equal to $z_k$ without $\mylitrl_x$ and with $\mylitrl_u$ as in $z_i$, moreover $z_r \diamond_x z_k$ contains as common equal literals the other two literals of $z_i$ (that were in $z_k$) and we can conclude that $z_r \diamond_x z_k$ is a 4\SAT{} clause with three common equal literals with 3\SAT{} clause $z_i$ and thus clause $z_r \diamond_x z_k$ is not needed since its coverage is strictly contained already in that of $z_i$ (clause subsumption).
%Moreover $z_r \notin z_{(j)}$, $z_r \notin z_{(i)}$ having respectively 2 and 0 common opposite literals and thus, with respect to $z_r$,
It follows that for all these $z_r$, the 4\SAT{} clause $z_k$ can be discarded.
\end{proof}

We removed cases $\beta, \gamma$ and $\delta$ thus 4\SAT{} clauses cannot be removed if there exists at least one $z_r \in z_{(k)}$, of type $\alpha$, with only one common opposite with its induced clause $z_j$ where both $z_i \diamond_u z_j = z_k$ and $z_r \diamond_x z_k$ are 4\SAT{}. To analyze these cases we need some further results.

\begin{MS_lemma}
\label{lemma_minimal_unsat_prb}
\opt{margin_notes}{\mynote{mbh.note: 'floating leaflets' p. 211}}%
All the $2^n$ \SAT{} problems with $n > 0$ literals made by the following $n + 1$ clauses: $n$ 1\SAT{} clauses $z_i = \mylitrl_i$ or $\myconjugate{\mylitrl}_i$ and one $n$\SAT{} clause $z_{n + 1} = \myconjugate{z}_1 \myconjugate{z}_2 \cdots \myconjugate{z}_n$ are unsatisfiable.
\end{MS_lemma}
\begin{proof}
We proceed by induction on $n$; for $n = 1$ the problem made by the two clauses \eg $z_1 = \mylitrl_1$ and $z_2 = \myconjugate{\mylitrl}_1$ is unsatisfiable. Let the proposition be true for $n - 1$ and we move to $n$ where we have \eg problem
$$
z_1 = \mylitrl_1, \; z_2 = \myconjugate{\mylitrl}_2, \ldots, \; z_{n-1} = \mylitrl_{n-1}, \; z_n = \myconjugate{\mylitrl}_n, \;\; z_{n+1} = \myconjugate{\mylitrl}_1 \mylitrl_2 \cdots \myconjugate{\mylitrl}_{n-1} \mylitrl_n
$$
then, by resolution, we can augment it with $z_n \diamond_n z_{n+1} = \myconjugate{\mylitrl}_1 \mylitrl_2 \cdots \myconjugate{\mylitrl}_{n-1}$ and the $n$ clauses $z_1, z_2, \ldots, z_{n-1}, z_n \diamond_n z_{n+1}$ form an unsatisfiable problem by induction hypothesis and thus also the initial $n$ problem is unsatisfiable.
%
%
%
%
%
%it is equivalent to the problem with $n - 1$ literals $z_1 = \mylitrl_1, z_2 = \myconjugate{\mylitrl}_2, \ldots, z_{n-1} = \mylitrl_{n-1}, z_n \diamond z_{n+1} = \myconjugate{\mylitrl}_1 \mylitrl_2 \cdots \myconjugate{\mylitrl}_{n-1}$ that is unsatisfiable by hypothesis.
\end{proof}
It easy to verify that the only possible proof of unsatisfiability by resolution is, apart from permutations of 1\SAT{} clauses $z_1, \ldots, z_n$, by
$$
(( \cdots ((z_{n+1} \diamond z_n) \diamond z_{n-1}) \cdots) \diamond z_1) = \epsilon \dotinformula
$$
All problems of this Lemma can be transformed in regular $n$\SAT{} problems ``immersing'' them in a larger space adding to each $1$\SAT{} clause $n-1$ new literals that transform them in $n$\SAT{} clauses. If these new literals are different from the initial $n$ and chosen so that they do not forbid previous resolution chain, unsatisfiability of the initial problem becomes unsatisfiability in a subspace of dimension $n$. For example for cases of $n=2$ or $3$ we could have
$$
\myconjugate{\mylitrl}_1 (\mylitrl_3), \mylitrl_2 (\mylitrl_4), \mylitrl_1 \myconjugate{\mylitrl}_2 \qquad \mbox{and} \qquad \myconjugate{\mylitrl}_1 (\mylitrl_4 \mylitrl_5), \mylitrl_2 (\mylitrl_5 \mylitrl_6), \mylitrl_3 (\mylitrl_4 \mylitrl_6), \mylitrl_1 \myconjugate{\mylitrl}_2 \myconjugate{\mylitrl}_3
$$
where, for clarity, we put in parentheses the added $n-1$ literals and resolution applied to these clauses give respectively $\mylitrl_3 \mylitrl_4$ and $\mylitrl_4 \mylitrl_5 \mylitrl_6$. A case of $n=2$ can also become a $3$\SAT{} problem, \eg
$$
\myconjugate{\mylitrl}_1 (\mylitrl_3 \mylitrl_4), \mylitrl_2 (\mylitrl_3 \mylitrl_4), \mylitrl_1 \myconjugate{\mylitrl}_2 (\mylitrl_5)
$$
and resolution gives $\mylitrl_3 \mylitrl_4 \mylitrl_5$. That this case is not so peregrine is proved by

\begin{MS_lemma}
\label{lemma_n4_case_alpha}
All cases $\alpha$ of $3$\SAT{} problems with $n = 5$ are instances of unsatisfiable problems of Lemma~\ref{lemma_minimal_unsat_prb} with $2$ or $3$ literals.
\end{MS_lemma}
\begin{proof}
With notation used in case $\alpha$ we observe that $z_j$ plays the role of the clause that contains more than one literal since it has only one common different literal with both $z_r$ and $z_i$ (since $z_i \in z_{(j)}$) that are necessarily different and this proves the thesis since $z_r, z_i$ and $z_j$ form an unsatisfiable problems of Lemma~\ref{lemma_minimal_unsat_prb} with $2$ literals and $3$ clauses. Moreover, as already remarked in description of case $\alpha$, $z_i$ and $z_r$ have necessarily two common equal literals, the added literals of previous example.

If there exists other clauses $z_s \in z_{(k)}$ of same case $\alpha$ and with same induced clause $z_j$, it follows that $z_s$ has one common opposite literal with $z_k$ that cannot be the same opposite of $z_r$ and $z_j$ because this would imply $z_s = z_r$ since the other two literals are necessarily those common equal with $z_i$. It follows that the common opposite literal of $z_s$ and $z_j$ must be different both from that of $z_i$ and $z_j$ (since $z_i \in z_{(j)}$) and from that of $z_r$ and $z_j$ and that also $z_s$ has two common equal literals with $z_i$. This proves two facts: the first is that there can be at most two clauses $z_r, z_s \in z_{(k)}$ of case $\alpha$ inducing the same clause $z_j$ since $z_j$ is a $3$\SAT{} clause; the second is that in this case $z_r, z_i, z_s$ and $z_j$ form an unsatisfiable problems of Lemma~\ref{lemma_minimal_unsat_prb} with $3$ literals and $4$ clauses. There can be other clauses in $z_{(k)}$ but, if of type $\alpha$, they must necessarily induce the other clause $z_i$ and the same proof applies also to them.
\end{proof}

\begin{MS_Proposition}
\label{prop_spinor_sum_O3}
\opt{margin_notes}{\mynote{mbh.note: 'floating leaflets' p. 203}}%
Any two linearly independent simple spinors of $\myClg{}{}{\R^{3,3}}$ of equal helicity can be summed.
\end{MS_Proposition}
\begin{proof}
By Proposition~\ref{prop_5_BudinichP_1989} two simple spinors of $\myClg{}{}{\R^{n,n}}$ can be summed if and only if the incidence of their associated $n$ dimensional null subspaces is $n-2$ that in our case is $1$ and we now prove that in $\myClg{}{}{\R^{3,3}}$ this condition is fulfilled for any two different simple spinors of equal helicity.

Given any simple spinor $\psi_i$ of $\myClg{}{}{\R^{3,3}}$ and its induced null subspace $M(\psi_i) = (\Identity, t_i)$ the incidence between any two of them is given by all vectors $u \in \R^3$ such that $(u, t_i u) = (u, t_j u)$, namely $t_i u = t_j u$ or $t_j^T t_i u = u$. But $t_i, t_j \in \OO{3}$ and spinors of same helicity means that $\det t_i = \det t_j$ and thus $\det t_j^T t_i = 1$ so that $t_j^T t_i \in \SO{3}$. It is well known that all \SO{3} elements different from identity, this being excluded for $t_j^T t_i$ by hypothesis of linearly independent simple spinors, have one eigenvalue $+1$ and thus there exists exactly one direction $u$ such that $t_j^T t_i u = u$ (the ``rotation axis'' of Euclidean space) that proves our proposition.
\end{proof}

Some remarks about this Proposition that, even if simple, is the cornerstone of the proof that $P = NP$. That any $\SO{3}$ element preserves a direction is a simple consequence of Cartan theorem for Euclidean space \cite[Theorem~5.15]{Porteous_1995} since in $\R^3$ the maximum even number of reflections composing a proper rotation is two. The second remark is that this Proposition applies also to $\myClg{}{}{\R^{2,2}}$ proved looking at \OO{2} elements as at the subgroup of \OO{3} that preserve a given direction, \eg $\mygen_{i}$.
\opt{margin_notes}{\mynote{mbh.note: also proved observing that in $\myClg{}{}{\R^{2,2}}$ the incidence necessary to sum two simple spinors is $0$ and is well known that all \SO{2} elements different from identity do not have $+1$ eigenvalues}}%
That Proposition does not hold in \OO{4} is proved by counterexample given by spinors $\frac{1}{\sqrt 2} (\Identity + \mygen_{1} \mygen_{2}) \psi_\Identity$ and $\frac{1}{\sqrt 2} (\Identity + \mygen_{3} \mygen_{4}) \psi_\Identity$ of incidence $n-4$. The third remark is that given any two simple spinors respecting the condition of the Proposition their $\myFockB$ basis expansion can be summed directly (duly rearranging normalization factor)
%\eg
%$$
%(\Identity + \mygen_{1} \mygen_{2}) \psi_\Identity \myssSum \mygen_{1} \mygen_{3} \psi_\Identity = (\Identity + \mygen_{1} \mygen_{2} + \mygen_{1} \mygen_{3}) \psi_\Identity = (\Identity + \mygen_{1} (\mygen_{2} + \mygen_{3}) ) \psi_\Identity
%$$
that will turn out to greatly simplify spinor addition in many instances.

Before proving a formal result we study one of the instances of case $\alpha$ to illustrate the principles of the sum; let
$$
%z_1 = \myconjugate{\mylitrl}_1 \quad
%z_2 = \mylitrl_1 \myconjugate{\mylitrl}_2 \quad
%z_3 = \mylitrl_1 \mylitrl_2 \myconjugate{\mylitrl}_3 \quad
z_i = \mylitrl_1 \mylitrl_2 \mylitrl_5 \quad
z_j = \mylitrl_3 \myconjugate{\mylitrl}_4 \myconjugate{\mylitrl}_5 \qquad \qquad
z_k := z_j \diamond_5 z_i = \mylitrl_1 \mylitrl_2 \mylitrl_3 \myconjugate{\mylitrl}_4 \quad
z_r = \mylitrl_1 \mylitrl_2 \mylitrl_4
$$
where $z_r$ and $z_j$ have only one common opposite literal, $\mylitrl_4$, and thus $z_r \diamond_4 z_j = \mylitrl_1 \mylitrl_2 \mylitrl_3 \myconjugate{\mylitrl}_5$ is another 4\SAT{} clause. Since $z_r \in z_{(k)}$ then
$$
z_r \diamond_4 z_k = z_r \diamond_4 (z_j \diamond_5 z_i) = (z_r \diamond_4 z_j) \diamond_5 z_i = \mylitrl_1 \mylitrl_2 \mylitrl_3
$$
how it is easy to check and the action of $\diamond$ eliminates, namely renders free, literals $\mylitrl_4$ and $\mylitrl_5$ but in both cases passing through a 4\SAT{} clause. How proved in Lemma~\ref{lemma_n4_case_alpha} clauses $z_r, z_i$ and $z_j$ form, in $\mylitrl_4$ and $\mylitrl_5$, an unsatisfiable problem of Lemma~\ref{lemma_minimal_unsat_prb} for $n = 2$.

On the other hand with spinor sums let
$$
\frac{1}{\sqrt 2} (\Identity + \mygen_{3} \mygen_{4}) \psi_\Identity \in {\cal T}_i \quad
\mygen_{4} \mygen_{5} \psi_\Identity \in {\cal T}_j \quad
\frac{1}{\sqrt 2} (\Identity + \mygen_{3} \mygen_{5}) \psi_\Identity \in {\cal T}_r
$$
and since they are all even helicity spinors of the three dimensional subspace $R_3 := \my_span{\mygen_{3}, \mygen_{4}, \mygen_{5}}$ by Proposition~\ref{prop_spinor_sum_O3} they can be summed obtaining
$$
\frac{1}{\sqrt 3} (\Identity + \mygen_{3} \mygen_{5} + \mygen_{3} \mygen_{4}) \psi_\Identity = \frac{1}{\sqrt 3} \mygen_{3} (\mygen_{3} + \mygen_{4} + \mygen_{5}) \psi_\Identity \in {\cal T}_r + {\cal T}_i
$$
where in last equality we wrote it as the product of two vectors and the normalization factor comes from the second vector (this is a case of sum of clauses with no common opposite literals). The frozen part of this sum represents a generalized clause we never met before: its involutory part is given by just $\mylitrl_1 \mylitrl_2$ while the \SO{2} term acts in subspace $\my_span{\mygen_{3}, \mygen_{4} + \mygen_{5}}$ associated to bivector $\mygen_{3} (\mygen_{4} + \mygen_{5})$ and not in one of the subspaces associated to base bivectors $\mygen_{i} \mygen_{j}$ but from the point of view of spinors there is nothing unusual this being a perfectly legal element of ${\cal T}_r + {\cal T}_i$ since it respects Proposition~\ref{prop_subspace_SSpinors}. To calculate its coverage we write it explicitly as $\frac{1}{\sqrt 3} (\psi_\Identity + \mygen_{3} \mygen_{5} \psi_\Identity + \mygen_{3} \mygen_{4} \psi_\Identity)$ and, being the first two literals $\mylitrl_1 \mylitrl_2$ frozen, each term covers exactly one element of the $\myFockB$ basis of $\myClg{}{}{\R^{5,5}}$ and coverage is thus $3 = 2^{n - 4} + 2^{n - 5}$ that respects our bound (this unusual coverage corresponds to the mean of the coverages of a 3\SAT{} and a 4\SAT{} clause).

Always by Proposition~\ref{prop_spinor_sum_O3} we can also add the element of ${\cal T}_j$ obtaining
$$
\frac{1}{2} (\Identity + \mygen_{4} \mygen_{5} + \mygen_{3} (\mygen_{4} + \mygen_{5})) \psi_\Identity \in ({\cal T}_r + {\cal T}_i) + {\cal T}_j
$$
%and collecting the last two terms in $\mygen_{3} (\mygen_{4} + \mygen_{5})$
and with $\Identity + \mygen_{4} \mygen_{5} = \mygen_{4} (\mygen_{4} + \mygen_{5})$ we can rewrite this spinor as the product of two vectors
$$
\frac{1}{2} (\mygen_{3} + \mygen_{4})(\mygen_{4} + \mygen_{5}) \psi_\Identity = \frac{1}{2} (\Identity + \mygen_{3} \mygen_{4}) (\Identity + \mygen_{4} \mygen_{5}) \psi_\Identity
$$
and in second equality we recognize two terms giving full \SO{3} coverage in $R_3$, an example of powerful Proposition~\ref{prop_spinor_sum_O3} in action: any \SO{3} element can be written in spinor form as the product of two vectors (whereas \eg for \SO{4} four vectors are needed). Expanding this spinor as before we can easily verify that its coverage is $2^{n - 3}$ that respects the bound. With even sum $({\cal T}_r + {\cal T}_i) + {\cal T}_j$ we thus find: the involutive part $\mylitrl_1 \mylitrl_2$, arriving here from clauses $z_i$ and $z_r$, together with an \SO{2} element acting in subspace $\my_span{\mygen_{3} + \mygen_{4}, \mygen_{4} + \mygen_{5}}$ of $R_3$ that covers all even bivectors of subspace $R_3$ and that contains also \SO{2} coverage of subspace $(\mygen_{4}, \mygen_{5})$ with involutive $\mylitrl_3$.
\opt{margin_notes}{\mynote{mbh.note: do we need a better explanation of coverage of $({\cal T}_r + {\cal T}_i) + {\cal T}_j$ ?}}%
%
%A subset of these cases contain the involutive part $\mylitrl_1 \mylitrl_2 \mylitrl_3$ with an even bivector in subspace $(\mygen_{4}, \mygen_{5})$ namely \SO{2} coverage of subspace $(\mygen_{4}, \mygen_{5})$. To
To complete the coverage of subspace $(\mygen_{4}, \mygen_{5})$ with odd \OO{2} elements we take $\mygen_{5} \psi_\Identity \in {\cal T}_r$ and $\mygen_{4} \psi_\Identity \in {\cal T}_i$ and by Proposition~\ref{prop_spinor_sum_O3}
$$
\frac{1}{\sqrt 2} (\mygen_{5} + \mygen_{4}) \psi_\Identity = \frac{1}{\sqrt 2} \mygen_{4} (\Identity + \mygen_{4} \mygen_{5}) \psi_\Identity \in {\cal T}_r + {\cal T}_i
$$
with involutive part $\mylitrl_1 \mylitrl_2$ and of coverage $2^{n - 3}$ and altogether we got full \OO{2} coverage of subspace $(\mygen_{4}, \mygen_{5})$ and thus we may conclude that we obtain clause $\mylitrl_1 \mylitrl_2 \mylitrl_3$, as with resolution, with the very significant difference that in every step we remained in coverage greater than $2^{n - 4}$ proving that in all cases of type $\alpha$ we reproduce resolution results but within the coverage bound. We just remark that we could also arrive at clause $\mylitrl_1 \mylitrl_2 \mylitrl_3$ covering the \SO{2} part in subspace $(\mygen_{4}, \mygen_{5})$ with $\frac{1}{\sqrt 2} (\Identity + \mygen_{4} \mygen_{5}) \psi_\Identity \in {\cal T}_i + {\cal T}_j$ but this sum has coverage $2^{n - 4}$ and moreover conceals the larger \SO{3} coverage of $({\cal T}_r + {\cal T}_i) + {\cal T}_j$.

We can resume this process observing that%
\opt{margin_notes}{\mynote{mbh.note: this could be put more formally \eg in a Corollary of Proposition~\ref{prop_spinor_sum_O3}}}%
, whenever we have three or more clauses $z_i$ with $\psi_i \in {\cal T}_i$ of same helicity, that contain all three bivectors of a three dimensional subspace, there always exists a sum of these spinors that cover \OO{3} of this subspace in given helicity and we do not need to actually do the spinor sums.

\begin{MS_lemma}
\label{lemma_case_n5}
All unsatisfiable $3$\SAT{} problems with $n = 5$ literals can be solved with spinor sum remaining in coverage greater than $2^{n - 4}$.
\end{MS_lemma}
\begin{proof}
We already proved that for all cases in which $\diamond$ produces at most $3$\SAT{} clauses to arrive at the empty clause also $\myssSum$ finds the empty clause. Moreover we proved that in all cases in which $\diamond$ produces a $4$\SAT{} clause $z_k = z_i \diamond z_j$ and all $z_r \in z_{(k)}$ are of types $\beta, \gamma$ or $\delta$, $z_k$ is not necessary to prove unsatisfiability. $z_k$ is needed only if any $z_r \in z_{(k)}$ is a $3$\SAT{} clause of type $\alpha$ and in Lemma~\ref{lemma_n4_case_alpha} we proved that these cases are instances of unsatisfiable problem described in Lemma~\ref{lemma_minimal_unsat_prb} so it remains to prove that the solution sketched in previous example is of general validity.

By Lemma~\ref{lemma_n4_case_alpha}, with notation used in that proof, there can be one or two clauses in $z_{(k)}$ inducing the same clause $z_j$ and we start proving this last case so that \eg
$$
z_i = \mylitrl_1 \mylitrl_2 \mylitrl_5 \quad
z_j = \mylitrl_3 \myconjugate{\mylitrl}_4 \myconjugate{\mylitrl}_5 \qquad
z_k = z_i \diamond_5 z_j = \mylitrl_1 \mylitrl_2 \mylitrl_3 \myconjugate{\mylitrl}_4 \quad
z_r = \mylitrl_1 \mylitrl_2 \mylitrl_4 \quad
z_s= \mylitrl_1 \mylitrl_2 \myconjugate{\mylitrl}_3
$$
where $\mylitrl_1 \mylitrl_2$ are the common equal literals between $z_i, z_r$ and $z_s$ while $z_j$ defines the three dimensional subspace $R_3 = \my_span{\mygen_{3}, \mygen_{4}, \mygen_{5}}$ and $z_j$ defines a parity in this subspace, even in this example. Clauses $z_i, z_r$ and $z_s$ have necessarily only one literal in $R_3$ and so all of them can contain two terms with same parity of $z_j$ in their respective ${\cal T}$ namely
$$
\frac{1}{\sqrt 2} (\Identity + \mygen_{3} \mygen_{4}) \psi_\Identity \in {\cal T}_i \quad
\mygen_{4} \mygen_{5} \psi_\Identity \in {\cal T}_j \quad
\frac{1}{\sqrt 2} (\Identity + \mygen_{3} \mygen_{5}) \psi_\Identity \in {\cal T}_r \quad
\frac{1}{\sqrt 2} \mygen_{3} (\mygen_{4} + \mygen_{5}) \psi_\Identity \in {\cal T}_s
$$
so that with Proposition~\ref{prop_spinor_sum_O3} we can easily form
$$
\frac{1}{2} (\Identity + \mygen_{3} \mygen_{4} + \mygen_{3} \mygen_{5} + \mygen_{4} \mygen_{5}) \psi_\Identity = \frac{1}{2} (\mygen_{3} + \mygen_{4})(\mygen_{4} + \mygen_{5}) \psi_\Identity = \frac{1}{2} (\Identity + \mygen_{3} \mygen_{4}) (\Identity + \mygen_{4} \mygen_{5}) \psi_\Identity \in ({\cal T}_r + {\cal T}_i) + {\cal T}_j
$$
giving full \SO{3} coverage in $R_3$ with involutive part given by $\mylitrl_1 \mylitrl_2$. For the odd part we start from \eg
$$
\frac{1}{\sqrt 2} (\mygen_{3} + \mygen_{4}) \psi_\Identity \in {\cal T}_i \quad
\frac{1}{\sqrt 2} (\mygen_{3} + \mygen_{5}) \psi_\Identity \in {\cal T}_r \quad
\mygen_{3} \mygen_{4} \mygen_{5} \psi_\Identity \in {\cal T}_s
$$
so that we get
$$
\frac{1}{2} (\mygen_{3} + \mygen_{4} + \mygen_{5} + \mygen_{3} \mygen_{4} \mygen_{5}) \psi_\Identity = \frac{1}{2} \mygen_{3} (\Identity + \mygen_{3} \mygen_{4} + \mygen_{3} \mygen_{5} + \mygen_{4} \mygen_{5}) \psi_\Identity \in ({\cal T}_r+ {\cal T}_i) + {\cal T}_s
$$
so that also the odd \OO{3} part of $R_3$ is covered and we deduce that the three literals of $\sup z_j$ are free and that spinor sum obtains, in coverage greater than $2^{n-4}$, $\mylitrl_1 \mylitrl_2$ like we get by resolution since
$$
((z_j \diamond z_i) \diamond z_r) \diamond z_s = \mylitrl_1 \mylitrl_2
$$
but through the $4$\SAT{} clause $z_j \diamond z_i$ of coverage $2^{n-4}$. The other 7 possible cases of clause $z_j$ are proved almost identically. Also the cases of only one clause $z_r \in z_{(k)}$ are proved similarly and like in previous example.
\end{proof}

\vspace{11cm}
\subsection*{Dedication}
I dedicate this paper to the memory of my father Paolo Budinich who passed away in November 2013 not before passing on to me his overwhelming enthusiasm for simple spinors, at the heart of this work. Twenty years ago we took our first steps together along this winding track \cite{BudinichP_2006}.
\opt{margin_notes}{\mynote{mbh.note: another version is here commented together with ChatGPT comments}}%

\newpage
%\vspace{-1cm}
%\opt{x,std,AACA}{% in all cases - but arXiv, JMP & TCS - standard BibTeX bibliography
%
%\bibliographystyle{plain} % or plain or.... see e.g.\ http://amath.colorado.edu/documentation/LaTeX/reference/faq/bibstyles.html#styles
%\bibliography{mbh}
%}
\opt{arXiv,JMP,TCS}{% only for arXiv, JMP & TCS we need to include here the L-A-S-T version of file .bbl
%
%
%\begin{thebibliography}{1}
%

%
%\end{thebibliography}
%
%
}

\opt{final_notes}{
\newpage
%\vspace{5.4cm}
%\include{Paper_notes}

\section{Good stuff removed from the paper}
\label{sec_Good stuff removed}

\noindent Here follows - \emph{in total chaos} - several more or less reasonable parts written during the birth of this work. As a rule of thumb older stuff is towards the end since I usually add 'new' stuff here on top.

%\vspace{-0.2cm}
\myseparation
%\vspace{-0.5cm}

We define $\inv{t}$ to be the \emph{involutory subspace} of $t$, namely the subspace of $\R^n$ spanned by the eigenvectors corresponding to $\pm 1$ eigenvalues, namely $\inv{t} = \ker(t + \Identity) \oplus \ker(t - \Identity)$; obviously $\inv{t}^\perp$ is the subspace corresponding to complex eigenvalues and
$$
\inv{t} \oplus \inv{t}^\perp = \ker(t + \Identity) \oplus \ker(t - \Identity) \oplus \inv{t}^\perp = \R^n
$$
and moreover the restrictions of $t$ to any couple of these subspaces commute.

Applying this definition to our case for any $t \in {\cal T}_j$ then $\mysbs z_j \subseteq \inv{t}$ where the inclusion can be strict since \inv{t} can contain other directions in subspace $\my_span{\mysbs z_j}^\perp$. Since all $\mysbs z_j$ are subsets of generators we will focus here on cases in which \inv{t} is made by subset of generators but in general it can contain any direction.

%\vspace{-0.2cm}
\myseparation
%\vspace{-0.5cm}

%\newpage
\subsection{An actual algorithm (``old'' 2024 version of section 8.2)}
\label{subsec_8_Simple_Spinors_SAT_algorithm}
We put at work the theory of previous section to outline an algorithm that tests unsatisfiability and we start characterizing $\varphi$ (\ref{formula_phi_full_sup}) and its associated succession (\ref{formula_phi_succession}).

\begin{MS_lemma}
\label{lemma_phi_k_chars}
Given $\varphi_k$ (\ref{formula_phi_succession}) then $p_{k+1}, p_{k+2}, \ldots, p_n \in M(\varphi_k)$ and thus $\inv{\varphi_k} = p_{k+1} q_{k+1} \; p_{k+2} q_{k+2} \cdots p_n q_n$. The corresponding $t_k \in \OO{n}$, such that $M(\varphi_k) = (\Identity, t_k)$, has $\inv{t_k} = \my_span{\mygen_{k+1}, \mygen_{k+2}, \ldots, \mygen_n}$ its non involutory part being $\my_span{\mygen_1, \mygen_2, \ldots, \mygen_k}$.
\end{MS_lemma}
\begin{proof}
That $p_{k+1}, p_{k+2}, \ldots, p_n \in M(\varphi_k)$ is clear from $\varphi_k$ definition (\ref{formula_phi_succession}) since in the product there are no terms with $\{\mygen_{k+1}, \mygen_{k+2}, \ldots, \mygen_n \}$ that can alter $\psi_\Identity = p_1 q_1 \; p_2 q_2 \cdots p_n q_n$ so we need just to prove that for $k>1$ all $v \in \my_span{\mygen_1, \mygen_2, \ldots, \mygen_k}$ are non involutory, namely that $t_k v \ne \pm v$ that we prove by induction on $k$: for $k=2$ it is well known that for any $t_2 \in \SO{2}$ $\inv{t_2} = \{ 0 \}$; supposing this true for $k-1$ we move to $\varphi_{k} = (\Identity + \mygen_{k-1} \mygen_{k}) \varphi_{k-1}$ and thus $t_k = t_\frac{\pi}{2} t_{k-1}$ with $t_\frac{\pi}{2} \in \SO{2}$ acting in $\my_span{\mygen_{k-1}, \mygen_k}$: should there exist $v$ such that $t_k v = \pm v$ we would have $t_{k-1} v = \pm t_{-\frac{\pi}{2}} v$ and we can write $v = \alpha \mygen_k + \beta \mygen_{k-1} + v_2$ with $v_2 \in \my_span{\mygen_1, \mygen_2, \ldots, \mygen_{k-2}}$ and thus we would have $t_{-\frac{\pi}{2}} (\alpha \mygen_k + \beta \mygen_{k-1} + v_2) = \alpha' \mygen_k + \beta' \mygen_{k-1} + v_2$ where $v_2$ is unmodified since $t_{-\frac{\pi}{2}}$ acts in $\my_span{\mygen_{k-1}, \mygen_k}$ and it should be equal to $t_{k-1} (\alpha \mygen_k + \beta \mygen_{k-1} + v_2) = \alpha \mygen_k + t_{k-1} (\beta \mygen_{k-1} + v_2)$ since $t_{k-1}$ acts in $\my_span{\mygen_1, \mygen_2, \ldots, \mygen_{k-1}}$ and thus to have equality we should have $\alpha = \pm \alpha'$ that, since $\inv{t_{-\frac{\pi}{2}}} = \{ 0 \}$, can be realized only if $\alpha = \beta = 0$ that would imply that $t_{k-1} v_2 = \pm v_2$ against induction hypothesis.
\end{proof}

\begin{MS_lemma}
\label{lemma_Givens_all_in_set}
\opt{margin_notes}{\mynote{mbh.note: commented here there is a stronger version of this Lemma that we do not actually need.}}%
Given $\varphi_k$ (\ref{formula_phi_succession}) then $\varphi_k \in \sum_{j \in J} {\cal T}_j$, $J \subseteq \{ 1, 2, \ldots, m \}$ being a subset of clauses, if and only if for all $l \in [1, k]$, $\varphi_l \in \sum_{j \in J} {\cal T}_j$, moreover $\sup \varphi_1 \subset \sup \varphi_2 \subset \cdots \subset \sup \varphi_{k}$ of sizes $1, 2, \ldots, 2^{k-1}$%
\opt{margin_notes}{\mynote{mbh.note: 'size' of $\varphi_i$ is not defined}}%
.
\end{MS_lemma}
\begin{proof}
Let $\varphi_k \in \sum_{j \in J} {\cal T}_j$, we just saw that $p_k q_k$ is a fixed coordinate of $\varphi_{k-1}$ and thus also in all $\psi_\lambda \in \sup \varphi_{k-1}$ and thus in $\sup \mygen_{k-1} \mygen_{k} \varphi_{k-1}$ the $k-$th coordinate is $q_k$: this proves that $\sup \varphi_{k-1} \subset \sup \varphi_k$
and that $|\sup \varphi_{k}| = 2 |\sup \varphi_{k-1}|$ and thus for any $l \in [1, k]$ $\sup \varphi_l \subset \sup \varphi_k$ that proves the result; the converse is immediate.%
\opt{margin_notes}{\mynote{mbh.note: could add a Corollary saying that $\det \varphi_k = \det \varphi_l \; \forall l$}}%
\end{proof}

\begin{MS_lemma}
\label{lemma_phi_k_in_T_j}
Given $\varphi_k$ (\ref{formula_phi_succession}) for $k > 1$ then $\varphi_k \in {\cal T}_j$ if and only if $\sup z_j \subseteq \{\mygen_{k+1}, \mygen_{k+2}, \ldots, \mygen_n \}$ with all generators of $z_j$ in positive form.
\end{MS_lemma}
\begin{proof}
$\varphi_k \in {\cal T}_j$ if and only if $M(z_j) \subseteq M(\varphi_k)$ and being the involutory part of $\varphi_k$ such that $\my_span{p_{k+1}, p_{k+2}, \ldots, p_n} \subseteq M(\varphi_k)$ it follows $M(z_j) \subseteq \my_span{p_{k+1}, p_{k+2}, \ldots, p_n}$ and thus the thesis; the converse is immediate.
\opt{margin_notes}{\mynote{mbh.note: should I add a prove that $\sup z_j \subseteq \inv{t}$ ? See incidence definition}}%
\end{proof}
\noindent We remark that the case $\varphi_1 = \psi_\Identity$ is different since in this case the condition is $\sup z_j \subseteq \{\mygen_1, \mygen_2, \ldots, \mygen_n \}$.

We are interested to study cases in which $\varphi_{k}$ belongs to a sum of ${\cal T}_j$ and we start with the simplest example, namely the sum of two clauses ${\cal T}_j + {\cal T}_l$. We purposely exclude the trivial case in which both $\varphi_{k}$ and $\varphi_{k-1}$ belongs to a single ${\cal T}_j$, by Lemma~\ref{lemma_phi_k_in_T_j} the case in which $\sup z_j \subseteq \{\mygen_{k+1}, \mygen_{k+2}, \ldots, \mygen_n \}$ all in positive form.

\begin{MS_lemma}
\label{lemma_phi_k_in_sum}
Given $\varphi_{k} = (\Identity + \mygen_{k-1} \mygen_{k}) \varphi_{k-1}$ and ${\cal T}_j$ and ${\cal T}_l$ such that $\varphi_{k-1} \left\{ \begin{array}{l} \in {\cal T}_j \\ \notin {\cal T}_l \end{array} \right.$ and $\mygen_{k-1} \mygen_{k} \varphi_{k-1} \left\{ \begin{array}{l} \notin {\cal T}_j \\ \in {\cal T}_l \end{array} \right.$ then $\varphi_{k} \in {\cal T}_j + {\cal T}_l$ with the involutory part containing $(z_j \cup z_l) \setminus \{ \mygen_{k} \}$ namely a set that contains all the generators of $z_j$ and $z_l$ but $\mygen_{k}$, all in positive form. We can thus associate to $\varphi_{k}$ a composed clause $(z_j \cup z_l) \setminus \{ \mygen_{k} \}$ that contains a number of generators strictly lower than $|\sup z_j| + |\sup z_l|$.
\end{MS_lemma}
\begin{proof}
By Lemma~\ref{lemma_phi_k_in_T_j} $\sup z_j \subseteq \{\mygen_{k}, \mygen_{k+1}, \ldots, \mygen_n \}$ and since $\mygen_{k-1} \mygen_{k} \varphi_{k-1} \notin {\cal T}_j$ by Lemma~\ref{lemma_eit_in_Tj} $\mygen_{k} \in \sup z_j$ in positive form and thus, since $\mygen_{k-1} \mygen_{k} \varphi_{k-1} \in {\cal T}_l$ then also $\mygen_{k} \in \sup z_l$ but in negative form. This does not violate Lemma~\ref{lemma_phi_k_in_T_j} that does not apply to $\mygen_{k-1} \mygen_{k} \varphi_{k-1}$. It follows that $z_j$ and $z_l$ are clauses with one common opposite generator $\mygen_{k}$, while they can have common equal, positive generators in $\{\mygen_{k+1}, \mygen_{k+2}, \ldots, \mygen_n \}$, and Propositions~\ref{prop_subspace_SSpinors} and \ref{prop_Tau_j_k} apply. For any $\psi = \alpha \psi_j + \beta \psi_l \in {\cal T}_{j} + {\cal T}_{l}$ ($\alpha, \beta \in \R$), $\psi = v_1 v_2 \psi_j$ with $v_1 \wedge v_2 \psi_j \in {\cal T}_{l}$ and moreover for $v_1 = \mygen_{k}, v_2 = \mygen_{k} - \mygen_{k-1}$ we get exactly $v_1 v_2 \varphi_{k-1} = (\Identity + \mygen_{k-1} \mygen_{k}) \varphi_{k-1}$. By Proposition~\ref{prop_subspace_SSpinors} $\my_span{v_1, v_2} = \my_span{\mygen_{k-1}, \mygen_{k}}$ identifies the non incident part of the two null subspaces induced by the simple spinors being summed while all other directions must be in common. It follows that all generators $(z_j \cup z_k) \setminus \{ \mygen_{k} \}$ must be present in the involutory part of the sum, their number being $|\sup z_j| + |\sup z_l| - 2$ or less if there are also common equal generators.
\end{proof}

We can thus easily generalize the set ${\cal T}_j$ (\ref{formula_cal_T_j_def2}) associated to clause $z_j$ to a set ${\cal T}_x$ associated to an arbitrary, given, involutory part $z_x$,
$$
{\cal T}_x := \{\psi \in \mySpinorS_s : M(z_x) \subseteq M(\psi)\} \subset \mySpinorS_s
$$
the involutory part typically being, like in previous Lemma, associated to a sum of simple spinors that builds $\varphi_{k}$
\begin{equation}
\label{formula_clause_composition}
z_x := (z_j \cup z_l) \setminus \{ \mygen_{k} \}
\end{equation}
that we define a \emph{composed} clause and gives the involutory part of $\alpha \psi_j + \beta \psi_l$. Proceeding to successive $\varphi_{k}$ we will use ${\cal T}_x$ while keeping separate track of the origin of the terms of the sum: $\varphi_k \in {\cal T}_{j} + {\cal T}_{l}$. We remark that given $\psi = \psi_j + \psi_l \in {\cal T}_{j} + {\cal T}_{l}$ when we want to iterate the procedure and add another simple spinor \eg $\phi \in {\cal T}_{r}$ to get $\phi + \psi$, Proposition~\ref{prop_subspace_SSpinors} applies again and thus necessarily $\phi + \psi = (\Identity + u_1 u_2) \psi = (\Identity + u_1 u_2) (\Identity + v_1 v_2) \psi_j$ and we can iterate previous Lemma to get that the involutions associated to $\phi + \psi$ will be $(((z_j \cup z_l) \setminus \{ \mygen_{k} \}) \cup z_r) \setminus \{ \mygen_{k+1} \}$ where we introduced this notation with brackets to indicate that this operation is \emph{not} associative and that in general $\phi + (\psi_j + \psi_l) \ne (\phi + \psi_j) + \psi_l$ since in general $(\phi + \psi_j)$ is not simple: for example let $\psi = \psi_j + \psi_k = \psi_\Identity + \mygen_{1} \mygen_{2} \psi_\Identity = (\Identity + \mygen_{1} \mygen_{2}) \psi_\Identity$ that is simple and we may add $\phi = \mygen_{3} \mygen_{4} \psi$ to get $\phi + \psi = (\Identity + \mygen_{3} \mygen_{4}) \psi = \mygen_{3} \mygen_{4} \psi + (\psi_\Identity + \mygen_{1} \mygen_{2} \psi_\Identity)$ where the parenthesis is strictly necessary since $\mygen_{3} \mygen_{4} \psi + \psi_\Identity$ is not a simple spinor since incidence of the two terms is $n-4$.

We remark that Lemma~\ref{lemma_phi_k_in_sum} is of general validity since it deals with \emph{two} simple spinors at the time and thus that we can apply it several times to form sum of more than two addends, each being associated to a plain or composed clause. It is simple to verify that composed clauses have all properties of real clauses like \eg Lemma~\ref{lemma_eit_in_Tj}, both being essentially identified by a given involutory part.
\opt{margin_notes}{\mynote{mbh.note: here follows commented an (unused) Corollary resuming Lemma~\ref{lemma_Givens_all_in_set}}}%
%
%\begin{MS_Corollary}
%\label{corollary_phi_k_in_sum}
%Given $\varphi_{k-1} \in {\cal T}_x$ then $\varphi_{k} \in {\cal T}_x + {\cal T}_y$ if and only if there exist a, plain or composed, clause $z_y$ that has $\mygen_{k}$ as common opposite generator with $z_x$.
%\end{MS_Corollary}
%\begin{proof}
%\end{proof}

\begin{MS_Proposition}
\label{prop_time_test}
Given $\varphi \in \mySpinorS_s$ (\ref{formula_phi_full_sup}) and $m$ sets ${\cal T}_j$ induced by clauses of a given \SAT{} problem, checking wether%
\opt{margin_notes}{\mynote{mbh.note: this proposition is probably flawed since there are cases in which we need $({\cal T}_j + {\cal T}_k) + ({\cal T}_l + {\cal T}_m)$, see 'floating leaflets' p. 103}}%
\begin{equation}
\label{formula_time_test}
\varphi \stackrel{?}{\in} \; \sum_{j = 1}^m {\cal T}_j
\end{equation}
can be done in polynomial time.
\end{MS_Proposition}
\begin{proof}
By Lemma~\ref{lemma_Givens_all_in_set} $\varphi \in \sum_{j = 1}^m {\cal T}_j$ if and only if all terms of the succession $\varphi_k$ (\ref{formula_phi_succession}) are in this set and thus our procedure to verify (\ref{formula_time_test}) starts from $\varphi_1 = \psi_\Identity$ and look after clause $z_j$ such that $\varphi_1 \in {\cal T}_j$ and continues increasing $k$ and searching a subset of clauses containing $\varphi_k$.

We start with $\varphi_1 = \psi_\Identity$ and in \bigO{n} we can find all clauses $z_j$ with all generators in positive form such that $\varphi_1 \in {\cal T}_j$ (obviously there may be many; should there be none by Lemma~\ref{lemma_Givens_all_in_set} $\varphi \notin \sum_{j = 1}^m {\cal T}_j$ and $\Identity$ is a solution for the given \SAT{} problem). For the next step $\varphi_2 = (\Identity + \mygen_1 \mygen_2) \varphi_1$: either $\mygen_1 \mygen_2 \varphi_1$ is also in ${\cal T}_j$, that thus contains also $\varphi_2$, or we have to find another clause $z_l$ such that $\mygen_1 \mygen_2 \varphi_1 \in {\cal T}_l$. Again: if $\varphi_1 \in {\cal T}_l$ then both $\varphi_1$ and $\varphi_2$ are in ${\cal T}_l$, if $\varphi_1 \notin {\cal T}_l$ then $\varphi_2 \in {\cal T}_j + {\cal T}_l$. We remark that $\varphi_2$ case is the only case (with respect to Lemma~\ref{lemma_phi_k_in_sum}) in which $z_j$ and $z_l$ may have also $2$ common opposite generators, see Proposition~\ref{prop_Tau_j_k}, but also in this case it is easy to define the associated composed clause $z_x := (z_j \cup z_l) \setminus \{ \mygen_{1}, \mygen_{2} \}$. If such a clause $z_l$ do not exist then $\mygen_1 \mygen_2 \varphi_1$ is a solution and we can stop here. At worst the time to find $\varphi_2 \in {\cal T}_x$ is \bigO{n^2} since we look for a couple of clauses $z_j$ and $z_l$ with one or two common opposite generators.

Before going to cases $\varphi_k$ with $k>2$ we remark that we will need composed clauses derived from repeated applications of Lemma~\ref{lemma_phi_k_in_sum} and we can start calculating, preliminarly and once for all, the composed clauses we can form with our sets ${\cal T}_j$ by repeated application of composition rule (\ref{formula_clause_composition}).

To do this we define an undirected graph of adjacency matrix $G$ with a node for each clause and an edge between clauses that have a common opposite generator. Let there exists an edge between clauses $z_j$ and $z_l$ with common opposite generator $\mygen_i$: this edge induces a composed clause given by $(z_j \cup z_l) \setminus \{ \mygen_i \}$. If moreover $z_l$ has an edge with $z_r$, with different common opposite generator $\mygen_j$, this second edge induces a further composed clause given by $((((z_j \cup z_l) \setminus \{ \mygen_i \}) \cup z_r) \setminus \{ \mygen_j \})$ since $\mygen_j \in \sup z_l$ is also $\mygen_j \in \sup ((z_j \cup z_l) \setminus \{ \mygen_i \})$.

This process can be iterated and it is simple to see that all composed clauses can be obtained in this fashion.
\opt{margin_notes}{\mynote{mbh.note: should mention difference between $n$ and $m$ or say assuming $\bigO{n} = \bigO{m}$}}%
The composed clauses obtained in $r$ iterations are given by matrix powers $G^r$ that is an \bigO{n^3} process. We remark that $G^r$ contains also illegally composed clauses in which the same common opposite generator is used more than once. The maximum power of the adjacency matrix is $n$ since there are at maximum $n$ different common opposite generators.

For any $r$, $G^r$ contains \bigO{n^2} composed clauses and so all possible composed clauses are \bigO{n^3} and to calculate them we repeated $n$ times an \bigO{n^3} matrix multiplication. In summary to build the \bigO{n^3} composed clauses we use \bigO{n^4} time but this process is done just once.

We are now ready to go to the generic case of $\varphi_k$ and we assume $\varphi_{k-1} \in {\cal T}_x$, $z_x$ being a plain or composed clause. There are only three mutually exclusive possibilities that we examine in succession:
\begin{enumerate}
\item $\mygen_{k-1} \mygen_{k} \varphi_{k-1} \in {\cal T}_x$, by Lemma~\ref{lemma_eit_in_Tj} this happens if and only if $\mygen_{k} \notin \sup z_x$ that can be easily verified in \bigO{n} simply searching $\sup z_x$;

\item there exists one plain or a composed clause $z_y$, such that $\mygen_{k-1} \mygen_{k} \varphi_{k-1} \in {\cal T}_y$. Since we are not in case $1$, $\mygen_{k-1} \mygen_{k} \varphi_{k-1} \notin {\cal T}_x$ and there are only two possibilities: if also $\varphi_{k-1} \in {\cal T}_y$ then both $\varphi_{k-1}$ and $\varphi_k$ are in ${\cal T}_y$ and this can be verified in \bigO{n} checking involutory parts of $z_y$ and of $\varphi_{k-1}$. If $\varphi_{k-1} \notin {\cal T}_y$ then we are in the conditions of Lemma~\ref{lemma_phi_k_in_sum} and to prove that $\varphi_k \in {\cal T}_x + {\cal T}_y$ we have to find $z_y$, plain or composed, such that $\mygen_{k-1} \mygen_{k} \varphi_{k-1} \in {\cal T}_y$ (we assume to be in cases with $k>2$ where Lemma~\ref{lemma_phi_k_in_sum} applies strictly). Any clause $z_y$, plain or composed, with the following characteristics: no generators in $\{ \mygen_1, \mygen_2, \ldots, \mygen_{k-1} \}$, generator $\mygen_k$ in negative form and any generator in $\{ \mygen_{k+1}, \mygen_{k+2}, \ldots, \mygen_n \}$ in positive form, constitutes a valid candidate. If such a $z_y$ exists it is necessarily in the set $\{ G, G^2, \ldots, G^n \}$ and can be found in \bigO{n^3};

\item if neither of previous cases occurred than we have a certificate that $\mygen_{k-1} \mygen_{k} \varphi_{k-1} \notin \sum_{j = 1}^m {\cal T}_j$ and thus $\varphi_{k}$ and by Lemma~\ref{lemma_Givens_all_in_set} also $\varphi$ and we can stop searching.
\end{enumerate}

So at every level checking for $\varphi_k$ is an \bigO{n^3} task and the process repeated for $n$ levels is at most an \bigO{n^4} task.
\end{proof}

\noindent In summary we proposed a proof-of-concept algorithm working for both $2$ and $3$\SAT{} problems. It is intriguing that it shows loose similarities with \emph{resolution}, when it composes two clause with one common opposite Boolean variable, and with \emph{transitive closure}, when it builds the graph of composed clauses.

%\vspace{-0.2cm}
\myseparation
%\vspace{-0.5cm}

%\newpage
\begin{MS_lemma}
\label{lemma_pi_qi}
Given any spinor $\psi \in \mySpinorS$ (simple or not) and any Witt basis element (\ref{formula_Witt_basis}) $x_i \in \{ p_i, q_i \}$ then $x_i \in M(\psi)$ if and only if $x_i \in M(\psi_\lambda)$ for all $\psi_\lambda \in \sup \psi$ (\ref{formula_support_spinor}).
\end{MS_lemma}
\begin{proof}
Let \eg $p_r \in M(\psi)$, this implies $p_r \psi = 0$ that with Fock basis expansion of $\psi$ (\ref{formula_Fock_basis_expansion}) becomes $\sum_\lambda \alpha_\lambda p_r \psi_\lambda = 0$. For any $\psi_\lambda$ such that $p_r \in M(\psi_\lambda)$, $p_r \psi_\lambda = 0$. Should there exist in $\psi$ expansion also terms such that $p_r \psi_\lambda \ne 0$, namely terms containing $q_r$, by multiplication by $p_r$ we would obtain $p_r \psi_\lambda := \psi_{\lambda'} \ne 0$ where $\psi_{\lambda'} \in \myFockB$ contains $p_r q_r$. Then $p_r \psi = 0$ would imply $\sum_{\lambda} \alpha_{\lambda} \psi_{\lambda'} = 0$, impossible to satisfy being Fock basis elements linearly independent. We must conclude that in $\psi$ expansion there are no terms such that $p_r \psi_\lambda \ne 0$ and thus the thesis. The converse is immediate.
\end{proof}

We remark that this property holds only for null vectors that are Witt basis elements how the following example shows for null vector $p_1 - q_2$
$$
(p_1 - q_2) (q_1 \; q_2 + p_1 q_1 \; p_2 q_2) = 0 \dotinformula
$$

\begin{MS_lemma}
\label{lemma_sum_psi_in_T}
Given two linearly independent simple spinors $\psi, \phi \in \mySpinorS_s$ such that $\psi + \phi \in \mySpinorS_s$ then $\psi + \phi \in {\cal T}_j$ for some $j$ if and only if both $\psi, \phi \in {\cal T}_j$.
\opt{margin_notes}{\mynote{mbh.note: can be generalized to $M(z) \subseteq M(\psi)$ to any $M(\sum_i \varphi_i = \psi)$ simple or not}}%
\end{MS_lemma}
\begin{proof}
Let $\psi, \phi \in {\cal T}_j$ then $M(z_j) \subseteq M(\psi), M(\phi)$ and thus $M(z_j) \subseteq M(\psi) \cap M(\phi)$ and by Proposition~\ref{prop_5_BudinichP_1989} $M(z_j) \subseteq M(\psi + \phi)$.
\opt{margin_notes}{\mynote{mbh.note: do we need $n > k+2$ ?}}%
Conversely let $M(z_j) \subseteq M(\psi + \phi)$ and since $M(z_j) = \my_span{x_1, \ldots, x_k}, x_i \in \{ p_i, q_i \}$ (\ref{formula_MTNS_M_psi}) it follows by Lemma~\ref{lemma_pi_qi} that for all $\psi_\lambda \in \sup{(\psi + \phi)}$ (\ref{formula_support_spinor}) we have $M(z_j) \subseteq M(\psi_\lambda)$ and thus also $M(z_j) \subseteq M(\psi), M(\phi)$.
\opt{margin_notes}{\mynote{mbh.note: see also last two paragraphs of section~\ref{subsec_8_Simple_Spinors_SAT_theory} and Lemma~\ref{lemma_phi_k_chars}}}%
\end{proof}

\opt{margin_notes}{\mynote{mbh.note: commented here there is an old, dubious, Lemma}}%
%
%\begin{MS_lemma}
%\label{lemma_sum_psi_in_union}
%Given simple spinors $\psi_j \in {\cal T}_j$ and $\psi_k \in {\cal T}_k$ such that $\psi_j + \psi_k = (\Identity + \mygen_{p} \mygen_{q}) \psi_j \in {\cal T}_j + {\cal T}_k$ then $\psi_j + \psi_k \in {\cal T}_j \cup {\cal T}_k$ if and only if $\psi_j + \psi_k \in {\cal T}_j \cap {\cal T}_k$ that is thus non empty (namely $z_j$ and $z_k$ have no common opposites).
%\end{MS_lemma}
%\begin{proof}
%%
%\opt{margin_notes}{\mynote{mbh.note: this is wrong: \eg $\psi_j$ and $\psi_k$ have common opposite $\mygen_{p} \in \sup z_k$}}%
%%
%Let $\psi_j + \psi_k \in {\cal T}_j \cup {\cal T}_k$, by Proposition~\ref{prop_Tau_j_k} the only possibility is that $z_j$ and $z_k$ have no common opposites and ${\cal T}_j \cap {\cal T}_k \ne \emptyset$. There are three mutually exclusive possibilities: the first is $\psi_j + \psi_k \in {\cal T}_j - {\cal T}_j \cap {\cal T}_k$, by Lemma~\ref{lemma_sum_psi_in_T} also $\psi_k \in {\cal T}_j - {\cal T}_j \cap {\cal T}_k$ that would imply the contradiction $\psi_k \notin {\cal T}_k$ that excludes this case, case $\psi_j + \psi_k \in {\cal T}_k - {\cal T}_j \cap {\cal T}_k$ is excluded similarly and the only remaining possibility is $\psi_j + \psi_k \in {\cal T}_j \cap {\cal T}_k$. The converse is immediate.
%\end{proof}

\begin{MS_lemma}
\label{lemma_v_psi}
Given simple spinor $\psi \in \mySpinorS_s$ and a vector $v$ such that $v^2 > 0$ and $| \sup v | > 1$, namely $v = v_i \mygen_{i} + v_j \mygen_{j} + \cdots$, then $v \psi \in \mySpinorS_s$ and for any $\mygen_{i} \in \sup v$ such that, for the corresponding $x_i \in \{ p_i, q_i \}$, $x_i \in M(\psi)$ then $x_i \notin M(v \psi)$.
\end{MS_lemma}
\begin{proof}
In given hypotheses by Lemma~\ref{lemma_same_support} $v \psi \in \mySpinorS_s$ (to be generalized) and given $\mygen_{i}$ with given characteristics from $x_i \in M(\psi)$ with Lemma~\ref{lemma_pi_qi} also $x_i \in M(\psi_\lambda)$ for all $\psi_\lambda \in \sup \psi$ then $v \psi = (v_i \mygen_{i} + v_j \mygen_{j} + \cdots) \sum_\lambda \alpha_\lambda \psi_\lambda = v_i \sum_\lambda \alpha_\lambda \mygen_{i} \psi_\lambda + v_j \sum_\lambda \alpha_\lambda \mygen_{j} \psi_\lambda + \cdots$ and in the final sum we find both $\psi_\lambda$ containing respectively $p_i q_i$ and $q_i$ and, by same Lemma, $x_i \notin M(v \psi)$.
\end{proof}

We remark that the hypothesis $x_i \in M(\psi)$ is essential to get $x_i \notin M(v \psi)$, for example let $\psi = v \psi_\Identity$ that surely has some $p_i \notin M(\psi)$ but $v \psi = v^2 \psi_\Identity$ has all $p_i \in M(v \psi)$.
\opt{margin_notes}{\mynote{mbh.note: there are also $x_i \in M(\psi), M(v \psi)$ but by Lemma~\ref{lemma_pi_qi} $\mygen_{i} \notin \sup v$}}%

\newpage
\begin{MS_Proposition}
\label{prop_clause_composition5}
Given simple spinors $\psi_j \in {\cal T}_j$ and $\psi_k \in {\cal T}_k$ such that $\psi_j + \psi_k \in \mySpinorS_s$ and thus $\psi_j + \psi_k = (\Identity + \mygen_{p} \mygen_{q}) \psi_j \in {\cal T}_j + {\cal T}_k$ with $z_j$ and $z_k$ with common opposite literal $\mygen_{p}$ and for some $\mygen_{q} \notin \sup z_j \cap \sup z_k$, then, given $\psi_l \in {\cal T}_l$ with common opposite literal $\mygen_{t} \ne \mygen_{p}$ with $\psi_j + \psi_k$, we can form
$$
(\psi_j + \psi_k) + \psi_l \in ({\cal T}_j + {\cal T}_k) + {\cal T}_l
$$
if and only if there exist $v \in \my_span{\sup (z_j + z_k) \cap \sup z_l - \{\mygen_{t}\}}^\perp$ such that $\mygen_{t} \wedge v (\psi_j + \psi_k) = \psi_l \in {\cal T}_l$, if moreover $\mygen_{p} \in \sup z_l$ then, necessarily, $v = \alpha \mygen_{t} + \frac{1}{2} (\mygen_{p} - \mygen_{q})$.
\end{MS_Proposition}
\begin{proof}
We first remark that $\mygen_{p}, \mygen_{q}, \mygen_{t}$ define a three dimensional subspace since by given hypotheses $\mygen_{p} \ne \mygen_{t}$, $\mygen_{p} \ne \mygen_{q}$ and from hypothesis that $\mygen_{t}$ is in the involutive part of $\psi_j + \psi_k$ (in turn implying $\mygen_{t} \in \sup z_j \cup \sup z_k$) we get also $\mygen_{q} \ne \mygen_{t}$ since, should $\mygen_{q} = \mygen_{t}$, by Lemma~\ref{lemma_v_psi} $\mygen_{t}$ could not be involutive in $\psi_j + \psi_k$.

Let $(\psi_j + \psi_k) + \psi_l \in ({\cal T}_j + {\cal T}_k) + {\cal T}_l$, by definition (\ref{formula_cal_T_j+k_def}) the spinor is simple and, given that $\psi_l$ and $\psi_j + \psi_k$ have common opposite literal $\mygen_{t} \ne \mygen_{p}$, by Proposition~\ref{prop_Tau_j_k} $(\psi_j + \psi_k) + \psi_l = \mygen_{t} v (\psi_j + \psi_k)$ with $v \in \my_span{\sup (z_j + z_k) \cap \sup z_l - \{\mygen_{t}\}}^\perp$ and
$$
(\psi_j + \psi_k) + \psi_l = (\mygen_{t} \cdot v + \mygen_{t} \wedge v) (\psi_j + \psi_k) = (\mygen_{t} \cdot v + \mygen_{t} \wedge v) (\Identity + \mygen_{p} \mygen_{q}) \psi_j \in ({\cal T}_j + {\cal T}_k) + {\cal T}_l
$$
with $\mygen_{t} \wedge v (\psi_j + \psi_k) = \mygen_{t} \wedge v (\Identity + \mygen_{p} \mygen_{q}) \psi_j = \psi_l \in {\cal T}_l$.

If $\mygen_{p} \in \sup z_l$, and thus $\mygen_{p}$ is in the involutive part of $\psi_l$, this imposes further constraints to $v$ and to simplify the notation we pose $v = \alpha \mygen_{t} + v'$ so that $\mygen_{t} \cdot v = \alpha$ and $\mygen_{t} \wedge v = \mygen_{t} v'$. From $\mygen_{t} v' (\psi_j + \psi_k) = \psi_l$ we observe that $\mygen_{t}$ is common opposite in $\psi_j + \psi_k$ and $\psi_l$ and thus that it cannot be in $M(\psi_j + \psi_k) \cap M(\psi_l)$ and neither $\mygen_{p}$ can be since it is non involutive in $\psi_j + \psi_k$ and involutive in $\psi_l$ so that the 'rotation' subspace must be given in this case by $\mygen_{t}, \mygen_{p}$ and thus that $\mygen_{t} v' (\psi_j + \psi_k) = \mygen_{t} v' (\Identity + \mygen_{p} \mygen_{q}) \psi_j = \psi_l$ and thus that, necessarily, $v' (\Identity + \mygen_{p} \mygen_{q}) = \mygen_{p}$, that right multiplied by $(\Identity + \mygen_{p} \mygen_{q})^{-1} = \frac{1}{2} (\Identity - \mygen_{p} \mygen_{q})$, gives $v' = \frac{1}{2} (\mygen_{p} - \mygen_{q})$ and finally $v = \alpha \mygen_{t} + \frac{1}{2} (\mygen_{p} - \mygen_{q})$.

The converse is immediate.
\end{proof}

\begin{MS_Corollary}
\label{corollary_clause_composition5}
Given $(\psi_j + \psi_k) + \psi_l \in ({\cal T}_j + {\cal T}_k) + {\cal T}_l$ then both ${\cal T}_k + {\cal T}_l$ and ${\cal T}_j + {\cal T}_l$ are defined.
\end{MS_Corollary}
\begin{proof}
Let suppose first $\mygen_{p} \notin \sup z_l$ then $\mygen_{t} v' (\psi_j + \psi_k) = \psi_l \in {\cal T}_l$ and by Lemma~\ref{lemma_sum_psi_in_T} both $\mygen_{t} v' \psi_j$ and $\mygen_{t} v' \psi_k$ are in ${\cal T}_l$ and thus also $(\Identity + \mygen_{t} v') \psi_j \in {\cal T}_j + {\cal T}_l$ and $(\Identity + \mygen_{t} v') \psi_k \in {\cal T}_k + {\cal T}_l$ are defined. In the other case $\mygen_{p} \in \sup z_l$, supposing $\psi_j$ and $\psi_l$ with $\mygen_{p}$ common opposite, then $\mygen_{t} v' (\Identity + \mygen_{p} \mygen_{q}) \psi_j = \frac{1}{2} \mygen_{t} (\mygen_{p} - \mygen_{q}) (\Identity + \mygen_{p} \mygen_{q}) \psi_j = \mygen_{t} \mygen_{p} \psi_j \in {\cal T}_l$ and thus $(\Identity + \mygen_{t} \mygen_{p}) \psi_j \in {\cal T}_j + {\cal T}_l$. Similarly $\mygen_{t} v' (\Identity - \mygen_{p} \mygen_{q}) \psi_k = \frac{1}{2} \mygen_{t} (\mygen_{p} - \mygen_{q}) (\Identity - \mygen_{p} \mygen_{q}) \psi_k = - \mygen_{t} \mygen_{q} \psi_k \in {\cal T}_l$ and thus $(\Identity - \mygen_{t} \mygen_{q}) \psi_k \in {\cal T}_k + {\cal T}_l$. In any case, by Corollary~\ref{coro_Tau_j_k}, both $z_j$, $z_l$ and $z_k$, $z_l$ must have $0, 1$ or $2$ common opposite literals.
\end{proof}

\newpage
\begin{MS_Proposition}
\label{prop_clause_composition3}
\opt{margin_notes}{\mynote{mbh.note: this version does not assume that $\psi_k$ and $\psi_l$ have $\mygen_{t}$ as common opposite}}%
Given simple spinors $\psi_j \in {\cal T}_j$ and $\psi_k \in {\cal T}_k$ such that $\psi_j + \psi_k \in \mySpinorS_s$ and thus $\psi_j + \psi_k = (\Identity + \mygen_{p} \mygen_{q}) \psi_j \in {\cal T}_j + {\cal T}_k$ with $z_j$ and $z_k$ with common opposite literal $\mygen_{p}$ and for some $\mygen_{q} \notin \sup z_j \cap \sup z_k$, then there exist $\psi_l \in {\cal T}_l$ such that
$$
(\psi_j + \psi_k) + \psi_l \in ({\cal T}_j + {\cal T}_k) + {\cal T}_l
$$
if and only if there exist $\mygen_{t} \ne \mygen_{u}$ such that $\mygen_{t} \mygen_{u} (\psi_j + \psi_k) \in {\cal T}_l$.
\end{MS_Proposition}
\begin{proof}
Let $(\psi_j + \psi_k) + \psi_l \in ({\cal T}_j + {\cal T}_k) + {\cal T}_l$, by definition (\ref{formula_cal_T_j+k_def}) the spinor is simple and thus by Proposition~\ref{prop_subspace_SSpinors} there must exist
\opt{margin_notes}{\mynote{mbh.note: but why $\mygen_{t} \ne \mygen_{u}$ and not $v_1 \ne v_2$ ?}}%
$\mygen_{t} \ne \mygen_{u}$ such that
$$
(\psi_j + \psi_k) + \psi_l = (\Identity + \mygen_{t} \mygen_{u}) (\psi_j + \psi_k) = (\Identity + \mygen_{t} \mygen_{u}) (\Identity + \mygen_{p} \mygen_{q}) \psi_j \in ({\cal T}_j + {\cal T}_k) + {\cal T}_l
$$
with $\mygen_{t} \mygen_{u} (\psi_j + \psi_k) = \psi_l$; the converse is immediate.
\end{proof}

We remark that with Lemma~\ref{lemma_sum_psi_in_T} $\mygen_{t} \mygen_{u} (\psi_j + \psi_k) \in {\cal T}_l$ if and only if both $\mygen_{t} \mygen_{u} \psi_j$ and $\mygen_{t} \mygen_{u} \psi_k$ are in ${\cal T}_l$ and thus also $(\Identity + \mygen_{t} \mygen_{u}) \psi_j \in {\cal T}_j + {\cal T}_l$ and $(\Identity + \mygen_{t} \mygen_{u}) \psi_k \in {\cal T}_k + {\cal T}_l$ are defined and, by Corollary~\ref{coro_Tau_j_k}, both $z_j$ and $z_l$ and $z_k$ and $z_l$ must have $0, 1$ or $2$ common opposite literals.

\begin{MS_Proposition}
\label{prop_clause_composition4}
\opt{margin_notes}{\mynote{mbh.note: this version assumes that $\psi_j$ and $\psi_k$ have $\mygen_{p}$ as unique common opposite and $\psi_k$ and $\psi_l$ have $\mygen_{t}$ as unique common opposite}}%
Given three simple spinors $\psi_i \in {\cal T}_i$ for $i \in \{ j, k, l \}$ such that there exist $\mygen_{p}, \mygen_{q}, \mygen_{u}$ and $\mygen_{t} \ne \mygen_{p}$ such that following hypotheses hold:
\begin{itemize}
\item $\psi_j + \psi_k = (\Identity + \mygen_{p} \mygen_{q}) \psi_j \in {\cal T}_j + {\cal T}_k$ with $\psi_j$ and $\psi_k$ simple spinors of common opposite literal $\mygen_{p}$ for $ \mygen_{q} \notin \sup z_j \cap \sup z_k$,
\item $\psi_k + \psi_l = (\Identity + \mygen_{t} \mygen_{u}) \psi_k \in {\cal T}_k + {\cal T}_l$ with $\psi_k$ and $\psi_l$ simple spinors of common opposite literal $\mygen_{t}$ for $ \mygen_{u} \notin \sup z_k \cap \sup z_l$,
\end{itemize}
then
$$
(\psi_j + \psi_k) + \psi_l = (\Identity + \mygen_{t} \mygen_{u}) (\psi_j + \psi_k) = (\Identity + \mygen_{t} \mygen_{u}) (\Identity + \mygen_{p} \mygen_{q}) \psi_j \in ({\cal T}_j + {\cal T}_k) + {\cal T}_l
$$
if and only if: $\mygen_{t} \mygen_{u} \psi_j \in {\cal T}_l$, $\mygen_{q} \ne \mygen_{t}$ and $\mygen_{p} \notin \sup z_l$.
\end{MS_Proposition}

\begin{proof}
Let $(\psi_j + \psi_k) + \psi_l = (\Identity + \mygen_{t} \mygen_{u}) (\psi_j + \psi_k) \in ({\cal T}_j + {\cal T}_k) + {\cal T}_l$ then $\mygen_{t} \mygen_{u} (\psi_j + \psi_k) = \psi_l$ and by Lemma~\ref{lemma_sum_psi_in_T} $\mygen_{t} \mygen_{u} \psi_j \in {\cal T}_l$ and thus $(\Identity + \mygen_{t} \mygen_{u}) \psi_j \in {\cal T}_j + {\cal T}_l$ and we remark that by Corollary~\ref{coro_Tau_j_k} necessarily $z_j$ and $z_l$ must have $0$, $1$ or $2$ common opposites. By second hypothesis $\mygen_{t} \in \sup z_l$ and it must be involutive also in $\psi_j + \psi_k$ that requires $\mygen_{q} \ne \mygen_{t}$. Moreover since both $\mygen_{t} \mygen_{u} \psi_j$ and $\mygen_{t} \mygen_{u} \psi_k \in {\cal T}_l$ and, for any $\mygen_{u}$, $\mygen_{p}$ is common opposite also between them, that implies $\mygen_{p} \notin \sup z_l$.

Conversely by second hypothesis $\mygen_{t} \mygen_{u} \psi_k \in {\cal T}_l$ and let $\mygen_{t} \mygen_{u} \psi_j \in {\cal T}_l$, by first hypothesis $\psi_j + \psi_k$ is a simple spinor that, since $\mygen_{q} \ne \mygen_{t}$, inherits $\mygen_{t}$ from $\psi_k$ in its involutive part (\ref{formula_clause_composition}) and by Lemma~\ref{lemma_same_support} $\mygen_{t} \mygen_{u} (\psi_j + \psi_k)$ is also a simple spinor that has $\mygen_{p}$ in its non involutive part that thus can be summed with $\psi_l$
\opt{margin_notes}{\mynote{mbh.note: is $\mygen_{p} \notin \sup z_l$ really needed ?}}%
that has also $\mygen_{p} \notin \sup z_l$ and thus with Lemma~\ref{lemma_sum_psi_in_T} $\mygen_{t} \mygen_{u} (\psi_j + \psi_k) \in {\cal T}_l$ and finally $(\Identity + \mygen_{t} \mygen_{u}) (\psi_j + \psi_k) = (\Identity + \mygen_{t} \mygen_{u}) (\Identity + \mygen_{p} \mygen_{q}) \psi_j \in ({\cal T}_j + {\cal T}_k) + {\cal T}_l$.
\opt{margin_notes}{\mynote{mbh.note: commented here there are two previous versions of this Proposition}}%
\end{proof}

This gives the rules to compose $\psi_j + \psi_k \in {\cal T}_j + {\cal T}_k$ with $\psi_k + \psi_l \in {\cal T}_k + {\cal T}_l$ into $(\psi_j + \psi_k) + \psi_l \in ({\cal T}_j + {\cal T}_k) + {\cal T}_l$. We remark that from $\mygen_{t} \mygen_{u} \psi_j \in {\cal T}_l$, with $\mygen_{t} \in \sup z_l$, we can deduce that if $\mygen_{t}$ and $\mygen_{u}$ are in $\sup z_j \cap \sup z_l$ they can only be common opposites and if $\mygen_{t}$ is common opposite it is necessarily common equal with $z_k$ (beyond that $\mygen_{t}$ can not be common equal also because then $z_j$ and $z_k$ would have $\mygen_{p}$ and $\mygen_{t}$ common opposites). Finally beyond $\mygen_{t}$ and $\mygen_{u}$ there can be no other common opposites between $z_j$ and $z_l$ while there can be common equals.
\opt{margin_notes}{\mynote{mbh.note: tentative notation: $j \stackrel{{\bf p} q}{\to} k \stackrel{{\bf t} u}{\to} l$}}%

\myseparation

.....and just complete (\ref{formula_cal_T_j_def2}) porting the definition of ${\cal T}_j$ in $\R(n)$.

\begin{MS_Proposition}
\label{prop_cal_t_j}
Given clause $z_j$ and the corresponding involution $\lambda_j$ induced by its literals then
\begin{equation}
\label{formula_cal_T_j_def3}
{\cal T}_j = \{ t := \left(\begin{array}{r r} \lambda_j & \\ & t' \end{array}\right) : t' \in \OO{n-k} \} \subseteq \R(n)
\end{equation}
where we supposed, for ease of notation and without loss of generality, that the literals occupy the first $k$ coordinates of spacelike space $\{0\} \times \R^n$.
\end{MS_Proposition}
\begin{proof}
The proof is immediate since for any $t' \in \OO{n-k}$ then $t \in \OO{n}$ and $M(z_j) \subseteq (\Identity, t )$ and conversely any $t \in \OO{n}$ satisfying this property must have this form.
\end{proof}

So from now on we move to the vectorial representation of \OO{n} in matrices of $\R(n)$ and we ultimately transpose our initial \SAT{} problem to the verification if two given $t, t' \in \OO{n} \myisom \R(n)$ are in the subset $\sum_{j = 1}^m {\cal T}_j \subseteq \R(n)$ generated by the \SAT{} problem and composed by a linear combination of elements from ${\cal T}_j$ (\ref{formula_cal_T_j_def3}).

\myseparation

\begin{MS_Proposition}
\label{prop_time_test_old3}
Given $\varphi \in \mySpinorS_s$ (\ref{formula_phi_full_sup}) and $m$ sets ${\cal T}_j$ induced by clauses checking wether
\begin{equation}
\label{formula_time_test_old3}
\varphi \stackrel{?}{\in} \; \sum_{j = 1}^m {\cal T}_j
\end{equation}
can be done in polynomial time.
\end{MS_Proposition}
\begin{proof}
By Lemma~\ref{lemma_Givens_all_in_set} $\varphi \in \sum_{j = 1}^m {\cal T}_j$ if and only if all terms of the succession $\varphi_k$ (\ref{formula_phi_succession}) are in this set and thus our procedure to verify (\ref{formula_time_test}) starts from $\varphi_1 = \psi_\Identity$ and look after clause $z_j$ such that $\varphi_1 \in {\cal T}_j$ and continues increasing $k$ and searching a subset of clauses containing $\varphi_k$.

We start with $\varphi_1 = \psi_\Identity$ and in \bigO{n} we can find all clauses $z_j$ with all generators in positive form such that $\varphi_1 \in {\cal T}_j$ (obviously there may be many; should there be none by Lemma~\ref{lemma_Givens_all_in_set} $\varphi \notin \sum_{j = 1}^m {\cal T}_j$ and $\Identity$ is a solution for the given \SAT{} problem). For the next step $\varphi_2 = (\Identity + \mygen_1 \mygen_2) \varphi_1$: either $\mygen_1 \mygen_2 \varphi_1$ is also in ${\cal T}_j$, that thus contains also $\varphi_2$, or we have to find another clause $z_l$ such that $\mygen_1 \mygen_2 \varphi_1 \in {\cal T}_l$. Again: if $\varphi_1 \in {\cal T}_l$ then both $\varphi_1$ and $\varphi_2$ are in ${\cal T}_l$, if $\varphi_1 \notin {\cal T}_l$ then $\varphi_2 \in {\cal T}_j + {\cal T}_l$. We remark that $\varphi_2$ case is the only case (with respect to Lemma~\ref{lemma_phi_k_in_sum}) in which $z_j$ and $z_l$ may have also $2$ common opposite generators, see Proposition~\ref{prop_Tau_j_k}, but also in this case it is easy to define the associated composed clause $z_x := z_j \cup z_l - \{ \mygen_{1}, \mygen_{2} \}$. If such a clause $z_l$ do not exist then $\mygen_1 \mygen_2 \varphi_1$ is a solution of \SAT{} problem and we can stop here. At worst the time to find $\varphi_2 \in {\cal T}_x$ is \bigO{n^2} since we look for a couple of clauses $z_j$ and $z_l$ with one or two common opposite generators.

Going to the generic case of $\varphi_k$ we assume for a moment $\varphi_{k-1} \in {\cal T}_x$, $z_x$ being a plain or composed clause. There are only three mutually exclusive possibilities that we examine in succession:
\begin{enumerate}
\item $\mygen_{k-1} \mygen_{k} \varphi_{k-1} \in {\cal T}_x$, by Lemma~\ref{lemma_eit_in_Tj} this happens if and only if $\mygen_{k} \notin \sup z_x$ that can be easily verified in \bigO{n} simply searching $\sup z_x$;

\item there exists one plain or a composed clause $z_y$, such that $\mygen_{k-1} \mygen_{k} \varphi_{k-1} \in {\cal T}_y$. Since we are not in case $1$, $\mygen_{k-1} \mygen_{k} \varphi_{k-1} \notin {\cal T}_x$ and there are only two possibilities: if also $\varphi_{k-1} \in {\cal T}_y$ then both $\varphi_{k-1}$ and $\varphi_k$ are in ${\cal T}_y$ and this can be verified in \bigO{n} checking involutory parts of $z_y$ and of $\varphi_{k-1}$. If $\varphi_{k-1} \notin {\cal T}_y$ then we are in the conditions of Lemma~\ref{lemma_phi_k_in_sum} and we deal with this case in the sequel;

\item if neither of previous cases occurred than we have a certificate that $\mygen_{k-1} \mygen_{k} \varphi_{k-1} \notin \sum_{j = 1}^m {\cal T}_j$ and thus $\varphi_{k}$ and by Lemma~\ref{lemma_Givens_all_in_set} also $\varphi$.
\end{enumerate}

The instances of case two that remain to examine are those that satisfy conditions of Lemma~\ref{lemma_phi_k_in_sum} and to prove that $\varphi_k \in {\cal T}_x + {\cal T}_y$ we have to find $z_y$, plain or composed, such that $\mygen_{k-1} \mygen_{k} \varphi_{k-1} \in {\cal T}_y$. We observe that Lemma~\ref{lemma_phi_k_in_sum} condition $\mygen_{k-1} \mygen_{k} \varphi_{k-1} \left\{ \begin{array}{l} \notin {\cal T}_j \\ \in {\cal T}_l \end{array} \right.$ is equivalent to $\prod_{i=1}^{k-2} (\Identity + \mygen_{i} \mygen_{i+1}) \mygen_{k} \psi_\Identity \left\{ \begin{array}{l} \notin {\cal T}_j \\ \in {\cal T}_l \end{array} \right.$: since $\mygen_{k-1} \mygen_{k} \varphi_{k-1} = \mygen_{k-1} \mygen_{k} \prod_{i=1}^{k-2} (\Identity + \mygen_{i} \mygen_{i+1}) \psi_\Identity$ and $\mygen_{k}$ commutes with the product and we can move it in front of $\psi_\Identity$ and, by Lemma~\ref{lemma_eit_in_Tj}, we can further multiply $\mygen_{k-1} \varphi_{k-1}$ by $\mygen_{k-1}$ that is in the non involutory part of $\varphi_{k-1}$ and thus it is not in $\sup z_j \cup \sup z_l$. With this in mind we step back to the case of $\varphi_1$ and beyond searching $z_j$ such that $\varphi_1 = \psi_\Identity \in {\cal T}_j$ we search for clauses that contain $\mygen_i \psi_\Identity$ for $i \in [3, n]$ that we indicate with ${\cal T}_j^i$ and, again, should one of them do not exist this would be a \SAT{} solution. Similarly, at next step, beyond $\varphi_2$ we look for clauses containing $\mygen_1 \mygen_2 \; \mygen_i \psi_\Identity$ for $i \in [3, n]$. When we have all these clauses we can calculate, for each $i \in [3, n]$, the sets containing $(\Identity + \mygen_1 \mygen_2) \mygen_i \psi_\Identity$ and their relative sets associated to composed clauses that we name ${\cal T}_x^i$. All in all building these sets is a \bigO{n^3} task since we need to repeat $n-2$ times an \bigO{n^2} search.

Back to the generic case of $\varphi_k$ and, supposing we updated sets ${\cal T}_x^i$ at any step, when we check $\varphi_k$ we already know that $\varphi_{k-1} \in {\cal T}_x$ with $z_x$ a simple or composed clause and from our sets we know already wether $\prod_{i=1}^{k-2} (\Identity + \mygen_{i} \mygen_{i+1}) \mygen_{k} \psi_\Identity \in {\cal T}_x^k$ that we saw it is equivalent to $ \mygen_{k-1} \mygen_{k} \varphi_{k-1} \in {\cal T}_x^k$ and thus we can conclude that $\varphi_k \in {\cal T}_x + {\cal T}_x^k$ and update the composed clause containing $\varphi_k$ to $z_x \cup z_x^k - \{ \mygen_k \}$.

Before moving to next $\varphi_{k+1}$ we need to update the lists of sets ${\cal T}_x^i$ for $i \in [k+1, n]$ (since we just used set ${\cal T}_x^k$) adding term $(\Identity + \mygen_{k-1} \mygen_{k})$ that we can do looking for clauses, plain or composed, that contain $\mygen_{k-1} \mygen_{k} \prod_{i=1}^{k-2} (\Identity + \mygen_{i} \mygen_{i+1}) \mygen_{i} \psi_\Identity$, namely a clause with no generators in $\{ \mygen_1, \mygen_2, \ldots, \mygen_{k-1} \}$, generators $\mygen_i$ and $\mygen_k$ in negative form and remaining generators all in positive form that again is an \bigO{n^3} task.
\opt{margin_notes}{\mynote{mbh.note: beware: the number of clauses is increased by composed clauses, even if the product $m k$ probably decreases. Is there substitution of $z_j$ and $z_l$ by $z_x$ ?}}%

So at every level checking for $\varphi_k$ and updating the list of clauses or sum of clauses started by $\mygen_i \psi_\Identity$ is an \bigO{n^3} task and so the process repeated for $n$ levels is at most an \bigO{n^4} task.
\end{proof}

\myseparation

\begin{MS_lemma}
\label{lemma_buh}
Given $\varphi_{k-1} \in {\cal T}_j$ and $\varphi_{k} = (\Identity + \mygen_{k-1} \mygen_{k}) \varphi_{k-1} \in {\cal T}_l$ then $\varphi_{k} \in {\cal T}_x$ where $z_x = z_j \cup z_l - \{ \mygen_{k} \}$, namely a new clause that contains all the generators of $z_j$ and $z_l$ but $\mygen_{k}$, and moreover the two clauses $z_j$ and $z_l$ may be replaced by $z_x$ in the sense that $z_x$ gives the same Fock basis coverage of ${\cal T}_j + {\cal T}_l$, namely ${\cal T}_j' \cup {\cal T}_l' = {\cal T}_x'$.
\end{MS_lemma}
\begin{proof}
By Proposition~\ref{prop_subspace_SSpinors} for any $\psi = \alpha \psi_j + \beta \psi_l \in {\cal T}_{j} + {\cal T}_{l}$ ($\alpha, \beta \in \R$), $\psi = v_1 v_2 \psi_j$ with $v_1 \wedge v_2 \psi_j \in {\cal T}_{l}$ and Proposition~\ref{prop_Tau_j_k} give the additional conditions on $v_1, v_2$ to be fulfilled that are $v_1 = \mygen_{k}, v_2 = \mygen_{k} - \mygen_{k-1}$. Since the non common part of $M(\psi_j)$ and $M(\psi_k)$ corresponds to $\my_span{v_1, v_2} = \my_span{\mygen_{k-1}, \mygen_{k}}$ and all other directions must be in common it follows that all generators $z_j \cup z_k - \{ \mygen_{k} \}$ must be in common, and thus in $\sup z_x$ proving the first part. For Fock basis coverage we observe that $\sup \varphi_{k} = \sup \varphi_{k-1} \cup \sup \mygen_{k-1} \mygen_{k}\varphi_{k-1}$ and since $\mygen_{k} \notin \sup z_x$ with $\psi \in {\cal T}_x$ we can cover both $\sup \varphi_{k-1}$, with fixed $p_k q_k$, and $\sup \mygen_{k-1} \mygen_{k} \varphi_{k-1}$, with fixed $q_k$.
\end{proof}

\noindent For example let $[z_1] = \mygen_{n}^2 \mygen_{n-1}^2 \mygen_{k} \Identity$ and $[z_2] = \mygen_{n-1}^2 \mygen_{n-2}^2 \mygen_{k}^2 \Identity$ then $[z_x] = \mygen_{n}^2 \mygen_{n-1}^2 \mygen_{n-2}^2 \Identity$.

\begin{MS_lemma}
\label{lemma_phi_incremental_buildup}
A given \SAT{} problem $S$ is unsatisfiable if and only if given $\varphi$ (\ref{formula_phi_full_sup}) then $\varphi \in \sum_{j \in J} {\cal T}_j$, $J \subseteq \{ 1, 2, \ldots, m \}$ being a subset of clauses, if and only if all terms of the succession $\varphi_k$ (\ref{formula_phi_succession}) are in $\sum_{j \in J} {\cal T}_j$ each term differing from the previous one by a case of one opposite variable of Proposition~\ref{prop_Tau_j_k}.
\end{MS_lemma}
\begin{proof}
Let $S$ be unsatisfiable then by Theorem~\ref{theorem_SAT_in_O(n)_4} $\varphi \in \sum_{j \in J} {\cal T}_j$ and by previous lemma all $\varphi_k \in \sum_{j \in J} {\cal T}_j$ and $\varphi_k = (\Identity + \mygen_{k-1} \mygen_{k}) \varphi_{k-1}$ and we already proved that this is the case of one opposite variable in Proposition~\ref{prop_Tau_j_k} with $v_1 = \mygen_{k}$ and $v_2 = \mygen_{k} - \mygen_{k-1}$. Conversely if there exist a subset of clauses $J \subseteq \{ 1, 2, \ldots, m \}$ that contain all terms of succession (\ref{formula_phi_succession}), each term differing by previous one by a common opposite generator, then by previous Lemma $\varphi \in \sum_{j \in J} {\cal T}_j$ and $S$ is unsatisfiable by Theorem~\ref{theorem_SAT_in_O(n)_4}.
\opt{margin_notes}{\mynote{mbh.note: commented here there is another Lemma with succession of subsets $J \subseteq J'$ that probably is wrong}}%
\end{proof}
%
%\noindent With some other conditions we can prove a similar result for the subsets of clauses that will be useful.
%
%\begin{MS_lemma}
%\label{lemma_eiejt}
%Given $t \in \sum_{j \in J} {\cal T}_j$ and any $t_{il}$ then $t_{il} t \in \sum_{j \in J'} {\cal T}_j$ with $J \subseteq J'$ if and only if $\mygen_{i} \mygen_{l} t \in \sum_{j \in J'} {\cal T}_j$.
%%
%\opt{margin_notes}{\mynote{mbh.note: log p. 792}}%
%%
%\end{MS_lemma}
%\begin{proof}
%Let $t_{il} t \in \sum_{j \in J'} {\cal T}_j$, by Proposition~\ref{prop_subspace_SSpinors} any $t_{il} t$ can be written as a sum of $t$ and $\mygen_{i} \mygen_{l} t$ and since by hypothesis $t \in \sum_{j \in J} {\cal T}_j$ it follows $\mygen_{i} \mygen_{l} t \in \sum_{j \in J'} {\cal T}_j$ with $J \subseteq J'$. Conversely given $\mygen_{i} \mygen_{l} t \in \sum_{j \in J_x} {\cal T}_j$ for the same reason we get $t_{il} t \in \sum_{j \in J \cup J_x} {\cal T}_j$ namely the thesis with $J \subseteq J \cup J_x := J'$.
%\end{proof}

We remark that one can start from a clause such that $\varphi_2 = (\Identity + \mygen_{1} \mygen_{2}) \psi_\Identity \in {\cal T}_j + {\cal T}_k$ and at every step find a subset $J_k$ such that $\varphi_k \in \sum_{j \in J_k} {\cal T}_j$ and at every step, given $\varphi_{k-1} \in \sum_{j \in J_{k-1}} {\cal T}_j$ either we find a subset $J_k$ that contains $\mygen_{k-1} \mygen_{k} \varphi_{k-1} \in \sum_{j \in J_k} {\cal T}_j$ and by previous Lemma $J_{k-1}$ and $J_k$ must have one opposite generator: $ \mygen_{k} $.

\myseparation

\noindent The rationale is that $\psi_j$ and $\psi_k$ induce a two dimensional linear subspace in $\mySpinorS_s$ and when we select this subspace for any $\psi$ in it all $\psi_\lambda \in \sup \psi$ represent covered assignments that renders $\myBooleanF$ the problem at hand.

This is a crucial point: in the discrete case we know that with the procedure of \emph{resolution}, given two clauses with one common opposite Boolean variables, \eg the first two clauses of (\ref{formula_SAT_std}): $(\mylitrl_1 \lor \myconjugate{\mylitrl}_2) \land (\mylitrl_2 \lor \mylitrl_3)$, we can \emph{add} to the \SAT{} problem the ``resolvent'' clause $(\mylitrl_1 \lor \mylitrl_3)$ but this clause is not equivalent to two given clauses since \eg $\mylitrl_1 \myconjugate{\mylitrl}_2 \myconjugate{\mylitrl}_3$ is covered by the two clauses but not from the resolvent clause. On the other hand in $z_j$ formalism the two clauses are $\myconjugate{\mylitrl}_1 \mylitrl_2$ and $\myconjugate{\mylitrl}_2 \myconjugate{\mylitrl}_3$ while $(\myconjugate{\mylitrl}_1 \mylitrl_2 + \myconjugate{\mylitrl}_2 \myconjugate{\mylitrl}_3) := \myconjugate{\mylitrl}_1 \myconjugate{\mylitrl}_3$ giving a Clifford algebra interpretation of the resolution rule.

In the continuous case, given two clauses $z_j$ and $z_k$ with one common opposite generator, we can \emph{replace} them with the composed clause $(z_j + z_k)$ since by Corollary~\ref{lemma_sum_equivalence} it guarantees the same Fock basis coverage and this works indifferently for $2$ and $3$\SAT{} problems.
\opt{margin_notes}{\mynote{mbh.note: most probably the two dimensional subspace of $\mySpinorS_s$ is NOT identified by $[z_j + z_k]$ that excludes ${\cal T}_{j} \cup {\cal T}_{k}$ and so we must find a different path...}}%

\myseparation

\begin{MS_Proposition}
\label{prop_time_test_old2}
Given $\varphi \in \mySpinorS_s$ (\ref{formula_phi_full_sup}) and $m$ sets ${\cal T}_j$ induced by clauses checking wether
\begin{equation}
\label{formula_time_test_old2}
\varphi \stackrel{?}{\in} \; \sum_{j = 1}^m {\cal T}_j
\end{equation}
can be done in polynomial time \eg \bigO{n^s}.
\end{MS_Proposition}
\begin{proof}
We give a procedure to verify (\ref{formula_time_test}); by Lemma~\ref{lemma_Givens_all_in_set} $\varphi \in \sum_{j = 1}^m {\cal T}_j$ if and only if for all terms of the succession $\varphi_k \in \sum_{j = 1}^m {\cal T}_j$ (\ref{formula_phi_succession}). We can thus start verifying whether $\varphi_1 = \psi_\Identity \in \sum_{j = 1}^m {\cal T}_j$ and continue incrementally up to $\varphi_{n}$: only if all $n$ successive $\varphi_k$ are in $\sum_{j = 1}^m {\cal T}_j$ we can conclude that $\varphi \in \sum_{j = 1}^m {\cal T}_j$, if any $\varphi_k$ fails the test this provides a certificate that $\varphi \notin \sum_{j = 1}^m {\cal T}_j$.
%Calling $J_k \subseteq \{ 1, \ldots, m\}$ the subset of clauses such that $t_k \in \sum_{j \in J_k} {\cal T}_j$, then by Lemma~\ref{lemma_eiejt} $J_1 \subseteq J_2 \subseteq \cdots \subseteq \{ 1, \ldots, m\}$.
%
\opt{margin_notes}{\mynote{mbh.note: commented there is the part with $J_1 \subseteq J_2 \subseteq \cdots \subseteq \{ 1, \ldots, m\}$ that is WRONG!}}%
In practice it is more practical, instead of using the full set of $m$ clauses, to start from $\varphi_1$ and look after a clause $z_j$ such that $\varphi_1 \in {\cal T}_j$ and at every step we look after a subset of the clauses $J_k$ such that $\varphi_k \in \sum_{j \in J_k} {\cal T}_j$.

We prove the proposition by induction on $k$; for $k = 1$, $\varphi_1 = \psi_\Identity$ and $J_1$ can contain any clause having only affirmative Boolean variables since for these clauses $M(z_j) \subseteq (\Identity, \Identity)$ (obviously there may be more than one; should there be none then $\varphi_1 \notin \sum_{j = 1}^m {\cal T}_j$ by Lemma~\ref{lemma_Givens_all_in_set} and, by the way, $\Identity$ would be a solution for the given \SAT{} problem). Looking for these clauses requires a time of \bigO{m}. For the induction step we prove that given $\varphi_{k-1} \in \sum_{j \in J_{k-1}} {\cal T}_j$, verifying whether there exists or not a set $J_k$ such that $\varphi_k = (\Identity + \mygen_{k-1} \mygen_{k}) \varphi_{k-1} \in \sum_{j \in J_k} {\cal T}_j$ requires polynomial time, \eg \bigO{n^{s-1}} and the complete test (\ref{formula_time_test}) will consequently take \bigO{n^s}.

By hypothesis $\varphi_{k-1} \in \sum_{j \in J_{k-1}} {\cal T}_j$ and $\varphi_{k} = (\Identity + \mygen_{k-1} \mygen_{k}) \varphi_{k-1} \in \sum_{j \in J_k} {\cal T}_j$ if and only if $\mygen_{k-1} \mygen_{k} \varphi_{k-1} \in \sum_{j \in X} {\cal T}_j$ and either we find $X$ and $\varphi_k = (\Identity + \mygen_{k-1} \mygen_{k}) \varphi_{k-1} \in \sum_{j \in J_k} {\cal T}_j$ with $J_k = J_{k-1} \cup X$ or we have a certificate that $\varphi_k \notin \sum_{j = 1}^m {\cal T}_j$ (and thus also $\varphi$). With Lemma~\ref{lemma_z_j+z_k} and (\ref{formula_t_in_T_j_form}) we can write
\begin{equation}
\label{formula_t_k-1}
\varphi_{k-1} = [k-1] \prod_{i=1}^{k-2} (\Identity + \mygen_{i} \mygen_{i+1}) \Identity
%\in {\cal T}_l + {\cal T}_o + {\cal T}_p + \cdots + {\cal T}_x + {\cal T}_y
\end{equation}
where $[k-1]$ is a shortcut for the involutory part $[[ \cdots [[z_l + z_o] + z_p] + \cdots ] + z_x]$. Given that the non involutory part of $\varphi_{k-1}$ covers $\{ \mygen_{1}, \mygen_{2}, \ldots, \mygen_{k-1} \}$ then $\inv{\varphi_{k-1}} = \{ \mygen_{k}, \mygen_{k+1}, \ldots, \mygen_{n} \}$ and we have to find whether there are clauses that contain $\mygen_{k-1} \mygen_{k} \varphi_{k-1}$ allowing us to go to next $\varphi_{k}$. Moreover since by hypothesis $\mygen_{k-1}$ is in the non involutory part of $\varphi_{k-1}$ it follows $\mygen_{k-1} \notin \sup ((k-1))$, where again $(k-1)$ is a shortcut for $(( \cdots ((z_l + z_o) + z_p) + \cdots ) + z_x)$, and we need to focus our attention on $\mygen_{k}$. There are only three mutually exclusive possibilities that we examine in succession:
\begin{enumerate}
\item $J_k = J_{k-1}$, namely $\mygen_{k-1} \mygen_{k} \varphi_{k-1} \in \sum_{j \in J_{k-1}} {\cal T}_j$. By Lemma~\ref{lemma_eit_in_Tj} this happens if and only if $\mygen_{k} \notin \sup ((k-1))$ that can be easily verified in \bigO{n} simply searching $\sup((k-1))$,

\item $J_k = J_{k-1} \cup X$ with $X \subseteq \{1, 2, \ldots, m \}$, namely exist one or more clauses such that $\mygen_{k-1} \mygen_{k} \varphi_{k-1} \in \sum_{j \in X} {\cal T}_j$. Since we are not in case $1$, $\mygen_{k-1} \mygen_{k} \varphi_{k-1} \notin \sum_{j \in J_{k-1}} {\cal T}_j$ and by Lemma~\ref{lemma_eit_in_Tj} $\mygen_{k} \in \sup ((k-1))$ and we tackle this case in the sequel,

\item If neither of previous cases occurred than we have a proof that $\mygen_{k-1} \mygen_{k} \varphi_{k-1} \notin \sum_{j = 1}^m {\cal T}_j$ and thus $\varphi_{k}$ and thus also $\varphi$.
\end{enumerate}

We study case $2$ in which $\mygen_{k-1} \mygen_{k} \varphi_{k-1} \in \sum_{j \in X} {\cal T}_j$ and we indicate with $(k) := (( \cdots (z_y + z_w) + z_z) + \cdots )$ the sum of clauses of $X = \{ y, w, z, \ldots \} \subseteq \{1, 2, \ldots, m \}$ that contain $\mygen_{k-1} \mygen_{k} \varphi_{k-1}$. In this case by Proposition~\ref{prop_subspace_SSpinors} the non incident part of the null subspaces induced by $\varphi_{k-1}$ and $\mygen_{k-1} \mygen_{k} \varphi_{k-1}$ is necessarily in $\my_span{\mygen_{k-1}, \mygen_{k}}$ and it follows by (\ref{formula_t_k-1}) that the rest must be identical and thus we necessarily must have
\begin{equation}
\label{formula_involutory_equality}
\left\{ \begin{array}{rlll}
\prod_{i=1}^{k-2} (\Identity + \mygen_{i} \mygen_{i+1}) \Identity & = & \prod_{i=1}^{k-2} (\Identity + \mygen_{i} \mygen_{i+1}) \Identity & \mbox{in} \; \{ \mygen_{1}, \mygen_{2}, \ldots, \mygen_{k-1} \} \\
{[k]} \mygen_{x'} \cdots \mygen_{z'} & = & \mygen_{k} [k-1] \mygen_{x} \cdots \mygen_{z} & \mbox{in} \; \{ \mygen_{k}, \mygen_{k+1}, \ldots \mygen_{n} \}
\end{array} \right.
%
%[k] \mygen_{x'} \cdots \mygen_{z'} : \prod_{i=1}^{k-2} (\Identity + \mygen_{i} \mygen_{i+1}) \Identity = \mygen_{k} [z_j] \mygen_{x} \cdots \mygen_{z} : \mygen_{k-1} \prod_{i=1}^{k-2} (\Identity + \mygen_{i} \mygen_{i+1}) \Identity
\end{equation}
where:
\begin{itemize}
\item we separated non involutory and involutory parts that have to be satisfied independently being in orthogonal subspaces, respectively $\{ \mygen_{1}, \mygen_{2}, \ldots, \mygen_{k-1} \}$ and $\{ \mygen_{k}, \mygen_{k+1}, \ldots \mygen_{n} \}$ of sizes $k-1$ and $n-k+1$,
\item $[k]$ stands for the involutory part of $(k)$, the sum of clauses of $X$ that contain $\mygen_{k-1} \mygen_{k} \varphi_{k-1}$, being in the involutory part $\sup ((k)) \subseteq \{ \mygen_{k}, \mygen_{k+1}, \ldots \mygen_{n} \}$,
%\item the involutory and non involutory parts are visually separated by $:$,
\item since $\mygen_{k-1}$ is in the non involutory part we should put it in first equation but since it would appear on both sides we eliminated it multiplying the equality by $\mygen_{k-1}$,
\item generators $\mygen_{x'} \cdots \mygen_{z'}$ are in $\sup ((k))^\perp \cap \{ \mygen_{k}, \mygen_{k+1}, \ldots \mygen_{n} \}$ and are all generators of $[k-1]$ that are not in $[k]$ whereas $\mygen_{x} \cdots \mygen_{z}$ are in $\sup ((k-1))^\perp \cap \{ \mygen_{k}, \mygen_{k+1}, \ldots \mygen_{n} \}$ and are all generators of $[k]$ that are not in $[k-1]$ and their role is to make equal the two terms following dictates of Lemma~\ref{lemma_z_j+z_k} the only opposite generator being $\mygen_{k}$ so for example if $[k-1] = [\mygen_{4} \mygen_{6}]$ and $[k] = [\mygen_{5} \mygen_{6} \mygen_{7}]$ then from (\ref{formula_involutory_equality}) we get $[\mygen_{5} \mygen_{6} \mygen_{7}] \mygen_{4} = [\mygen_{4} \mygen_{6}] \mygen_{5} \mygen_{7}$.
\end{itemize}
By hypothesis $\mygen_{k} \in \sup ((k-1))$, so in $[k] \mygen_{x'} \cdots \mygen_{z'}$ it is opposite to that in $(k-1)$ and it can appear either in $\sup ((k))$ or in $\mygen_{x'} \cdots \mygen_{z'}$. If $\mygen_{k} \notin \sup ((k)$ by Lemma~\ref{lemma_eit_in_Tj} $\sum_{j \in X} {\cal T}_j$ contains not only $\mygen_{k-1} \mygen_{k} \varphi_{k-1}$ but also $\varphi_{k-1}$.
% and we show that case never happens if at every step we choose the most 'hospitable' clause.
%
\opt{margin_notes}{\mynote{mbh.note: commented part on ``hospitable'' clauses, what follows uses this hypothesis and is known to be WRONG!!}}%

\myseparation

We start with the simplest case in which at level $k-2$ only one set $J_{k-1}$ existed: in this case clearly $\mygen_{k}$ can not be in $\mygen_{x'} \cdots \mygen_{z'}$ because this would imply that there is another set $X$ beyond $J_{k-1}$ that contains $\varphi_{k-1}$ against hypothesis of unicity of $J_{k-1}$. In cases in which more than one $J_{k-1}$ exist it is sufficient to choose the most hospitable of them. In this case would $\mygen_{k}$ be in $\mygen_{x'} \cdots \mygen_{z'}$ this would imply that $X$, containing both $\mygen_{k-1} \mygen_{k} \varphi_{k-1}$ and $\varphi_{k-1}$ would be more hospitable of $J_{k-1}$ against hypothesis.
\opt{margin_notes}{\mynote{mbh.note: the argument of most hospitable clause holds only for a single clause $z_y$, not in general for $z_y + z_w + \cdots$ and probably also the case of single common opposite fails...}}%
We must conclude that the only possibility is $\mygen_{k} \in \sup ((k))$ and thus, being $\mygen_{k-1}$ in the non involutory part $\{ \mygen_{1}, \mygen_{2}, \ldots, \mygen_{k-1} \}$, whereas $\mygen_{k}$ is common opposite on the two sides, we are in the case of one opposite Boolean variable of Proposition~\ref{prop_Tau_j_k}.

In involutory space $\{ \mygen_{k}, \mygen_{k+1}, \ldots \mygen_{n} \}$ of size $n-k+1$ there can be at most $n-k+1 - |\sup ((k-1))|$ available places to accomodate $\mygen_{x} \cdots \mygen_{z}$ and $n-k+1 - |\sup ((k))|$ available places to accomodate $\mygen_{x'} \cdots \mygen_{z'}$.

The simplest case is $X = \{y\}$ and sum $(k)$ reduces to $z_y$ and by (\ref{formula_cal_T_j_def2}) and (\ref{formula_t_in_T_j_form}) $\mygen_{k-1} \mygen_{k} \varphi_{k-1} \in {\cal T}_y$ if and only if $M(z_y) \subseteq (\Identity, \mygen_{k-1} \mygen_{k} [k-1] \mygen_{x} \cdots \mygen_{z} \prod_{i=1}^{k-2} (\Identity + \mygen_{i} \mygen_{i+1}) \Identity)$ that from the non involutory part requires $\sup z_y \subseteq \{ \mygen_{k}, \mygen_{k+1}, \ldots \mygen_{n} \}$ while the involutory part of (\ref{formula_involutory_equality}) requires
$$
[z_y] \mygen_{x'} \cdots \mygen_{z'} = \mygen_{k} [k-1] \mygen_{x} \cdots \mygen_{z}
$$
with $\mygen_{k} \in \sup z_y$ and we can verify if one or more clauses with these characteristics exist scanning the clauses in time \bigO{m}.

If there are no single clauses containing $\mygen_{k-1} \mygen_{k} \varphi_{k-1}$ we move to the next case of $X = \{y, w\}$ where the sum $(k) = (z_y + z_w)$ and thus, again by (\ref{formula_cal_T_j_def2}) and (\ref{formula_t_in_T_j_form}), $\mygen_{k-1} \mygen_{k} t_{k-1} \in {\cal T}_y + {\cal T}_w$ if and only if $M(z_y + z_w) \subseteq (\Identity, \mygen_{k-1} \mygen_{k} [k-1] \mygen_{x} \cdots \mygen_{z} \prod_{i=1}^{k-2} (\Identity + \mygen_{i} \mygen_{i+1}) \Identity)$ and also in this case the non involutory part requires $\sup (z_y + z_w) \subseteq \{ \mygen_{k}, \mygen_{k+1}, \ldots \mygen_{n} \}$ and since by Lemma~\ref{lemma_z_j+z_k} $\sup (z_y + z_w) = \sup z_y \cup \sup z_w - C$ and in this case the involutory parts of $\sup z_y$ and $\sup z_w$ could have parts both in $\{ \mygen_{1}, \mygen_{2}, \ldots, \mygen_{k-1} \}$ and in $\{ \mygen_{k}, \mygen_{k+1}, \ldots \mygen_{n} \}$ and we show that both parts are necessarily not empty. We already proved that necessarily $\mygen_{k} \in \sup (z_y + z_w)$ and thus $\mygen_{k} \in \sup z_y \cup \sup z_w$ and so $\mygen_{k}$ appears in at least one of $z_y$ and $z_w$. On the other hand would $\sup z_y \cap \{ \mygen_{1}, \mygen_{2}, \ldots, \mygen_{k-1} \} = \sup z_w \cap \{ \mygen_{1}, \mygen_{2}, \ldots, \mygen_{k-1} \} = \emptyset$ since $\mygen_{k} \in \sup z_y \cup \sup z_w$ at least one of the two clauses contains $\mygen_{k}$ and this would be enough for this clause to contain $\mygen_{k-1} \mygen_{k} t_{k-1}$ and we would be in previous case of $X$ made of just one clause. At the same time it is not possible that only one of the two clauses has a generator in $\{ \mygen_{1}, \mygen_{2}, \ldots, \mygen_{k-1} \}$ because the two clauses could not be summed and thus the only possibility is that both clauses have at least one generator in $\{ \mygen_{1}, \mygen_{2}, \ldots, \mygen_{k-1} \}$ that must disappear when the clauses are summed and are thus in set $C$ of Lemma~\ref{lemma_z_j+z_k}.

Moreover the involutory part in $\{ \mygen_{k}, \mygen_{k+1}, \ldots \mygen_{n} \}$ must be identical in all clauses of the sum $(k)$ because only identical Boolean variables in addends remain involutory as remarked just after Corollary~\ref{coro_sum_support} and in Lemma~\ref{lemma_z_j+z_k}. It follows that any clause $z$ entering the sum must satisfy
\begin{equation}
\label{formula_clause_nec_condition}
M(z) \cap \{ \mygen_{k}, \mygen_{k+1}, \ldots \mygen_{n} \} \subseteq (\Identity, \mygen_{k-1} \mygen_{k} [k-1])
\end{equation}
and whereas the involutory part of $z$ in $\{ \mygen_{k}, \mygen_{k+1}, \ldots \mygen_{n} \}$ has to be satisfied exactly, other involutory parts of $z$ in $\{ \mygen_{1}, \mygen_{2}, \ldots \mygen_{k-1} \}$ must disappear, with Lemma~\ref{lemma_z_j+z_k}, when summing the clauses to make room for $\prod_{i=1}^{k-2} (\Identity + \mygen_{i} \mygen_{i+1}) \Identity$.

It follows that in case $X = \{y, w\}$ we have to search one or more couples of clauses with these characteristics and this can be done in time \bigO{m^2}.

In more general case to prove that $\mygen_{k-1} \mygen_{k} t_{k-1} \in \sum_{j \in \{y, w, z, \dots \}} {\cal T}_j$ we need to prove that with clauses of $\{y, w, z, \dots \}$ we can build $\prod_{i=1}^{k-2} (\Identity + \mygen_{i} \mygen_{i+1}) \Identity$: clearly a problem like (\ref{formula_time_test}) in subspace $\{ \mygen_{1}, \mygen_{2}, \ldots \mygen_{k-1} \}$ and it is easy to show that this recursive definition can bring to an exponential time of solution but we now show that this is not the case.

We already remarked that being $\mygen_{k}$ in both involutory parts of (\ref{formula_involutory_equality}), in the left side it could be either in $\sup ((k))$ or in $\mygen_{x'} \cdots \mygen_{z'}$ and let us suppose that it is in $\mygen_{x'} \cdots \mygen_{z'}$: this implies that we can multiply by $\mygen_{k}$ the left side and, by Lemma~\ref{lemma_eit_in_Tj}, we remain in the same sum of clauses $(k)$ and thus we would be in a case in which both $\mygen_{k-1} \mygen_{k} t_{k-1}$ and $t_{k-1}$ are in the same sum of clauses. We would thus be in a different instance of case $1$ the only difference being given by the two set of clauses: $J_{k-1}$ or $\{y, w, z, \dots \}$: this is certainly possible but there would be no difference from previously analyzed case $1$ given that the Fock basis is a standard basis of a linear space and it is irrelevant which set of clauses gives coverage. In other words we are interested in cases in which our sum is in ${\cal T}_y + {\cal T}_w - {\cal T}_y \cup {\cal T}_w$ since only for these cases the coverage of the Fock basis doubles.
\opt{margin_notes}{\mynote{mbh.note: is it possible that there are clauses such that $\mygen_{k} \in \mygen_{x'} \cdots \mygen_{z'}$ but not such that $\mygen_{k} \in \sup ((k))$ ??}}%

We thus focus on cases such that $\mygen_{k} \in \sup ((k))$ and thus in the sum $t_k = (\Identity + \mygen_{k-1} \mygen_{k}) t_{k-1}$, $\mygen_{k}$ is \emph{necessarily} a common opposite Boolean variable. For all these cases we thus have clauses for which necessarily $\mygen_{k} \in \sup z_y$ and thus the remaining part in $\mygen_{i} \in \{ \mygen_{1}, \mygen_{2}, \ldots \mygen_{k-1} \}$ have $k-1$ Boolean variables and if \eg the initial clauses were a 3\SAT{} the subproblem generated in $\{ \mygen_{1}, \mygen_{2}, \ldots \mygen_{k-1} \}$ form a 2\SAT{} problem for which we are looking for an unsatisfiability solution that corresponds to the possibility to build $\prod_{i=1}^{k-2} (\Identity + \mygen_{i} \mygen_{i+1}) \Identity$ in $\{ \mygen_{1}, \mygen_{2}, \ldots \mygen_{k-1} \}$ that is well known to be polynomial thus concluding the proof.
\end{proof}

\myseparation

The second technicality is to write $t \in {\cal T}_j$ (\ref{formula_cal_T_j_def3}) as $t = [z_j] t'$ where:
\begin{itemize}
\item
\opt{margin_notes}{\mynote{mbh.note: here spinorial and vectorial representations are badly mixed}}%
$[z_j] \in \O1{n}$ is the involution obtained from clause $z_j$ completed with all 1 in coordinates in $(\sup z_j)^\perp$, namely $[z_j] = \mygen_{r} \ldots \mygen_{s} \mygen_{t}^2 \ldots \mygen_{u}^2 \Identity$ where $\mygen_{r} \ldots \mygen_{s}$ stand for the Boolean variables in negative forms of $z_j$ and $\mygen_{t}^2 \ldots \mygen_{u}^2$ (all equal to $1$) are the 'placeholders' for the affirmative Boolean variables of $z_j$ put there just to remind us their position in $\sup z_j$,
\item $t' \in \OO{n-k}$ acts in subspace $(\sup z_j)^\perp$.
\opt{margin_notes}{\mynote{mbh.note: commented here there is this formalism written as a Lemma practically identical to Proposition~\ref{prop_cal_t_j}}}%
\end{itemize}

%\begin{MS_lemma}
%\label{lemma_t_j_form}
%Given clause
%%
%\opt{margin_notes}{\mynote{mbh.note: this Lemma is practically equal to Proposition~\ref{prop_cal_t_j}}}%
%%
%$z_j$ then $t \in {\cal T}_j$ if and only if it can be written as $t = [z_j] t'$ where:
%\begin{itemize}
%\item $[z_j] \in \O1{n}$ is the involution obtained from clause $z_j$ completed with all 1 in coordinates in $(\sup z_j)^\perp$, namely $[z_j] = \mygen_{r} \ldots \mygen_{s} \mygen_{t}^2 \ldots \mygen_{u}^2 \Identity$ where $\mygen_{r} \ldots \mygen_{s}$ stand for the Boolean variables in negative forms of $z_j$ and $\mygen_{t}^2 \ldots \mygen_{u}^2$ (all equal to $1$) are the 'placeholders' for the affirmative Boolean variables of $z_j$ put there just to remind us their position in $\sup z_j$,
%\item $t' \in \OO{n-k}$ acts in subspace $(\sup z_j)^\perp$.
%\end{itemize}
%\end{MS_lemma}
%\begin{proof}
%The proof is immediate since $M(z_j) \subseteq (\Identity, [z_j])$ and for any $t' \in \OO{n-k}$ then $t \in \OO{n}$ and $M(z_j) \subseteq (\Identity, t)$ and conversely any $t \in \OO{n}$ satisfying this property can be put in the given form.
%\end{proof}

We remark that $t' \in \OO{n-k}$ has no restrictions but we will mostly be interested in cases tailored to our problem and usually $t'$ will have the form
$$
t' = \mygen_{x} \cdots \mygen_{z} \prod_i (\Identity + \mygen_{i} \mygen_{i+1}) \Identity
$$
with $\mygen_{x} \cdots \mygen_{z}$ generators of $(\sup z_j)^\perp$ and $\prod_i$ is extended to terms such that $\mygen_{i}, \mygen_{i+1} \in (\sup z_j)^\perp$. Thus the form of $t \in {\cal T}_j$ will frequently be
\begin{equation}
\label{formula_t_in_T_j_form}
t = [z_j] \mygen_{x} \cdots \mygen_{z} \prod_i (\Identity + \mygen_{i} \mygen_{i+1}) \Identity
\end{equation}
where $\inv {t} = [z_j] \mygen_{x} \cdots \mygen_{z}$ and $\inv {t}^\perp = \prod_i (\Identity + \mygen_{i} \mygen_{i+1}) \Identity$ the two subspaces being orthogonal and since $M(z_j) \subseteq (\Identity, t)$ the inclusion must be necessarily in the involutory part and thus $M(z_j) \subseteq (\Identity, [z_j] \mygen_{x} \cdots \mygen_{z})$, moreover the two parts commute. For example if $[z_j] = \mygen_{4} \mygen_{5}^2 \Identity$ then $t = [z_j] \mygen_{6} \prod_{i=1}^{2} (\Identity + \mygen_{i} \mygen_{i+1}) \Identity \in {\cal T}_j$ and has $\det t = 1$. In summary we just gave another equivalent definition of ${\cal T}_j$ by means of involution $[z_j]$.

\begin{MS_lemma}
\label{lemma_z_j+z_k}
Given two clauses $z_j, z_k$ with their induced sets ${\cal T}_{j}, {\cal T}_{k}$ then $\psi \in {\cal T}_{j} + {\cal T}_{k} - {\cal T}_{j} \cup {\cal T}_{k}$ if and only if it can take the just defined form where the involutory part generated by clauses $z_j$ and $z_k$ is given by a composed clause we name $(z_j + z_k)$ such that $\sup (z_j + z_k) = \sup z_j \cup \sup z_k - C$ where $C$ contains the common opposite generators of $z_j$ and $z_k$ and, if there are none, the two generators of Proposition~\ref{prop_Tau_j_k}. In other words, in our formalism, $t \in {\cal T}_{j} + {\cal T}_{k} - {\cal T}_{j} \cup {\cal T}_{k}$ if and only if it can be written as $t = [z_j + z_k] t'$.
\opt{margin_notes}{\mynote{mbh.note: surely WRONG for case of one common opposite generator, should be checked in other cases and corrected in this one}}%
\end{MS_lemma}
\begin{proof}
By Proposition~\ref{prop_subspace_SSpinors} for any $\psi = \alpha \psi_j + \beta \psi_k \in {\cal T}_{j} + {\cal T}_{k}$ ($\alpha, \beta \in \R$), $\psi = v_1 v_2 \psi_j$ with $v_1 \wedge v_2 \psi_j \in {\cal T}_{k}$ and Proposition~\ref{prop_Tau_j_k} give the additional conditions on $v_1, v_2$ to be fulfilled. Since the non common part of $M(\psi_j)$ and $M(\psi_k)$ corresponds to $\my_span{v_1, v_2}$ and all other directions must be in common it follows that all generators $\mygen_{i} \in \sup z_j \cup \sup z_k - C$ must be in common, and thus in the involutory part of $[z_j + z_k]$ proving the thesis.
\end{proof}
\noindent For example let $[z_1] = \mygen_{1} \mygen_{2} \Identity$ and $[z_2] = \mygen_{2}^2 \mygen_{3}^2 \Identity$ then $[z_1 + z_2] = \mygen_{1} \mygen_{3}^2 \Identity$ with $C = \{ \mygen_{2} \}$
\opt{margin_notes}{\mynote{mbh.note: should I write $[z_1 + z_2] = \mygen_{1} \mygen_{3}^2 (\Identity + \mygen_{1} \mygen_{2}) \Identity $ here ?}}%
while if $[z_2] = \mygen_{3} \mygen_{4} \Identity$ then $[z_1 + z_2]$ can be: $\mygen_{1} \mygen_{3} \Identity$, $\mygen_{1} \mygen_{4} \Identity$, $\mygen_{2} \mygen_{3} \Identity$ or $\mygen_{2} \mygen_{4} \Identity$.

\begin{MS_Corollary}
\label{lemma_sum_equivalence}
Given %
\opt{margin_notes}{\mynote{mbh.note: this Corollary is manifestly FALSE, see 'floating leaflets' pp. 190, 193}}%
two clauses $z_j, z_k$ such that the composition $(z_j + z_k)$ is possible then the two clauses can be replaced by $(z_j + z_k)$ that guarantees the same Fock basis coverage, namely $({\cal T}_{j} + {\cal T}_{k}) \cap \O1{n} = {\cal T}_{j}' \cup {\cal T}_{k}'$.
\end{MS_Corollary}
\begin{proof}
This is a direct consequence of Lemma~\ref{lemma_sum_intersection} but we give another proof: for any $\psi = \alpha \psi_j + \beta \psi_k \in {\cal T}_{j} + {\cal T}_{k}$ ($\alpha, \beta \in \R$) by Corollary~\ref{coro_sum_support} $\sup \psi \subset {\cal T}_{j}' \cup {\cal T}_{k}'$ and setting either $\alpha = 0$ or $\beta = 0$ we cover ${\cal T}_{j} \cup {\cal T}_{k}$ while the part ${\cal T}_{j} + {\cal T}_{k} - {\cal T}_{j} \cup {\cal T}_{k}$ is covered by Lemma~\ref{lemma_z_j+z_k}.
\opt{margin_notes}{\mynote{mbh.note: isn't this simply a direct consequence of Lemma~\ref{lemma_sum_intersection} ? Commented another (long and dubious) proof}}%
\end{proof}
%\begin{proof}
%(old proof) Given $z_j, z_k$ for any $\psi \in {\cal T}_{j} + {\cal T}_{k}$ by Corollary~\ref{coro_sum_support} $\sup \psi \subset {\cal T}_{j}' \cup {\cal T}_{k}'$ while by Lemma~\ref{lemma_z_j+z_k} we know that $t = [z_j + z_k] t'$. Conversely given $t = [z_j + z_k] t'$ we know that $\sup (z_j + z_k) = \sup z_j \cup \sup z_k - C$, $C$ containing one or two generators according to Proposition~\ref{prop_Tau_j_k}. If there are zero or two opposite generators in $z_j, z_k$ let $C = \{\mygen_{x}, \mygen_{y}\}$ then we know that the non involutive part $t'$ contains term $(\Identity + \mygen_{x} \mygen_{y})$ and so we can express any spinor in $\my_span{{\cal T}_{j}' \cup {\cal T}_{k}'}$ but not in $\my_span{{\cal T}_{j}'}$ if $\psi \notin {\cal T}_{j} \cup {\cal T}_{k}$. If $C = \{\mygen_{x}\}$ then we know that the non involutive part $t'$ contains term $(\Identity + \mygen_{x} v)$ with $v \in \my_span{\sup z_j \cap \sup z_k}^\perp$ so we can express any spinor with $\mygen_{x}$ in both forms and thus also in this case in $\my_span{{\cal T}_{j}' \cup {\cal T}_{k}'}$.
%\end{proof}

\noindent The rationale is that $\psi_j$ and $\psi_k$ induce a two dimensional linear subspace in $\mySpinorS_s$ and when we select this subspace for any $\psi$ in it all $\psi_\lambda \in \sup \psi$ represent covered assignments that renders $\myBooleanF$ the problem at hand.

So in previous example we can replace the two clauses $[z_1] = \mygen_{1} \mygen_{2} \Identity$ and $[z_2] = \mygen_{2}^2 \mygen_{3}^2 \Identity$ with the single clause $[z_1 + z_2] = \mygen_{1} \mygen_{3}^2 \Identity$ that guarantees the same Fock basis coverage, or in second case, by any clause of the four $\mygen_{1} \mygen_{3} \Identity$, $\mygen_{1} \mygen_{4} \Identity$, $\mygen_{2} \mygen_{3} \Identity$ or $\mygen_{2} \mygen_{4} \Identity$ always giving same basis coverage.

This is a crucial point: in the discrete case we know that with the procedure of \emph{resolution}, given two clauses with one common opposite Boolean variables, \eg the first two clauses of (\ref{formula_SAT_std}): $(\mylitrl_1 \lor \myconjugate{\mylitrl}_2) \land (\mylitrl_2 \lor \mylitrl_3)$, we can \emph{add} to the \SAT{} problem the ``resolvent'' clause $(\mylitrl_1 \lor \mylitrl_3)$ but this clause is not equivalent to two given clauses since \eg $\mylitrl_1 \myconjugate{\mylitrl}_2 \myconjugate{\mylitrl}_3$ is covered by the two clauses but not from the resolvent clause. On the other hand in $z_j$ formalism the two clauses are $\myconjugate{\mylitrl}_1 \mylitrl_2$ and $\myconjugate{\mylitrl}_2 \myconjugate{\mylitrl}_3$ while $(\myconjugate{\mylitrl}_1 \mylitrl_2 + \myconjugate{\mylitrl}_2 \myconjugate{\mylitrl}_3) := \myconjugate{\mylitrl}_1 \myconjugate{\mylitrl}_3$ giving a Clifford algebra interpretation of the resolution rule.

In the continuous case, given two clauses $z_j$ and $z_k$ with one common opposite generator, we can \emph{replace} them with the composed clause $(z_j + z_k)$ since by Corollary~\ref{lemma_sum_equivalence} it guarantees the same Fock basis coverage and this works indifferently for $2$ and $3$\SAT{} problems.
\opt{margin_notes}{\mynote{mbh.note: most probably the two dimensional subspace of $\mySpinorS_s$ is NOT identified by $[z_j + z_k]$ that excludes ${\cal T}_{j} \cup {\cal T}_{k}$ and so we must find a different path...}}%

\begin{MS_Corollary}
\label{lemma_k_of_z_j+z_k}
Given two clauses $z_j, z_k$ of $|\sup z_i | = l_i$ and their composed clause $(z_j + z_k)$ then $|\sup (z_i + z_k) | = | \sup z_j \cup \sup z_k | - 1$ in case of one common opposite generator being $|\sup (z_i + z_k) | = | \sup z_j \cup \sup z_k | - 2$ in all other cases.
\end{MS_Corollary}
\begin{proof}
The reason to use $| \sup z_j \cup \sup z_k |$ is that all common equal generators appear only once in $\sup (z_i + z_k)$. In case of one common opposite generators it can be not in $\sup (z_i + z_k)$, in all other cases two generators of $\sup z_j \cup \sup z_k$ are not in $\sup (z_i + z_k)$.
\end{proof}

It follows that every time we add two clauses both the number of clauses $m$ decreases to $m-1$ and also the total support of clauses $\sum_ i l_i$ strictly decreases being reduced by at least $1$.

We remark that given $\psi = \psi_j + \psi_k \in {\cal T}_{j} + {\cal T}_{k}$ if we want to iterate the procedure and add another simple spinor \eg $\phi \in {\cal T}_{l}$ to get $\phi + \psi$ Proposition~\ref{prop_subspace_SSpinors} applies again and thus necessarily $\phi + \psi = (\Identity + u_1 u_2) \psi = (\Identity + u_1 u_2) (\Identity + v_1 v_2) \psi_j$ and we can iterate previous Lemma to get that the involutions associated to $\phi + \psi$ will be $[[z_j + z_k] + z_l]$ where we introduced this notation with square brackets to indicate that this sum in \emph{not} associative and that in general $\phi + (\psi_j + \psi_k) \ne (\phi + \psi_j) + \psi_k$ since in general $(\phi + \psi_j)$ is not simple: for example let $\psi = \psi_\Identity + \mygen_{1} \mygen_{2} \psi_\Identity = (\Identity + \mygen_{1} \mygen_{2}) \psi_\Identity$ is simple and we may add $\phi = \mygen_{3} \mygen_{4} \psi$ to get $\phi + \psi = \mygen_{3} \mygen_{4} \psi + \psi = \mygen_{3} \mygen_{4} \psi + (\psi_\Identity + \mygen_{1} \mygen_{2} \psi_\Identity)$ where the parenthesis is strictly necessary since $\mygen_{3} \mygen_{4} \psi + \psi_\Identity$ is not a simple spinor since incidence is $n-4$.

We remark that Lemma~\ref{lemma_z_j+z_k} is of general validity since it deals with \emph{two} simple spinors at the time and thus that the involutory part of a sum of more than two addends will be indicated by $[[ \cdots [[z_l + z_o] + z_p] + \cdots ] + z_x] $. Thus composed clauses continue to have all properties of real clauses like \eg Lemma~\ref{lemma_eit_in_Tj}.

\myseparation

\section{(OLD) An unsatisfiability test based on \OO{n} representations: a theoretical result}
\label{sec_Simple_Spinors_SAT_rank_theory}
We
\opt{margin_notes}{\mynote{mbh.note: this is the last version that attempted to use rank on linear space $\R(n)$ but after submission to TCS I realized that pivotal Theorem~\ref{theorem_SAT_in_R(n)} is a necessary but not sufficient condition for unsatisfiability; see flying leaflets pp. $\beta \lambda - \beta \pi$}}%
already remarked \cite{Budinich_2019} that an unsatisfiability test exploiting Proposition~\ref{prop_SAT_in_On1} offers no advantages with respect to standard algorithms and thus we focus our attention to the continuous setting of Theorem~\ref{theorem_SAT_in_O(n)} where we need to prove that all $t \in \OO{n}$ are in $\sum_{j = 1}^m {\cal T}_j$. By quoted isomorphisms (\ref{formula_simple_spinors}) to any $t \in \OO{n}$ corresponds the simple spinor $\psi_t \in \mySpinorS_s$ such that $M(\psi_t) = (\Identity, t)$ and this spinor can be expanded in Fock basis $\myFockB$ (\ref{formula_Fock_basis_expansion}) and we define its \emph{support} $\sup \psi_t \subseteq \myFockB$ as the subset of Fock basis elements used in (\ref{formula_Fock_basis_expansion}) so that $\psi_t \in \my_span{\sup \psi_t}$. So any $t \in \OO{n}$ induces $\sup \psi_t$, namely a set of $\lambda \in \O1{n}$ that in turn can be seen as a set of Boolean assignments (\ref{formula_bijection_Mn_O^n(1)}). Applying this to our case, given any $t \in {\cal T}_j$, since ${\cal T}_j \cap \O1n = {\cal T}_j'$ \cite[Lemma 1]{Budinich_2019} then in this case
$$
\sup \psi_t \subseteq {\cal T}_j' \quad \implies \quad {\cal T}_j \subset \my_span{{\cal T}_j'}
$$
provided we identify $\psi_\lambda \in \myFockB$ with corresponding $\lambda \in \O1{n}$ (more precisely $M(\psi_\lambda) = (\Identity, \lambda)$). In other words $\sup \psi_t$ give the set of Boolean assignments induced by $t$ that make the problem unsatisfiable.

Our \SAT{} problem is defined in the Clifford algebra of $\R^{n,n}$ and precisely in the spinorial representation of $\OO{n}$ and since $\myClg{}{}{\R^{n,n}} \myisom \R(2^n)$, \SAT{} is a problem in the algebra of real matrices of dimension $2^n \times 2^n$
\opt{margin_notes}{\mynote{mbh.note: doubly check spinorial vs vectorial representations of \OO{n}, see log pp. 736, 759, M\_532}}%
and, from the computational side, the situation looks problematic. The turning point is that the spinorial representation of $\OO{n}$ is equivalent to its vectorial representation corresponding to the much more manageable and familiar algebra of real $n \times n$ matrices $\R(n)$.

In a nutshell to any $t \in \OO{n}$ corresponds an orthogonal matrix $T \in \R(n)$ of the vectorial representation and the action of $T$ on $u \in \R^n$ is given by $T u$. In the spinorial representation of $t$ in $\myClg{}{}{\R^{n,n}} \myisom \R(2^n)$ the action of $t$ on $u$ is given by
$$
(-1)^k v_1 v_2 \cdots v_k \; u \; (v_1 v_2 \cdots v_k)^{-1}
$$
for some $v_1, v_2, \ldots v_k \in \{0\} \times \R^n$, linearly independent and with $k \le n$, this being nothing else than the Cartan theorem for Euclidean spaces in disguise \cite[Theorem~5.15]{Porteous_1995}: vectors $v_1, v_2, \ldots v_k$ give the directions of the $k \le n$ hyperplane reflections in which isometries of $\OO{n}$ can be decomposed.

Through equivalence between the representations of $\OO{n}$ and exploiting commutative diagram of Section~\ref{sec_SAT_in_O(n)}, our results can be equally formulated either in $\R(2^n)$ or in $\R(n)$ (and in particular Theorem~\ref{theorem_SAT_in_O(n)}) but we do not insist on this and just complete (\ref{formula_cal_T_j_def2}) porting the definition of ${\cal T}_j$ in $\R(n)$.

\begin{MS_Proposition}
\label{prop_cal_t_j_old}
Given clause $z_j$ and the corresponding involution $\lambda_j$ induced by its literals then
\begin{equation}
\label{formula_cal_T_j_def3_old}
{\cal T}_j = \{ t := \left(\begin{array}{r r} \lambda_j & \\ & t' \end{array}\right) : t' \in \OO{n-k} \}
\end{equation}
where we supposed, for ease of notation and without loss of generality, that the literals occupy the first $k$ coordinates of spacelike space $\{0\} \times \R^n$.
\end{MS_Proposition}
\begin{proof}
The proof is immediate since for any $t' \in \OO{n-k}$ then $t \in \OO{n}$ and $M(z_j) \subseteq (\Identity, t )$ and conversely any $t \in \OO{n}$ satisfying this property must have this form.
\end{proof}

So from now on we move to the vectorial representation of \OO{n} in matrices of $\R(n)$ and we need to prove that any $t \in \OO{n}$ can be built by a linear combination of elements from ${\cal T}_j$ (\ref{formula_cal_T_j_def3_old}).

\begin{MS_Proposition}
\label{prop_matrix_algebra_coverage}
Given the set of orthogonal matrices of \OO{n} in $\R(n)$ they can form a basis of $\R(n)$ seen as a linear space of dimension $n^2$.
\end{MS_Proposition}
\begin{proof}
Result is obvious for $n=1$, for $n=2$ we define the \OO{2} isometries:
$$
t_\theta = \left(\begin{array}{r r} \cos \theta & - \sin \theta \\ \sin \theta & \cos \theta \end{array}\right) \qquad a_\theta = \left(\begin{array}{r r} \cos \theta & \sin \theta \\ \sin \theta & - \cos \theta \end{array}\right)
$$
then
$$
\begin{array}{ll}
\left(\begin{array}{r r} 1 & 0 \\ 0 & 0 \end{array}\right) = \frac{1}{2} (t_0 + a_0) &
\left(\begin{array}{r r} 0 & 0 \\ 0 & 1 \end{array}\right) = \frac{1}{2} (t_0 + a_\pi) \\
\left(\begin{array}{r r} 0 & 0 \\ 1 & 0 \end{array}\right) = \frac{1}{2} (t_\frac{\pi}{2} + a_\frac{\pi}{2}) &
\left(\begin{array}{r r} 0 & 1 \\ 0 & 0 \end{array}\right) = \frac{1}{2} (t_{\frac{3}{2} \pi} + a_\frac{\pi}{2})
\end{array}$$
and the proposition is true also for $n=2$. For generic $n$ we can obtain the $n^2$ matrices $B$ all $0$ with the exception of a single $b_{i j} = 1$ considering \OO{2} isometries acting in subspaces $\mygen_i \mygen_j$ and applying same procedure used in case of $n = 2$ with the foresight of filling the diagonal of addends respectively with $1$ and $-1$ so that it cancels out in the sums.
\end{proof}

\noindent For example for $n = 4$ and for $\theta = \frac{\pi}{2}$
$$
\left(\begin{array}{c c c c} 0 & 0 & 0 & 0 \\ 0 & 0 & 0 & 0 \\ 0 & 1 & 0 & 0 \\ 0 & 0 & 0 & 0 \end{array}\right) = \frac{1}{2} \left[ \left(\begin{array}{c c c c} 1 & 0 & 0 & 0 \\ 0 & \cos \theta & - \sin \theta & 0 \\ 0 & \sin \theta & \cos \theta & 0 \\ 0 & 0 & 0 & 1 \end{array}\right) + \left(\begin{array}{c c c c} -1 & 0 & 0 & 0 \\ 0 & \cos \theta & \sin \theta & 0 \\ 0 & \sin \theta & - \cos \theta & 0 \\ 0 & 0 & 0 & -1 \end{array}\right) \right]
$$
and this result is not surprising in view of the fact that we can think at \OO{n} as at a compact, disconnected real manifold of dimension $n (n-1)/2$ immersed in the linear space of dimension $n^2$ of $\R(n)$.
\begin{MS_Corollary}
\label{coro_base_matrix_algebra}
The $n (n-1)/2$ isometries $t_\theta$ and $a_\theta$, acting in subspace $\my_span{\mygen_{i}, \mygen_{j}}$, with respective diagonals filled with $1$ and $-1$, all in all $n(n-1)$ \OO{n} elements, form a basis of $\R(n)$.
\end{MS_Corollary}

We can thus generalize Theorem~\ref{theorem_SAT_in_O(n)} to

\begin{MS_theorem}
\label{theorem_SAT_in_R(n)}
\opt{margin_notes}{\mynote{mbh.note: Theorem disproved p. $\beta \pi$ !!!}}%
A given \SAT{} problem $S$ in $\myClg{}{}{\R^{n,n}}$ (\ref{formula_SAT_EFB_2}) with $n$ Boolean variables is unsatisfiable if and only if the isometries induced by its $m$ clauses (\ref{formula_cal_T_j_def3_old}) form a cover for $\R(n)$:
\begin{equation}
\label{formula_SAT_in_R(n)}
\my_span{\sum_{j = 1}^m {\cal T}_j} = \R(n) \dotinformula
\end{equation}
\end{MS_theorem}
\begin{proof}
Let $S$ be unsatisfiable, then by Theorem~\ref{theorem_SAT_in_O(n)} we can cover all simple spinors and thus all orthogonal matrices of $\R(n)$ and thus with Proposition~\ref{prop_matrix_algebra_coverage} also the whole linear space of $\R(n)$.
\opt{margin_notes}{\mynote{mbh.note: here there is the difficulty of getting ${\cal T}_j = \my_span{{\cal B}_j}$}}%
Conversely given a set of clauses that covers linear space of $\R(n)$ we can also cover all orthogonal matrices and thus \OO{n} and the problem $S$ is unsatisfiable by Theorem~\ref{theorem_SAT_in_O(n)}.
\end{proof}

From the theoretical point of view Theorems~\ref{theorem_SAT_in_O(n)} and \ref{theorem_SAT_in_R(n)} are almost identical but from the computational point of view things change significantly since $\OO{n}$ is a compact, disconnected real manifold of dimension $n (n-1)/2$ and is not easy to establish whether $\sum_{j = 1}^m {\cal T}_j$ form a cover of this manifold. On the other hand the linear space of $\R(n)$ is a good, old linear space and it is much easier to verify if it is covered by $\sum_{j = 1}^m {\cal T}_j$.

To do this we define the set of matrices ${\cal B}_j \subset \R(n)$ associated to clause $z_j$ made of $k$ literals, obtained applying Proposition~\ref{prop_matrix_algebra_coverage} to set ${\cal T}_j$ (\ref{formula_cal_T_j_def3_old}), namely the set of $(n-k)^2$ basis matrices of $\R(n)$, seen as a linear space of dimension $n^2$. For example for $n=4$ and $z_j = \mylitrl_1 \myconjugate{\mylitrl}_2$, let $\lambda_j = \left(\begin{array}{c c} 1 & 0 \\ 0 & -1 \end{array}\right)$, then
$$
{\cal B}_j = \left\{
\left(\begin{array}{c c c} \lambda_j \\ & 1 & 0 \\ & 0 & 0 \end{array}\right),
\left(\begin{array}{c c c} \lambda_j \\ & 0 & 1 \\ & 0 & 0 \end{array}\right),
\left(\begin{array}{c c c} \lambda_j \\ & 0 & 0 \\ & 1 & 0 \end{array}\right),
\left(\begin{array}{c c c} \lambda_j \\ & 0 & 0 \\ & 0 & 1 \end{array}\right)
\right\}
$$
are the $4$ matrices induced by $z_j$ in linear space of $\R(4)$ of dimension $16$.

\begin{MS_Proposition}
\label{prop_T_j_vectors}
\opt{margin_notes}{\mynote{mbh.note: here there are the difficulties: disproved p. $\beta \pi$ !!!}}%
For any clause $z_j$ then ${\cal T}_j \subset \my_span{{\cal B}_j}$ and $ \my_span{{\cal T}_j} = \my_span{{\cal B}_j}$.
\end{MS_Proposition}
\begin{proof}
By Proposition~\ref{prop_cal_t_j_old} any $t \in {\cal T}_j$ can be written in the form (\ref{formula_cal_T_j_def3_old}) with $t' \in \OO{n-k}$ and any of these $t'$ can be expressed as a linear combination of the vectors of ${\cal B}_j$ that proves the first statement,
the second is a direct consequence of Proposition~\ref{prop_matrix_algebra_coverage}.
\end{proof}

\begin{MS_Corollary}
\label{coro_sum_span}
Given \SAT{} problem $S$ and the sets ${\cal T}_j$ and ${\cal B}_j$ induced by its clauses, then
$$
\sum_{j = 1}^m {\cal T}_j = \my_span{\cup_{j = 1}^m {\cal B}_j} \dotinformula
$$
\end{MS_Corollary}
\begin{proof}
$\sum_{j = 1}^m {\cal T}_j$ contains (allowed) linear combinations of elements of ${\cal T}_j$ and since ${\cal T}_j = \my_span{{\cal B}_j}$ the thesis follows immediately.
\end{proof}

\begin{MS_theorem}
\label{theorem_SAT_in_R(n)_basis}
A given \SAT{} problem $S$ in $\myClg{}{}{\R^{n,n}}$ (\ref{formula_SAT_EFB_2}) with $n$ Boolean variables is unsatisfiable if and only if $\cup_{j = 1}^m {\cal B}_j$ covers the basis of linear space $\R(n)$ of dimension $n^2$.
\end{MS_theorem}
\begin{proof}
Let $S$ be unsatisfiable, then by Theorem~\ref{theorem_SAT_in_R(n)} $\sum_{j = 1}^m {\cal T}_j$ covers $\R(n)$ seen as a linear space and Corollary~\ref{coro_sum_span} proves the thesis. Conversely if $\cup_{j = 1}^m {\cal B}_j$ contains a full basis of $\R(n)$ then $S$ is unsatisfiable by Theorem~\ref{theorem_SAT_in_R(n)}.
\end{proof}

\begin{MS_Corollary}
\label{coro_P_equal_NP}
A given \SAT{} problem $S$ in $\myClg{}{}{\R^{n,n}}$ (\ref{formula_SAT_EFB_2}) with $n$ Boolean variables is unsatisfiable if and only if
$$
\rank \cup_{j = 1}^m {\cal B}_j = n^2 \dotinformula
$$
\end{MS_Corollary}

\noindent Since rank calculation of vectors of $\R^n$ is \bigO{n^3} this indicates that the unsatisfiability test that exploits Theorem~\ref{theorem_SAT_in_R(n)_basis} requires \bigO{n^6}. Very preliminary tests made with a symbolic manipulation program confirm this finding.

\bigskip

This result is theoretically interesting but probably there are better algorithms in sight based on the fact that there are $t \in \OO{n}$ with $\sup \psi_t$ of $2^{n-1}$ $\myFockB$ elements and thus two tests $\psi_t \stackrel{?}{\in} \sum_{j = 1}^m {\cal T}_j$ can give a certificate of unsatisfiability: these ideas will be developed in a forthcoming paper.

\myseparation

With Theorem~\ref{theorem_SAT_in_O(n)} We underline that this formulation has no combinatorics since $\myFockB$ is a proper basis of the linear space of spinors $\mySpinorS$ and expansion (\ref{formula_Fock_basis_expansion}) is unique and that if an element of the Fock basis $\psi_\lambda \notin \sum_{j = 1}^m {\cal T}_j$ necessarily it is a solution of the \SAT{} problem.

Summarizing the main ingredient in this recipe for \SAT{} is the equivalence between the spinorial and vectorial representations of $\OO{n}$ respectively in matrix algebras $\R(2^n)$ and $\R(n)$. We will also show how the expansion of a simple spinor in the Fock basis is deeply intertwined with bivectors of $\myClg{}{}{\R^{n,n}}$ and the corresponding Lie algebra that, in the vectorial representation, correspond to an expansion of $t \in \OO{n}$ in terms of $\SO{2}$ elements, namely to the Givens expansion of $t$.

We already defined $\psi_\Identity = p_1 q_1 \; p_2 q_2 \; p_3 q_3 \cdots p_n q_n$ the \emph{reference} spinor; clearly $\psi_\Identity \in \myFockB$ and $M(\psi_\Identity) = \my_span{p_{1}, p_{2}, \ldots, p_{n}} = P = (\Identity, \Identity)$, then we can prove \cite{Budinich_2012}:
\begin{MS_Proposition}
\opt{margin_notes}{\mynote{mbh.note: log pp. 775, 776}}%
\label{prop_all_SSpinors}
All simple spinors $\psi$ of $\myClg{}{}{\R^{n,n}}$ can be written as
$$
\psi = v_1 v_2 \cdots v_k \psi_\Identity
$$
for some $v_1, v_2, \ldots v_k \in \{0\} \times \R^n$, linearly independent and with $k \le n$.
\end{MS_Proposition}
This is nothing else than the Cartan theorem for Euclidean spaces \cite[Theorem~5.15]{Porteous_1995} in disguise: the vectors $v_1, v_2, \ldots v_k$ give the directions of spacelike $\{0\} \times \R^n$ that give the $k \le n$ reflections in which any isometry of $\OO{n}$ can be decomposed. We just hint that expanding each couple $v_i v_{i+1} = v_i \cdot v_{i+1} + v_i \wedge v_{i+1}$ we can expand $\psi$ and moreover expanding each $v_i \wedge v_{i+1}$ in bivector basis $\mygen_{k} \mygen_{l}$
\opt{margin_notes}{\mynote{mbh.note: from \cite[p. 2129]{BudinichP_1989} The Lie algebra spin(g) of the group Spin(g) can be identified with the subspace [V, V] of Cl(g) spanned by all the commutators [u,v] where $u, v \in V$.}}%
$$
v_i \wedge v_{i+1} = \left( \sum_k \alpha_{i, k} \mygen_{k} \right) \wedge \left( \sum_l \alpha_{i+1, l} \mygen_{l} \right) = \sum_{k,l; \, l > k} (\alpha_{i, k} \alpha_{i+1, l} - \alpha_{i, l} \alpha_{i+1, k}) \mygen_{k} \mygen_{l}
$$
we finally arrive at $\psi$ expansion (\ref{formula_Fock_basis_expansion}) since all $\mygen_{i_1} \mygen_{i_2} \cdots \mygen_{i_r} \psi_\Identity \in \myFockB$.

\begin{MS_lemma}
\opt{margin_notes}{\mynote{mbh.note: log p. 783}}%
\label{lemma_sum_intersection_old}
Given a \SAT{} problem let $J$ be any non empty subset of the $m$ clauses, then
$$
\left( \sum_{j \in J} {\cal T}_j \right) \cap \O1n = \cup_{j \in J} {\cal T}_j'
$$
\end{MS_lemma}
\begin{proof}
We already know that $ {\cal T}_j \cap \O1n = {\cal T}_j'$ \cite[Lemma 1]{Budinich_2019} so we just need to prove that $({\cal T}_{j} + {\cal T}_{k}) \cap \O1n = {\cal T}_j' \cup {\cal T}_k'$ but this follows trivially from the definition of ${\cal T}_{j} + {\cal T}_{k}$ (\ref{formula_cal_T_j+k_def}) and from the fact that $\myFockB$ is a proper basis of $\mySpinorS$ and thus any element of ${\cal T}_{j} + {\cal T}_{k}$ is necessarily in $\my_span{{\cal T}_j' \cup {\cal T}_k'}$.
\end{proof}

\begin{MS_Corollary}
\label{coro_sum_support_old}
Given $t \in \OO{n}$ such that $t \in \sum_{j \in J} {\cal T}_j$ where $J$ is any non empty subset of the $m$ clauses and its corresponding simple spinor $\psi_t$ such that $M(\psi_t) = (\Identity, t)$, then
$$
\sup \psi_t \subseteq \cup_{j \in J} {\cal T}_j' \dotinformula
$$
\end{MS_Corollary}

We recall some standard properties of $\R^n$ and of its group of isometries $\OO{n}$: it is well known that any $t \in \OO{n}$ can only have $3$ kind of eigenvalues: $\pm 1$ and couples of complex conjugates and that all eigenvectors corresponding to different eigenvalues are reciprocally orthogonal. Moreover any $t \in \OO{n}$ having only $\pm 1$ eigenvalues is an \emph{involution}: $t^2 = \Identity$ and $t = t^T$, a prominent example being the $2^n$ involutions $\lambda \in \O1{n}$.

It follows that any $t \in \OO{n}$ defines univocally three reciprocally orthogonal subspaces of $\R^n$ corresponding respectively to eigenvectors of $\pm 1$ eigenvalues and to their orthogonal complement and at least one of these subspaces have non null dimensions. These subspaces are in close connection with the Wall parametrization of $t \in \OO{n}$ \cite[Chapter 11]{Taylor_1992} but we do not insist on this here.

We define $\inv{t}$ to be the \emph{involutory subspace} of $t$, namely the subspace of $\R^n$ spanned by the eigenvectors corresponding to $\pm 1$ eigenvalues, namely $\inv{t} = \ker(t + \Identity) \oplus \ker(t - \Identity)$; obviously $\inv{t}^\perp$ is the subspace corresponding to complex eigenvalues and
$$
\inv{t} \oplus \inv{t}^\perp = \ker(t + \Identity) \oplus \ker(t - \Identity) \oplus \inv{t}^\perp = \R^n
$$
and moreover the two $t$ restricted to $\inv{t}$ and $\inv{t}^\perp$ commute.

Applying this definition to our case for any $t \in {\cal T}_j$ then $\my_span{\sup z_j} \subseteq \inv{t}$ where the inclusion can be strict since \inv{t} can contain other directions in subspace $\my_span{\sup z_j}^\perp$. Since for our problem all $\sup z_j$ are subsets of generators we will focus here on cases in which \inv{t} is made by subset of generators but in general it can contain any $\{0\} \times \R^n$ direction.

\begin{MS_lemma}
\label{lemma_t_j_form}
Given clause $z_j$ then $t \in {\cal T}_j$ if and only if it can be written as $t = [z_j] t'$ where:
\begin{itemize}
\item $[z_j] \in \O1{n}$ is the involution obtained from clause $z_j$ completed with all 1 in coordinates in $(\sup z_j)^\perp$, namely $[z_j] = \mygen_{r} \ldots \mygen_{s} \mygen_{t}^2 \ldots \mygen_{u}^2 \Identity$ where $\mygen_{r} \ldots \mygen_{s}$ stand for the Boolean variables in negative forms of $z_j$ and $\mygen_{t}^2 \ldots \mygen_{u}^2$ (all equal to $1$) are the 'placeholders' for the affirmative Boolean variables of $z_j$ put there just to remind us their position in $\sup z_j$,
\item $t' \in \OO{n-k}$ acts in subspace $(\sup z_j)^\perp$.
\end{itemize}
\end{MS_lemma}
\begin{proof}
The proof is immediate since $M(z_j) \subseteq (\Identity, [z_j])$ and for any $t' \in \OO{n-k}$ then $t \in \OO{n}$ and $M(z_j) \subseteq (\Identity, t)$ and conversely any $t \in \OO{n}$ satisfying this property can be put in the given form.
\end{proof}

Since $M(z_j) \subseteq (\Identity, t)$ the inclusion must be necessarily in the involutory part and thus $M(z_j) \subseteq (\Identity, [z_j])$, moreover the involutory and non involutory parts commute. In summary we just gave another equivalent definition of ${\cal T}_j$ by means of involution $[z_j]$.

\myseparation

To simplify the proof of the induction step we remark that with (\ref{formula_t_in_T_j_form}) for any clause we may write
$$
[z_j] \prod_{i=1}^{r_j} (\Identity + \mygen_{i} \mygen_{i+1}) \Identity \in {\cal T}_j
$$
where $r_j$ is the maximum number of consecutive terms contained in the product and that in general is different for each clause. Let the clause $z_l$ be such that $r_l \ge r_j$ for all other clauses $z_j$ (in practice is the clause whose lowest index in $\sup z_l$ is not inferior to similar indexes in all clauses) and this clause alone contains the first $r_l$ terms of $t$ (\ref{formula_time_test}). This clause thus contains $t_{k-1}$ with the first $k-1 = r_l$ terms of the induction proof and we need to proceed to
$$
t_{k} = (\Identity + \mygen_{k-1} \mygen_{k}) t_{k-1}
$$
by Lemma~\ref{lemma_eiejt} $(\Identity + \mygen_{k-1} \mygen_{k}) t_{k-1} \in \sum_{j \in J_k} {\cal T}_j$ if and only if $\mygen_{k-1} \mygen_{k} t_{k-1} \in \sum_{j \in J_k} {\cal T}_j$ and either we find $J_{k-1} \subseteq J_k$ and $t_k = (\Identity + \mygen_{k-1} \mygen_{k}) t_{k-1} \in \sum_{j \in J_k} {\cal T}_j$ or we have a certificate that $t_k \notin \sum_{j = 1}^m {\cal T}_j$ (and thus also $t$).

\myseparation

Bt hypothesis $t_{k-1} \in \sum_{j \in J_{k-1}} {\cal T}_j$ and with Lemma~\ref{lemma_z_j+z_k} we can write
\begin{equation}
\label{formula_t_k-1_old}
t_{k-1} = [z_j] \prod_{i=1}^{k-2} (\Identity + \mygen_{i} \mygen_{i+1}) \Identity
%\in {\cal T}_l + {\cal T}_o + {\cal T}_p + \cdots + {\cal T}_x + {\cal T}_y
\end{equation}
and given that the non involutory part of $t_{k-1}$ covers $\{ \mygen_{1}, \mygen_{2}, \ldots, \mygen_{k-1} \}$ then $\inv{t_{k-1}} = \{ \mygen_{k}, \mygen_{k+1}, \ldots, \mygen_{n} \}$ and we have to find whether there are clauses that contain $\mygen_{k-1} \mygen_{k} t_{k-1}$ allowing us to go to next $t_{k}$. Moreover since by hypothesis $\mygen_{k-1}$ is in the non involutory part of $t_{k-1}$ it follows $\mygen_{k-1} \notin \sup z_j$ and we need to focus our attention on $\mygen_{k}$. There are only two possibilities: either there are no clauses that contain $\mygen_{k-1} \mygen_{k} t_{k-1}$ and thus also $t$ or there exist one or more clauses $J_k \subseteq \{1, 2, \ldots , m\}$ such that $\mygen_{k-1} \mygen_{k} t_{k-1} \in \sum_{j \in J_k} {\cal T}_j$.

\myseparation

By hypothesis $\mygen_{k} \in \sup z_j$, so in $[k] \mygen_{x'} \cdots \mygen_{z'}$ it is opposite to that in $z_j$ and it could appear either in $\sup ((k))$ or in $\mygen_{x'} \cdots \mygen_{z'}$. If $\mygen_{k} \notin \sup ((k)$ by Lemma~\ref{lemma_eit_in_Tj} we would have that $\sum_{j \in J_k} {\cal T}_j$ contains not only $\mygen_{k-1} \mygen_{k} t_{k-1}$ but also $t_{k-1}$ but this is impossible since it would contain also $(\Identity + \mygen_{k-1} \mygen_{k}) t_{k-1}$ against hypothesis that $z_j$ is the clause that contains the largest non involutory part so we must conclude that $\mygen_{k} \in \sup ((k))$.

\myseparation

\begin{MS_Proposition}
\label{prop_time_test_old}
(OLD) Given $t \in \OO{n}$ and $m$ sets ${\cal T}_j$ induced by clauses checking wether
\begin{equation}
\label{formula_time_test_old}
t = \prod_{i=1}^{n-1} (\Identity + \mygen_{i} \mygen_{i+1}) \Identity \; \stackrel{?}{\in} \; \sum_{j = 1}^m {\cal T}_j
\end{equation}
can be done in polynomial time \eg \bigO{n^s}.
\end{MS_Proposition}
\begin{proof}
We give a procedure to verify (\ref{formula_time_test}); by Lemma~\ref{lemma_Givens_all_in_set} $t \in \sum_{j = 1}^m {\cal T}_j$ if and only if for all terms of the succession $t_k \in \sum_{j = 1}^m {\cal T}_j$. We can thus start verifying whether $t_1 = \Identity \in \sum_{j = 1}^m {\cal T}_j$ and continue incrementally up to $t_{n}$: only if all $n$ successive $t_k$ are in $\sum_{j = 1}^m {\cal T}_j$ we can conclude that $t \in \sum_{j = 1}^m {\cal T}_j$, if any $t_k$ fails the test this provides a certificate that $t \notin \sum_{j = 1}^m {\cal T}_j$.
\opt{margin_notes}{\mynote{mbh.note: do we need to convert $t \to \psi_t$ here ?}}%
Calling $J_k \subseteq \{ 1, \ldots, m\}$ the subset of clauses such that $t_k \in \sum_{j \in J_k} {\cal T}_j$, then by Lemma~\ref{lemma_eiejt} $J_1 \subseteq J_2 \subseteq \cdots \subseteq \{ 1, \ldots, m\}$.

We prove the proposition by induction on $k$; for $k = 1$, $t_1 = \Identity$ and $J_1$ can contain any clause having only affirmative Boolean variables since for these clauses $M(z_j) \subseteq (\Identity, \Identity)$ (obviously there may be more than one; should there be none then $t \notin \sum_{j = 1}^m {\cal T}_j$ by Lemma~\ref{lemma_Givens_all_in_set} and, by the way, $\Identity$ would be a solution for the given \SAT{} problem). Looking for these clauses requires a time of \bigO{m}. For the induction step we prove that given $t_{k-1} \in \sum_{j \in J_{k-1}} {\cal T}_j$, verifying whether there exists or not a superset $J_{k-1} \subseteq J_k$ such that $t_k = (\Identity + \mygen_{k-1} \mygen_{k}) t_{k-1} \in \sum_{j \in J_k} {\cal T}_j$ requires polynomial time, \eg \bigO{n^{s-1}} and the complete test (\ref{formula_time_test}) will consequently take \bigO{n^s}.

We thus assume we are given a certain $t_{k-1}$ together with a subset of clauses $J_{k-1}$ such that $t_{k-1} \in \sum_{j \in J_{k-1}} {\cal T}_j$. By Lemma~\ref{lemma_eiejt} $(\Identity + \mygen_{k-1} \mygen_{k}) t_{k-1} \in \sum_{j \in J_k} {\cal T}_j$ if and only if $\mygen_{k-1} \mygen_{k} t_{k-1} \in \sum_{j \in J_k} {\cal T}_j$ and either we find $J_{k-1} \subseteq J_k$ and $t_k = (\Identity + \mygen_{k-1} \mygen_{k}) t_{k-1} \in \sum_{j \in J_k} {\cal T}_j$ or we have a certificate that $t_k \notin \sum_{j = 1}^m {\cal T}_j$ (and thus also $t$).

By hypothesis $t_{k-1} \in \sum_{j \in J_{k-1}} {\cal T}_j$ and with Lemma~\ref{lemma_z_j+z_k} we can write
\begin{equation}
\label{formula_t_k-1_old}
t_{k-1} = [k-1] \mygen_{x} \cdots \mygen_{z} \prod_{i=1}^{k-2} (\Identity + \mygen_{i} \mygen_{i+1}) \Identity
%\in {\cal T}_l + {\cal T}_o + {\cal T}_p + \cdots + {\cal T}_x + {\cal T}_y
\end{equation}
where $[k-1]$ is a shortcut for the involutory part $[[ \cdots [[z_l + z_o] + z_p] + \cdots ] + z_x]$. Given that the non involutory part of $t_{k-1}$ covers $\{ \mygen_{1}, \mygen_{2}, \ldots, \mygen_{k-1} \}$ then $\inv{t_{k-1}} = \{ \mygen_{k}, \mygen_{k+1}, \ldots, \mygen_{n} \}$ and we have to find whether there are clauses that contain $\mygen_{k-1} \mygen_{k} t_{k-1}$ allowing us to go to next $t_{k}$. Moreover since by hypothesis $\mygen_{k-1}$ is in the non involutory part of $t_{k-1}$ it follows $\mygen_{k-1} \notin \sup ((k-1))$, where again $(k-1)$ is a shortcut for $(( \cdots ((z_l + z_o) + z_p) + \cdots ) + z_x)$, and we need to focus our attention on $\mygen_{k}$. There are only three possibilities that we examine in succession:
\begin{enumerate}
\item $J_k = J_{k-1}$, namely $\mygen_{k-1} \mygen_{k} t_{k-1} \in \sum_{j \in J_{k-1}} {\cal T}_j$ that by Lemma~\ref{lemma_eit_in_Tj} happens if and only if $\mygen_{k} \notin \sup ((k-1))$ that can be easily verified in \bigO{n} simply searching $\sup((k-1))$,

\item $J_k = J_{k-1} \cup \{y, w, z, \ldots\}$, namely exist one or more clauses such that $\mygen_{k-1} \mygen_{k} t_{k-1} \in \sum_{j \in \{y, w, z, \ldots\}} {\cal T}_j$ since we are not in case $1$ and, by Lemma~\ref{lemma_eit_in_Tj}, $\mygen_{k} \in \sup ((k-1))$ and we study this case in the sequel,

\item If neither of previous cases occurred than we have a proof that $\mygen_{k-1} \mygen_{k} t_{k-1} \notin \sum_{j = 1}^m {\cal T}_j$ and thus $t_{k}$ and thus also $t$.
\end{enumerate}

We study case $2$ in which $\mygen_{k-1} \mygen_{k} t_{k-1} \in \sum_{j \in \{y, w, z, \ldots\}} {\cal T}_j$ and we indicate with $(k) := (( \cdots (z_y + z_w) + z_z) + \cdots )$ the sum of clauses that contain $\mygen_{k-1} \mygen_{k} t_{k-1}$. By Proposition~\ref{prop_subspace_SSpinors} the non incident part of $t_{k-1}$ and $\mygen_{k-1} \mygen_{k} t_{k-1}$ is necessarily in $\my_span{\mygen_{k-1}, \mygen_{k}}$ and it follows by (\ref{formula_t_k-1}) that the rest must be necessarily identical and thus we necessarily have
\opt{margin_notes}{\mynote{mbh.note: probably I can bring $\mygen_{k-1}$ in the non involutory part}}%
\begin{equation}
\label{formula_involutory_equality_old}
[k] \mygen_{x'} \cdots \mygen_{z'} \prod_{i=1}^{k-2} (\Identity + \mygen_{i} \mygen_{i+1}) \Identity = \mygen_{k-1} \mygen_{k} [k-1] \mygen_{x} \cdots \mygen_{z} \prod_{i=1}^{k-2} (\Identity + \mygen_{i} \mygen_{i+1}) \Identity
\end{equation}
and since the non involutory part is in $\{ \mygen_{1}, \mygen_{2}, \ldots, \mygen_{k-1} \}$ it follows $\sup ((k)) \subseteq \{ \mygen_{k}, \mygen_{k+1}, \ldots \mygen_{n} \}$. Since $\mygen_{k} \in \sup ((k-1))$, in the left hand side of (\ref{formula_involutory_equality}) $\mygen_{k}$ is opposite to that in $[k-1]$ in the right side and, being $\mygen_{k}$ in both involutory parts, in the left side it could appear either in $\sup ((k))$ or in $\mygen_{x'} \cdots \mygen_{z'}$.

Since $\mygen_{k-1} \mygen_{k} t_{k-1}$ is contained in sum $(k)$, of involutory part $[k] \mygen_{x'} \cdots \mygen_{z'}$ (\ref{formula_involutory_equality}), with $[k]$ calculated from the addends by repeated application of Lemma~\ref{lemma_z_j+z_k} and this involutory part must be identical in all terms because only identical Boolean variables in addends remain involutory by Lemma~\ref{lemma_z_j+z_k}. It follows that any clause $z_y$ entering the sum must satisfy
\begin{equation}
\label{formula_clause_nec_condition_old}
M(z_y) \cap \{ \mygen_{k}, \mygen_{k+1}, \ldots \mygen_{n} \} \subseteq (\Identity, \mygen_{k-1} \mygen_{k} [k-1] \mygen_{x} \cdots \mygen_{z} \Identity)
\end{equation}
and the involutory part of $z_y$ in $\{ \mygen_{k}, \mygen_{k+1}, \ldots \mygen_{n} \}$ has to be satisfied exactly whereas possible involutory parts of $z_y$ in $\{ \mygen_{1}, \mygen_{2}, \ldots \mygen_{k-1} \}$ must vanish when summing the clauses to make room for $\prod_{i=1}^{k-2} (\Identity + \mygen_{i} \mygen_{i+1}) \Identity$.

In simple case $J_k = J_{k-1} \cup \{y\}$, the sum $(k)$ reduces to $z_y$ that must satisfy the two conditions $\sup z_y \subseteq \{ \mygen_{k}, \mygen_{k+1}, \ldots \mygen_{n} \}$ and $M(z_y) \subseteq (\Identity, \mygen_{k-1} \mygen_{k} [k-1] \mygen_{x} \cdots \mygen_{z} \Identity)$ and, like in case $1$, we can verify if there exists (at least) one clause with these characteristics scanning the clauses in time \bigO{m}.

In more general case to prove that $\mygen_{k-1} \mygen_{k} t_{k-1} \in \sum_{j \in \{y, w, z, \dots \}} {\cal T}_j$ we need to prove that with clauses of $\{y, w, z, \dots \}$ we can build $\prod_{i=1}^{k-2} (\Identity + \mygen_{i} \mygen_{i+1}) \Identity$: clearly a problem like (\ref{formula_time_test}) in subspace $\{ \mygen_{1}, \mygen_{2}, \ldots \mygen_{k-1} \}$ and it is easy to show that this recursive definition can bring to an exponential time of solution but we now show that this is not the case.

We already remarked that being $\mygen_{k}$ in both involutory parts of (\ref{formula_involutory_equality}), in the left side it could be either in $\sup ((k))$ or in $\mygen_{x'} \cdots \mygen_{z'}$ and let us suppose that it is in $\mygen_{x'} \cdots \mygen_{z'}$: this implies that we can multiply by $\mygen_{k}$ the left side and, by Lemma~\ref{lemma_eit_in_Tj}, we remain in the same sum of clauses $(k)$ and thus we would be in a case in which both $\mygen_{k-1} \mygen_{k} t_{k-1}$ and $t_{k-1}$ are in the same sum of clauses. We would thus be in a different instance of case $1$ the only difference being given by the two set of clauses: $J_{k-1}$ or $\{y, w, z, \dots \}$: this is certainly possible but there would be no difference from previously analyzed case $1$ given that the Fock basis is a standard basis of a linear space and it is irrelevant which set of clauses gives coverage. In other words we are interested in cases in which our sum is in ${\cal T}_y + {\cal T}_w - {\cal T}_y \cup {\cal T}_w$ since only for these cases the coverage of the Fock basis doubles.
\opt{margin_notes}{\mynote{mbh.note: is it possible that there are clauses such that $\mygen_{k} \in \mygen_{x'} \cdots \mygen_{z'}$ but not such that $\mygen_{k} \in \sup ((k))$ ??}}%

We thus focus on cases such that $\mygen_{k} \in \sup ((k))$ and thus in the sum $t_k = (\Identity + \mygen_{k-1} \mygen_{k}) t_{k-1}$, $\mygen_{k}$ is \emph{necessarily} a common opposite Boolean variable. For all these cases we thus have clauses for which necessarily $\mygen_{k} \in \sup z_y$ and thus the remaining part in $\mygen_{i} \in \{ \mygen_{1}, \mygen_{2}, \ldots \mygen_{k-1} \}$ have $k-1$ Boolean variables and if \eg the initial clauses were a 3\SAT{} the subproblem generated in $\{ \mygen_{1}, \mygen_{2}, \ldots \mygen_{k-1} \}$ form a 2\SAT{} problem for which we are looking for an unsatisfiability solution that corresponds to the possibility to build $\prod_{i=1}^{k-2} (\Identity + \mygen_{i} \mygen_{i+1}) \Identity$ in $\{ \mygen_{1}, \mygen_{2}, \ldots \mygen_{k-1} \}$ that is well known to be polynomial thus concluding the proof.
\end{proof}

\myseparation

As already remarked for all clauses of $J_{k-1}$, $\mygen_{k} \in \inv{t_{k-1}}$ whereas for any $y \in J_k - J_{k-1}$ necessarily $\mygen_{k}$ must appear in the involutory part in opposite form this being a necessary condition for selecting these clauses in other words a necessary condition for any clause $y \in J_k - J_{k-1}$ is
\begin{equation}
\label{formula_clause_nec_condition_old2}
M(z_y) \cap \{ \mygen_{k}, \mygen_{k+1}, \ldots \mygen_{n} \} \subseteq (\Identity, \mygen_{k-1} \mygen_{k} [k-1] \mygen_{x} \cdots \mygen_{z} \Identity)
\end{equation}
since this is the part common to all addends since we have to sum it to $t_{k-1}$. The non involutory part is in $\{ \mygen_{1}, \mygen_{2}, \ldots \mygen_{k-1} \}$ and must contain $\prod_{i=1}^{k-2} (\Identity + \mygen_{i} \mygen_{i+1}) \Identity$.

So for example if $J_k - J_{k-1} = \{y\}$ $z_y$ must satisfy the two conditions $\sup z_y \subseteq \{ \mygen_{k}, \mygen_{k+1}, \ldots \mygen_{n} \}$ and $M(z_y) \subseteq (\Identity, \mygen_{k} [k-1] \mygen_{x} \cdots \mygen_{z} \Identity)$ and (\ref{formula_clause_nec_condition}) becomes
$$
M(z_y) \cap \{ \mygen_{k}, \mygen_{k+1}, \ldots \mygen_{n} \} = M(z_y) \subseteq (\Identity, \mygen_{k} [k-1] \mygen_{x} \cdots \mygen_{z} \Identity)
$$
giving previous condition for one clause case.

For $J_k - J_{k-1} = \{y, w\}$ and such that neither $z_y$ nor $z_w$ alone contain $\mygen_{k-1} \mygen_{k} t_{k-1}$, namely in ${\cal T}_y + {\cal T}_w - {\cal T}_y \cup {\cal T}_w$, this implies that both have at least one Boolean variable in $\{ \mygen_{1}, \mygen_{2}, \ldots \mygen_{k-1} \}$ and necessarily it is the same Boolean variable, otherwise they can not be summed and this implies that there exist $\mygen_{i} \in \{ \mygen_{1}, \mygen_{2}, \ldots \mygen_{k-1} \}$ that appear in opposite form in the two clauses that can be summed together since we have shown that to fit for $(\Identity + \mygen_{i} \mygen_{i+1})$ then $\mygen_{i}$ must be a common opposite Boolean variables in the two clauses (example of page 808').

\myseparation

To complete the proof that a given \SAT{} problem is unsatisfiable we have to test that also $\mygen_{i} \psi_t \in \sum_{j = 1}^m {\cal T}_j$ (\ref{formula_SAT_in_O(n)_4}) but this is easily done as outlined in Proposition~\ref{prop_time_test} starting with $t_0 = \mygen_{1} \Identity$ instead of $t_0 = \Identity$.

We remark that this algorithm treats indifferently $2-$ and $3\SAT{}$ problems essentially because all $t \in \OO{n}$ can be always decomposed in a succession of plane rotations.
%and that there are indications that in the $2\SAT{}$ case it results to be very similar to the well known polynomial algorithm of this case.

\myseparation
old pieces...
\myseparation

So starting from the set of compatible clauses $z_y$ we start from $(\Identity + \mygen_{1} \mygen_{2})$ and either there are two clauses that contain this term or we are over and thus we proceed by all $\mygen_{i} \in \{ \mygen_{1}, \mygen_{2}, \ldots \mygen_{k-1} \}$. At every step we look after one or two clauses with given characteristics and this is thus an \bigO{m^2} process.

\myseparation

The simplest case is $J_k = J_{k-1} \cup \{y\}$ where the condition reduces to $M(z_y) \subseteq (\Identity, \mygen_{k-1} \mygen_{k} t_{k-1})$ (if more than one clause fulfills this condition we can freely choose any of them) and $M(z_y)$ is necessarily in the involutory part:
$$
M(z_y) \subseteq (\Identity, \mygen_{k} [[ \cdots [[z_l + z_o] + z_p] + \cdots ] + z_x] \mygen_{x} \cdots \mygen_{z} \Identity)
$$
while $\mygen_{k}$ may or may not be in $\sup z_y$. These conditions are easily verified just scanning the clauses and thus in \bigO{m}. In this case the part of $\inv{t_k}$ generated by clauses is updated from $[[ \cdots [[z_l + z_o] + z_p] + \cdots ] + z_x]$ to $[[ \cdots [[z_l + z_o] + z_p] + \cdots ] + z_x] + z_y]$.
%
% old proof of old case 2
%and only if $\mygen_{k-1} \mygen_{k} t_{k-1} \in {\cal T}_y$ that implies that $\{ \mygen_{k-1}, \mygen_{k} \} \cap \sup z_y = \emptyset$ that guarantees that for any $t \in {\cal T}_y$ also $\mygen_{k-1} \mygen_{k} t \in {\cal T}_y$ and, moreover that also $ t_{k-1} \in {\cal T}_y$ and this happens again if $\{ \mygen_{1}, \ldots \mygen_{k-1} \} \cap \sup z_y = \emptyset$ and thus that $\{ \mygen_{1}, \ldots \mygen_{k} \} \cap \sup z_y = \emptyset$

It remains the case in which several clauses have to be added and thus for all single clauses $z_y$, $M(z_y) \nsubseteq (\Identity, \mygen_{k-1} \mygen_{k} t_{k-1})$ (\ie either $\{ \mygen_{1}, \mygen_{2}, \ldots \mygen_{k-1} \} \cap \sup z_y \ne \emptyset$ or they differ in the involutory part) since otherwise we would have been in previous simpler case.

We analyze first the case in which we suppose there exist exactly two clauses $z_y, z_w$ such that $M(z_y + z_w) \subseteq (\Identity, \mygen_{k-1} \mygen_{k} t_{k-1})$: to find them we just need to build a list of all couples of clauses
\opt{margin_notes}{\mynote{mbh.note: all couples or just the couples of 'unused' clauses ? can I prove $[[ \cdots [[z_l + z_o] + z_p] + \cdots ] + z_x]$ contain all Fock basis covered by already summed clause ? it could also be unessential, I take anyhow \emph{all} couples}}%
that have elements in ${\cal T}_y + {\cal T}_w$ and for each we calculate corresponding $[z_y + z_w]$ and if it satisfies (\ref{formula_involutory_equality}) we are done. This is clearly an (\bigO{m^2}) procedure and in this case the part of $\inv{t_k}$ generated by clauses is updated from $[[ \cdots [[z_l + z_o] + z_p] + \cdots ] + z_x]$ to $[[ \cdots [[z_l + z_o] + z_p] + \cdots ] + z_x] + [z_y + z_w]]$.

We can resume this process in spinor space writing $\psi_y + \psi_w = \psi$ where $\psi = \mygen_{k-1} \mygen_{k} \psi_{k-1}$ with obvious meanings, but since all these spinors have relative incidence $n-2$ we may also write $\psi_y + \psi_w = (\Identity + \mygen_{r} \mygen_{s}) \psi_y = \psi$, with which, observing that $1/2 (\Identity - \mygen_{r} \mygen_{s}) (\Identity + \mygen_{r} \mygen_{s}) = \Identity$ we get $\psi_y = 1/2 (\Identity - \mygen_{r} \mygen_{s}) \psi$ that simply expresses the fact that $\psi_y, \psi_w$ and $\psi$ all belong to a two dimensional linear subspace of spinor space and so any two of them can give the third.

To deal with the case in which there are more than two clauses it is simple to prove that previous relations generalize to $(( \cdots (\psi_y + \psi_w) + \cdots ) + \psi_z) = \psi$ that can be written as $(\Identity + \mygen_{r} \mygen_{s}) \cdots (\Identity + \mygen_{t} \mygen_{u}) \psi_y = \psi$ or, forgetting normalizing factor $2^{-r}$, as $\psi_y = (\Identity - \mygen_{r} \mygen_{s}) \cdots (\Identity - \mygen_{t} \mygen_{u}) \psi$. This means that such clause we can find $\mygen_{k-1} \mygen_{k} \psi_{k-1}$ in ${\cal T}_y + {\cal T}_w + {\cal T}_z + \cdots$ if and only if, starting from $\mygen_{k-1} \mygen_{k} \psi_{k-1}$ and 'moving' in steps of $(\Identity - \mygen_{r} \mygen_{s})$ we arrive at a single clause ${\cal T}_y$. At any move starting from $\mygen_{k-1} \mygen_{k} \psi_{k-1}$ we are thus in a clause or in a sum of clauses. At every step we can calculate the modified involutory part of $(\Identity - \mygen_{t} \mygen_{u}) \psi$ that we have to find in a clause or in a sum of clauses and we have already seen that the first step is done in (\bigO{m^2}) and ...

they necessarily should be in form of $[[z_y + z_w] + z_z]$ and I would know that $[[z_y + z_w] + z_z] \mygen_{x'} \cdots \mygen_{z'} = \mygen_{k} [[ \cdots [[z_l + z_o] + z_p] + \cdots ] + z_x] \mygen_{x} \cdots \mygen_{z}$ and knowing the final form I can look after any clause $z_z$ having incidence $n-2$ with $\mygen_{k-1} \mygen_{k} t_{k-1}$ and if I find it I can bring $z_z$ on the other side, as outlined above, and look after clauses $[z_l + z_o]$ that are equal to $z_z + \mygen_{k-1} \mygen_{k} t_{k-1}$ and so on and thus making at most $k$ steps of (\bigO{m^2}), all in all an (\bigO{m^3}) process.
\opt{margin_notes}{\mynote{mbh.note: it remains to be proved that the order of summing the clauses is irrelevant since we have a real basis}}%

\myseparation

We underline that this procedure (of generating all \mysnqG{} elements of \mysetM{} at the beginning of Section~\ref{sec_O(n)_MTNS}) is remarkably similar to that of forming \SAT{} problems: we must choose, for each of $n$ Boolean variables, whether we take it in plain or complemented form and this similarity is no chance since there is an one to one correspondance between the two procedures \cite{Budinich_2019}. The other way round as any spinor can be seen as the action of a succession of creation and destruction operators applied to a vacuum state, any logical expression can be built by the action of creation and destruction operators on any initial state.
\opt{margin_notes}{\mynote{mbh.note: to be developed}}%

\myseparation

\begin{MS_lemma}
\label{lemma_t_j_form_old}
(OLD) Given clause $z_j$ then $t \in {\cal T}_j$ if and only if it can be written in form
\begin{equation}
\label{formula_t_j}
t = [z_j] \mygen_{x} \cdots \mygen_{z} \prod_i (\Identity + \mygen_{i} \mygen_{i+1})
\end{equation}
where:
\begin{itemize}
\item $[z_j] \in \O1{n}$ is the involution obtained from clause $z_j$ completed with all 1 in coordinates in $(\sup z_j)^\perp$, namely $[z_j] = \mygen_{r} \ldots \mygen_{s} \mygen_{t}^2 \ldots \mygen_{u}^2 \Identity$ where $\mygen_{r} \ldots \mygen_{s}$ stand for the Boolean variables in negative forms of $z_j$ and $\mygen_{t}^2 \ldots \mygen_{u}^2$ (all equal to $1$) are the 'placeholders' for the affirmative Boolean variables of $z_j$ put there just to remind their position in $\sup z_j$,
\item $\mygen_{x} \cdots \mygen_{z}$ are generators of $(\sup z_j)^\perp$,
\item the product $\prod_i$ is extended to the subset of products such that $\mygen_{i}, \mygen_{i+1} \in (\sup z_j)^\perp$,
\end{itemize}
moreover $\inv {t} = [z_j] \mygen_{x} \cdots \mygen_{z}$ and $\inv {t}^\perp = \prod_i (\Identity + \mygen_{i} \mygen_{i+1}) \Identity$ the two subspaces being obviously orthogonal and since $M(z_j) \subseteq (\Identity, t)$ the inclusion must be necessarily in the involutory part and thus $M(z_j) \subseteq (\Identity, [z_j] \mygen_{x} \cdots \mygen_{z})$, moreover the two parts commute.
\end{MS_lemma}
\begin{proof}
... simple ... see also Proposition~\ref{prop_cal_t_j}
\end{proof}

\myseparation

\noindent We will need a similar result that collapses the various cases
\opt{margin_notes}{\mynote{mbh.note: probably also previous Proposition is if and only if}}%
\begin{MS_Corollary}
\label{coro_Tau_j_k_2}
Given two clauses $z_j, z_k$ with their induced sets ${\cal T}_{j}, {\cal T}_{k}$ then $\psi \in {\cal T}_{j} + {\cal T}_{k}$ and $\psi \notin {\cal T}_{j} \cup {\cal T}_{k}$ if and only if the incidence of any couple of $M(\psi)$, $M(\psi_j)$ and $M(\psi_k)$ is $n-2$ and if there exist $\mygen_{i}, \mygen_{l}$ such that $\mygen_{i} \mygen_{l} \psi_j \in {\cal T}_{k}$.
\end{MS_Corollary}
\begin{proof}
The result follows immediately from Propositions~\ref{prop_5_BudinichP_1989}, \ref{prop_subspace_SSpinors} and \ref{prop_Tau_j_k}.
\end{proof}

\myseparation

We need to characterize the form of the partial products $t_k$ of $t$ (\ref{formula_t_full_sup})
\begin{MS_lemma}
\label{lemma_t_(j)_form}
Given $t_k$ together with a subset of clauses $J_k \subseteq \{ 1, 2, \ldots, m \}$ such that $t_k \in \sum_{j \in J_k} {\cal T}_j$ so that we can express it as a sum of various parts each belonging to a subset ${\cal T}_j$
$$
t_k = \prod_{i=1}^{k-1} (\Identity + \mygen_{i} \mygen_{i+1}) \Identity = \sum_{j \in J_{k}} t_{(j)} \qquad \mbox{and} \qquad t_{(j)} \in {\cal T}_j
$$
then the form of terms $t_{(j)}$ is
\begin{equation}
\label{formula_t_(j)}
t_{(j)} = [z_j] \prod_x \mygen_{x} \prod_i (\Identity + \mygen_{i} \mygen_{i+1}) \Identity \quad \implies \quad t_{(j)} \in \OO{n}
\end{equation}
where:
\begin{itemize}
\item $[z_j] \in \O1{n}$ is the involution obtained from clause $z_j$ completed with all 1 in coordinates in $(\sup z_j)^\perp$, namely $[z_j] = \mygen_{r} \ldots \mygen_{s} \mygen_{t}^2 \ldots \mygen_{u}^2 \Identity$ where $\mygen_{r} \ldots \mygen_{s}$ stand for the Boolean variables in negative forms of $z_j$ and $\mygen_{t}^2 \ldots \mygen_{u}^2$ (all equal to $1$) are the 'placeholders' for the affirmative Boolean variables of $z_j$ put there just to remind their position in $\sup z_j$,
\item $\prod_x \mygen_{x}$ are generators of $(\sup z_j)^\perp$ which number is odd when $\det [z_j] = -1$ so that $\det ([z_j] \prod_x \mygen_{x}) = \det t_{(j)} = 1$,
\item the product $\prod_i$ is extended to the subset of products of $t_k$ such that $\mygen_{i}, \mygen_{i+1} \in (\sup z_j)^\perp$,
\item $\inv {t_{(j)}} = [z_j] \prod_x \mygen_{x}$ and $\inv {t_{(j)}}^\perp = \prod_i (\Identity + \mygen_{i} \mygen_{i+1}) \Identity$ the two subspaces being obviously orthogonal and since $M(z_j) \subseteq (\Identity, t_{(j)})$ the inclusion must be necessarily in the involutory part and thus $M(z_j) \subseteq (\Identity, [z_j] \prod_x \mygen_{x})$, moreover referring to different, orthogonal, subspaces the two parts commute.
\end{itemize}
\end{MS_lemma}

\begin{proof}
We prove the proposition by induction on $k$: for $k=1$, $t_1 = \Identity$ and since by hypothesis the subset of clauses $J_k$ allows to express $t_k$, by Lemma~\ref{lemma_Givens_all_in_set}, in $J_k$ we find also all partial products $t_1 , t_2, \ldots, t_k$ and thus there exists $j \in J_k$ such that $t_1 = \Identity = [z_j] \in {\cal T}_j$ that proves case $k=1$.

For the induction step let $t_{k-1} \in \sum_{j \in J_{k}} {\cal T}_j$ with $t_{(j)}$ as in (\ref{formula_t_(j)}) and
$$
t_k = (\Identity + \mygen_{k-1} \mygen_{k}) t_{k-1} = \sum_{j \in J_{k}} (\Identity + \mygen_{k-1} \mygen_{k}) t_{(j)}
$$
and for the terms of the sum there are two possibilities: for those such that $\mygen_{k-1} \mygen_{k} t_{(j)} \in {\cal T}_j$ then to the new version of $t_{(j)}$ we can add term $(\Identity + \mygen_{k-1} \mygen_{k})$ to $\inv {t_{(j)}}^\perp$ and its form is as in (\ref{formula_t_(j)}) and there is nothing more to prove.

For the other terms $\mygen_{k-1} \mygen_{k} t_{(j)} \notin {\cal T}_j$ but by hypothesis there exists either ${\cal T}_l$ or a sum, \eg ${\cal T}_l + {\cal T}_o + \cdots + {\cal T}_y$, that contains $\mygen_{k-1} \mygen_{k} t_{(j)}$ and thus we need to verify form (\ref{formula_t_(j)}) for these terms only. We start from the case in which there exists a single $l \in J_k$ such that $\mygen_{k-1} \mygen_{k} t_{(j)} \in {\cal T}_l$ and thus
$$
\mygen_{k-1} \mygen_{k} t_{(j)} = \mygen_{k-1} \mygen_{k} [z_j] \prod_x \mygen_{x} \prod_i (\Identity + \mygen_{i} \mygen_{i+1}) = [z_l] \prod_y \mygen_{y} \prod_o (\Identity + \mygen_{o} \mygen_{o+1})
$$
\opt{margin_notes}{\mynote{mbh.note: is this TRUE ???}}%
since by induction hypothesis also $t_{(l)}$ has form (\ref{formula_t_(j)}). Given the orthogonality of involutory and non involutory parts, it must hold separately
$$
\mygen_{k-1} \mygen_{k} [z_j] \prod_x \mygen_{x} = [z_l] \prod_y \mygen_{y} \qquad \mbox{and} \qquad \prod_i (\Identity + \mygen_{i} \mygen_{i+1}) = \prod_o (\Identity + \mygen_{o} \mygen_{o+1})
$$
with two remarks: the first equality \emph{do not} imply a similar equality between clauses $z_j$ and $z_l$ and second equality requires that \emph{all} terms of $\prod_i (\Identity + \mygen_{i} \mygen_{i+1})$ appear also in $\prod_o (\Identity + \mygen_{o} \mygen_{o+1})$ because by Lemma~\ref{lemma_Givens_all_in_set} we can remove any number of terms to $\prod_o (\Identity + \mygen_{o} \mygen_{o+1})$ remaining within ${\cal T}_l$.

There are now two possibilities: if $\{ \mygen_{k-1}, \mygen_{k} \} \cap \sup z_l = \emptyset$ then, since we can multiply by $\mygen_{k-1} \mygen_{k}$ we get that also $t_{(j)} \in {\cal T}_l$ and so $t_{(l)}$ can be updated adding term $(\Identity + \mygen_{k-1} \mygen_{k})$ to $\inv {t_{(l)}}^\perp$ and its form is as in (\ref{formula_t_(j)}) while $t_{(j)}$ is untouched and the proposition is proved in this case. Otherwise if $\{ \mygen_{k-1}, \mygen_{k} \} \cap \sup z_l \neq \emptyset$ this is not possible and the non involutory part of $t_{(l)}$ remains untouched while if one of $\mygen_{k-1}, \mygen_{k}$ is not in $\sup z_l$ it is added to $\prod_y \mygen_{y}$ and the proposition is proved also in this case. We remark that only in this latter case $(\Identity + \mygen_{k-1} \mygen_{k}) t_{(j)} = t_{(j)} + t_{(l)} \notin {\cal T}_j \cup {\cal T}_l$.

%and if both $\mygen_{k-1}, \mygen_{k} \in \sup z_l$ then, since $[z_l] = z_l \Identity$ necessarily $[z_l] = \mygen_{1} \mygen_{2} \Identity$ and thus $t_{(l)} = [z_l]$ whereas if \eg only $\mygen_{1} \in \sup z_l$ and thus $[z_l] = \mygen_{1} \Identity$ and $t_{(l)} = [z_l] \mygen_{2}$ and the proposition is thus proved for $k = 1$.
%
%
%
% then, since $\mygen_{1} \mygen_{2} \Identity \in {\cal T}_l$, it follows that $M(z_l) \subseteq (\Identity, \mygen_{1} \mygen_{2} \Identity)$ and if both $\mygen_{1}, \mygen_{2} \in \sup z_l$ then, since $[z_l] = z_l \Identity$ necessarily $[z_l] = \mygen_{1} \mygen_{2} \Identity$ and thus $t_{(l)} = [z_l]$ whereas if \eg only $\mygen_{1} \in \sup z_l$ and thus $[z_l] = \mygen_{1} \Identity$ and $t_{(l)} = [z_l] \mygen_{2}$ and the proposition is thus proved for $k = 1$.
%
%
%
%
%
%
%and thus, again like in the case of $k=1$, necessarily either $[z_l] = \mygen_{k-1} \mygen_{k} [z_j]$ or \eg $[z_l] = \mygen_{k-1} [z_j]$ and in any case $t_{(l)}$ remains unchanged; in both cases the form is that of (\ref{formula_t_(j)}) and the thesis is proved.

It remains the case $\mygen_{k-1} \mygen_{k} t_{(j)} \in {\cal T}_l + {\cal T}_o + \cdots + {\cal T}_y$: we start remarking that to have $\mygen_{k-1} \mygen_{k} t_{(j)} = t_{(l)} + t_{(o)} \in {\cal T}_l + {\cal T}_o$ necessarily by Corollary~\ref{coro_Tau_j_k_2} the incidence of any two of the null subspaces $(\Identity, \mygen_{k-1} \mygen_{k} t_{(j)})$, $(\Identity, t_{(l)})$ and $(\Identity, t_{(o)})$ must be $n-2$ and they can differ either in the involutory part or in the non involutory since the difference in non involutory part is surely even and so it can be either $0$ or $2$. If the non common subspace is in the involutory parts then the non involutory parts of the two terms must be equal. Thus $\mygen_{k-1} \mygen_{k} t_{(j)} = t_{(l)} + t_{(o)}$ may only happen if the non involutory parts of $t_{(l)}$ and $t_{(o)}$ are equal and both contain all the terms of $t_{(j)}$ but one, let it be $(\Identity + \mygen_{r-1} \mygen_{r})$ and moreover $[z_o] = \mygen_{r-1} \mygen_{r} [z_l]$ so that $\mygen_{k-1} \mygen_{k} t_{(j)} = (\Identity + \mygen_{r-1} \mygen_{r})t_{(l)} = t_{(l)} + t_{(o)} \in {\cal T}_l + {\cal T}_o$. If the non common subspace is in the non involutory parts then the involutory parts of the two terms must be equal and $\mygen_{k-1} \mygen_{k} t_{(j)} = t_{(l)} + t_{(o)}$ may only happen if the non involutory parts of $t_{(l)}$ and $t_{(o)}$ differ by just one factor, let it be $(\Identity + \mygen_{r-1} \mygen_{r})$ and thus also in this case $\mygen_{k-1} \mygen_{k} t_{(j)} = (\Identity + \mygen_{r-1} \mygen_{r}) t_{(l)} = t_{(l)} + t_{(o)} \in {\cal T}_l + {\cal T}_o$ since by Lemma~\ref{lemma_eiejt} $ \mygen_{r-1} \mygen_{r} t_{(l)} \in {\cal T}_o$. This reasoning can be iterated to any number of addends that, incidentally, can be no more than the terms appearing in the non involutory part of $\mygen_{k-1} \mygen_{k} t_{(j)}$. Once written in this way the form of the various terms used to build $\mygen_{k-1} \mygen_{k} t_{(j)}$ can be easily shown to have the required form as done in previous simpler case of $\mygen_{k-1} \mygen_{k} t_{(j)} \in {\cal T}_l$.
%
%
%\myseparation
%
%
%
%
%We prove the proposition by induction on $k$: for $k=2$, $t_2 = (\Identity + \mygen_{1} \mygen_{2}) \Identity$ and by hypothesis there is $j \in \{ 1, \ldots, m\}$ such that $\Identity = [z_j] \in {\cal T}_j$, now there are two possibilities: in the first case also $\mygen_{1} \mygen_{2} \Identity \in {\cal T}_j$ and then $t_1 \in {\cal T}_j$ and thus $t_1 = t_{(j)} = [z_j] (\Identity + \mygen_{1} \mygen_{2}) \Identity$. In the other case $\mygen_{1} \mygen_{2} \Identity \notin {\cal T}_j$ and by hypothesis there exists also $l \in \{ 1, \ldots, m\}$ such that $\mygen_{1} \mygen_{2} \Identity \in {\cal T}_l$ and thus $t_1 \in {\cal T}_j + {\cal T}_l$ and $t_{(j)} = [z_j] \Identity$ whereas for $t_{(l)}$ there are two more possibilities: if $\{ \mygen_{1}, \mygen_{2} \} \cap \sup z_l = \emptyset$ then also $\Identity \in {\cal T}_l$ and thus $t_{(l)} = [z_l] (\Identity + \mygen_{1} \mygen_{2}) \Identity$. If on the contrary $\{ \mygen_{1}, \mygen_{2} \} \cap \sup z_l \ne \emptyset$ then, since $\mygen_{1} \mygen_{2} \Identity \in {\cal T}_l$, it follows that $M(z_l) \subseteq (\Identity, \mygen_{1} \mygen_{2} \Identity)$ and if both $\mygen_{1}, \mygen_{2} \in \sup z_l$ then, since $[z_l] = z_l \Identity$ necessarily $[z_l] = \mygen_{1} \mygen_{2} \Identity$ and thus $t_{(l)} = [z_l]$ whereas if \eg only $\mygen_{1} \in \sup z_l$ and thus $[z_l] = \mygen_{1} \Identity$ and $t_{(l)} = [z_l] \mygen_{2}$ and the proposition is thus proved for $k = 1$.
%
%
%
%
%
%
%
%
%
%
\end{proof}

\myseparation

Old Proof of Proposition~\ref{prop_time_test}.

\begin{proof}
We give a procedure to verify (\ref{formula_time_test}); by Lemma~\ref{lemma_Givens_all_in_set} $t \in \sum_{j = 1}^m {\cal T}_j$ if and only if for all terms of the succession $t_k \in \sum_{j = 1}^m {\cal T}_j$. We can thus start verifying whether $t_1 = \Identity \in \sum_{j = 1}^m {\cal T}_j$ and continue incrementally up to $t_{n}$: only if all $n$ successive $t_k$ are in $\sum_{j = 1}^m {\cal T}_j$ we can conclude that $t \in \sum_{j = 1}^m {\cal T}_j$, if any $t_k$ fails the test this provides a certificate that $t \notin \sum_{j = 1}^m {\cal T}_j$.
\opt{margin_notes}{\mynote{mbh.note: do we need to convert $t \to \psi_t$ here ?}}%
Calling $J_k \subseteq \{ 1, \ldots, m\}$ the subset of clauses such that $t_k \in \sum_{j \in J_k} {\cal T}_j$, then by Lemma~\ref{lemma_eiejt} $J_1 \subseteq J_2 \subseteq \cdots \subseteq \{ 1, \ldots, m\}$.

We prove the proposition by induction on $k$; for $k = 1$, $t_1 = \Identity$ and $J_1$ can contain any clause having only affirmative Boolean variables since for these clauses $M(z_j) \subseteq (\Identity, \Identity)$ (obviously there may be more than one; should there be none then $t \notin \sum_{j = 1}^m {\cal T}_j$ by Lemma~\ref{lemma_Givens_all_in_set}). Looking for these clauses requires a time of \bigO{m}. For the induction step we prove that given $t_{k-1} \in \sum_{j \in J_{k-1}} {\cal T}_j$, verifying whether there exists or not a superset $J_{k-1} \subseteq J_k$ such that $t_k = (\Identity + \mygen_{k-1} \mygen_{k}) t_{k-1} \in \sum_{j \in J_k} {\cal T}_j$ requires polynomial time, \eg \bigO{n^{s-1}} and the complete test (\ref{formula_time_test}) will clearly take \bigO{n^s}.

We thus assume we are given a certain $t_{k-1}$ together with a subset of clauses $J_{k-1}$ such that $t_{k-1} \in \sum_{j \in J_{k-1}} {\cal T}_j$. By Lemma~\ref{lemma_eiejt} $(\Identity + \mygen_{k-1} \mygen_{k}) t_{k-1} \in \sum_{j \in J_k} {\cal T}_j$ if and only if $\mygen_{k-1} \mygen_{k} t_{k-1} \in \sum_{j \in J_k} {\cal T}_j$ and either we find $J_{k-1} \subseteq J_k$ and $t_k = (\Identity + \mygen_{k-1} \mygen_{k}) t_{k-1} \in \sum_{j \in J_k} {\cal T}_j$ or we have a certificate that $t_k \notin \sum_{j = 1}^m {\cal T}_j$ (and thus also $t$).

With Lemma~\ref{lemma_t_(j)_form}
$$
t_{k-1} = \sum_{j \in J_{k-1}} t_{(j)} \qquad t_{(j)} \in {\cal T}_j \qquad M(z_j) \subseteq (\Identity, t_{(j)})
$$
with $t_{(j)}$ given by (\ref{formula_t_(j)}) (we remark that the expanded form of $t_{(j)}$ could be prohibitively expensive to calculate since it could contain \bigO{2^n} Fock basis elements but we will use just unexpanded form (\ref{formula_t_(j)})) so that
$$
\mygen_{k-1} \mygen_{k} t_{k-1} = \sum_{j \in J_{k}} \mygen_{k-1} \mygen_{k} t_{(j)} = \sum_{j \in J_{k}} \mygen_{k-1} \mygen_{k} [z_j] \prod_i (\Identity + \mygen_{i} \mygen_{i+1})
$$
and we can easily check whether $\mygen_{k-1} \mygen_{k} t_{(j)}$ belongs or not to ${\cal T}_j$ applying Lemma~\ref{lemma_eit_in_Tj}: $\mygen_{k-1} \mygen_{k} t_{(j)} \in {\cal T}_j$ if and only if $\{ \mygen_{k-1}, \mygen_{k} \} \cap \sup z_j = \emptyset$.
%
%If for all $j \in J_{k-1}$ we have $\mygen_{k-1} \mygen_{k} t_{(j)} \in {\cal T}_j$ this gives a sufficient condition to have $J_{k-1} =J_k$. This conditions is not necessary since it may happen that $\mygen_{k-1} \mygen_{k} t_{(j)} \in {\cal T}_{l}$ so that we can conclude that: $(\Identity + \mygen_{k-1} \mygen_{k}) t_{k-1} \in \sum_{j \in J_k} {\cal T}_j$ if and only if for any $j \in J_{k-1}$ then $\mygen_{k-1} \mygen_{k} t_{(j)} \in {\cal T}_{l}$ for at least one $l \in \{ 1, 2, \ldots, m \}$.

We subdivide the various $\mygen_{k-1} \mygen_{k} t_{(j)}$ into two subsets: those $j \in J_{k-1}$ such that $\mygen_{k-1} \mygen_{k} t_{(j)} \in {\cal T}_j$ and for them we need to do nothing and the complementary subset of $j \in J_{k-1}$ such that $\mygen_{k-1} \mygen_{k} t_{(j)} \notin {\cal T}_j$. For each of these latter cases there are only two possibilities:
\begin{itemize}
\item we prove that in the set of all clauses $\{ 1, \ldots, m\}$ there exists a clause $z_{l}$ or a sum of clauses in which we may express $\mygen_{k-1} \mygen_{k} t_{(j)}$ (note that $l$ can be in $J_{k-1}$ or not) and thus that $t_k = (\Identity + \mygen_{k-1} \mygen_{k}) t_{k-1} \in \sum_{j \in J_k} {\cal T}_j$ with $J_{k-1} \subseteq J_k$;
\item we prove that for at least one $\mygen_{k-1} \mygen_{k} t_{(j)}$ there are no clauses that contain it and thus that $t_k \notin \sum_{j = 1}^m {\cal T}_j$ (and thus also $t$).
\end{itemize}

In any case given $\mygen_{k-1} \mygen_{k} t_{(j)} \notin {\cal T}_j$ we need a procedure that looks after a clause $z_{l}$ or a sum of clauses that contain $\mygen_{k-1} \mygen_{k} t_{(j)}$.

We start with the case of a single clause $z_l$ such that $\mygen_{k-1} \mygen_{k} t_{(j)} \in {\cal T}_{l}$ and both $t_{(j)}$ and $t_{(l)}$ are as in (\ref{formula_t_(j)}): if $\{ \mygen_{k-1}, \mygen_{k} \} \cap \sup z_l = \emptyset$ then the form of non involutory part of $t_{(l)}$ must be identical to that of $t_{(j)}$ while if $\{ \mygen_{k-1}, \mygen_{k} \} \cap \sup z_l \neq \emptyset$ beyond this it must also be true that $\mygen_{k-1} \mygen_{k} [z_j] = [z_l]$. In both cases this is a \bigO{m} process since we have just to search within clauses and if the same search is repeated for all clauses $z_j$ such that $\mygen_{k-1} \mygen_{k} t_{(j)} \notin {\cal T}_j$ is thus a \bigO{m^2} process.

We examine the case in which $\mygen_{k-1} \mygen_{k} t_{(j)}$ is in a sum of ${\cal T}$'s and we start from simplest case $\mygen_{k-1} \mygen_{k} t_{(j)} \in {\cal T}_l + {\cal T}_o$. By Corollary~\ref{coro_Tau_j_k_2} this happens if and only if the incidence of any two of the three null subspaces $(\Identity, \mygen_{k-1} \mygen_{k} t_{(j)})$, $(\Identity, t_{(l)})$ and $(\Identity, t_{(o)})$ is $n-2$ and, as mentioned in the proof of Lemma~\ref{lemma_t_(j)_form}, we can consider separately the involutory and non involutory parts. In this search $\mygen_{k-1} \mygen_{k} t_{(j)}$ is given and we look for viable clauses $z_l$ and $z_o$: for them the involutory part is frozen while, by Lemma~\ref{lemma_Givens_all_in_set}, we can remove any number of terms from the non involutory part and they remain in their respective sets ${\cal T}_l$ and ${\cal T}_o$. In other words we can build a list of clauses that can have incidence $n-2$ with $\mygen_{k-1} \mygen_{k} t_{(j)}$ and, if among them there are two having also reciprocal incidence $n-2$, our problem is solved and we can proceed; in this case we have found $\mygen_{k-1} \mygen_{k} t_{(j)} = (\Identity + \mygen_{r-1} \mygen_{r}) t_{(l)} = t_{(l)} + \mygen_{r-1} \mygen_{r} t_{(l)} \in {\cal T}_l + {\cal T}_o$. If such clauses do not exist we have a certificate that $\mygen_{k-1} \mygen_{k} t_{(j)}$ is not in $\sum_{j = 1}^m {\cal T}_j$ (and also get a partial hint for a solution of the given problem). In this case the search of two clauses that contain $\mygen_{k-1} \mygen_{k} t_{(j)}$ is a \bigO{m^2} process and the same search repeated for all clauses is thus a \bigO{m^3} process.

We are left with the case $\mygen_{k-1} \mygen_{k} t_{(j)} \in {\cal T}_l + {\cal T}_o + {\cal T}_p + \cdots + {\cal T}_x + {\cal T}_y$ and by remarks made in the proof of Lemma~\ref{lemma_t_(j)_form} the maximum number of addends $r$ is the number of terms appearing in the non involutory part of $\mygen_{k-1} \mygen_{k} t_{(j)}$. In this case we are looking for
$$
\mygen_{k-1} \mygen_{k} t_{(j)} = (\Identity + \mygen_{r-1} \mygen_{r}) t_{(l)} = (\Identity + \mygen_{r-1} \mygen_{r}) (\Identity + \mygen_{r-1} \mygen_{r}) t_{(p)} = \cdots = \prod_i (\Identity + \mygen_{i} \mygen_{i+1})
$$
where the last term can contain up to all terms of the non involutory part of $t_{(j)}$. If this is true than we will have a succession of clauses such that $\mygen_{k-1} \mygen_{k} t_{(j)} \in {\cal T}_l + ({\cal T}_o + ({\cal T}_p + \cdots + ({\cal T}_x + {\cal T}_y)) \cdots )$ since it is easy to check that the sum (\ref{formula_cal_T_j+k_def}) is commutative but \emph{not} associative. To look after these clauses it is better to order lexicographically all the clauses with respect to the terms contained in their non involutory parts, a \bigO{m \log m} process, and then we have to look after two clause ${\cal T}_x + {\cal T}_y$ having incidence $n-2r$ with $\mygen_{k-1} \mygen_{k} t_{(j)}$ and reciprocal incidence $n-2$ and then for a third clause having incidence $n-2(r-1)$ with $\mygen_{k-1} \mygen_{k} t_{(j)}$ and incidence $n-2$ with $t_{(x)} + t_{(y)}$ until we arrive at clause ${\cal T}_l$ having incidence $n-2$ both with $\mygen_{k-1} \mygen_{k} t_{(j)}$ and with previous sum. In any fashion this a polynomial process since to look at the first couple is an \bigO{m^2} process and then again we look for another clause that can be summed to the first two, again an \bigO{m^2} process and so the whole search is a \bigO{m^3} process and the same search repeated for all clauses is thus a \bigO{m^4} process.
\opt{margin_notes}{\mynote{mbh.note: do we need here a part proving that if there are no clauses satisfying the request the same elements can't be found in any other combination of clauses ? a sketch of this part is here commented.}}%
%
%It remains to be proved that if there are no clauses satisfying $\mygen_{k-1} \mygen_{k} t_{(j)} \in {\cal T}_l + {\cal T}_o + \cdots + {\cal T}_y$ then the same elements can't be found in any other combination of clauses. Let us suppose the contrary namely that we do not find any such clause (or sum of clauses) but in any case $\mygen_{k-1} \mygen_{k} t_{(j)}$ can be expressed in Fock basis and the given hypothesis would violate Lemma~\ref{lemma_sum_intersection_old} since Fock basis is a full basis of a linear space. In any case $\mygen_{k-1} \mygen_{k} t_{(j)}$ has the same non involutory part of $t_{(j)}$ with a different involutory part and if such element exists it must necessarily be in ${\cal T}_l + {\cal T}_o + \cdots + {\cal T}_y$ with the characteristics expressed above.
\end{proof}

\myseparation

The Givens expansion of isometries of $\OO{n}$ \cite{Hoffman_Raffenetti_Ruedenberg_1972} in a nutshell:
\opt{margin_notes}{\mynote{mbh.note: log p. 780.3}}%
\begin{MS_Proposition}
\label{prop_Givens_expansion}
Any $t \in \OO{n}$ can be decomposed in a succession of $g \le {\bino{n}{2}}$ $\SO{2}$ rotations $t_{i l}$ (with $t_{i l} \ne \pm \Identity_2$) acting in subspaces $\my_span{\mygen_{i}, \mygen_{l}}$ namely
\begin{equation}
\label{formula_Givens_expansion}
t = \mygen_{x} \prod_{i, l; l > i} t_{i l}
\end{equation}
where the ordering of the $g$ $t_{i l}$ can be freely chosen and where $\mygen_{x}$ accounts for cases of $\det t = -1$.
\opt{margin_notes}{\mynote{mbh.note: clearly missing cases of $\det t = -1$ and of 'frozen' subspace \inv{t}.}}%
\end{MS_Proposition}

\noindent Exploiting the bijection between $\mySpinorS_s$ and $\OO{n}$ this proposition can be transposed in spinor language. We warn the reader that in what follows we will freely switch between isometries and simple spinors, vectorial and spinorial representation of \OO{n} trying to follow the least treackerous path. % arduous track ?
So for example the same reflection with respect to $\mygen_{i}$ composed with isometry $t$ is written in vectorial (matricial) form as $(\Identity - 2 \mygen_{i} \mygen_{i}^T) t$ and $\mygen_{i} \psi_t$ in spinorial form. The notation advantage we get in spinorial form is paid loosing the more familiar meaning of matrix formalism.

\begin{MS_Proposition}
\label{prop_SSpinors_Givens_expansion}
Any $\psi \in \mySpinorS_s$ can be written as a Clifford product of $g \le {\bino{n}{2}}$ bivectors of subspaces $\my_span{\mygen_{i}, \mygen_{l}}$ namely
\begin{equation}
\label{formula_Givens_expansion_spinor}
\psi = \mygen_{x} \left( \prod_{i, l; l > i} \mygen_{i} v_l \right) \psi_\Identity
\end{equation}
where $v_l \in \my_span{\mygen_{i}, \mygen_{l}}$, $v_l \ne \mygen_{i}$ and the ordering of the $g$ bivectors $\mygen_{i} v_l$ can be freely chosen and where $\mygen_{x}$ is present only in cases of $\det \psi = -1$.
\end{MS_Proposition}

\noindent We remark that while this Propositions stands for the Givens expansion of $t \in \OO{n}$, Proposition~\ref{prop_all_SSpinors} refers to a close relative: the Householder expansion both having as common ancestor the remarkable Cartan theorem.

\myseparation

Given $u \in \OO{n}$ let $\inv{u} = \my_span{\mygen_{r}, \mygen_{s}, \ldots, \mygen_{t}}$: we can associate to these generators a fictitious clause $z_x$ with $\sup z_x = \{\mygen_{r}, \mygen_{s}, \ldots, \mygen_{t} \}$ and $M(z_x) \subseteq (\Identity, u)$ generalizing also clauses definition (\ref{formula_cal_T_j_def2}) to:
\begin{equation}
\label{formula_cal_T_x_def}
{\cal T}_x = \{ t \in \OO{n} : \inv{u} \subseteq \inv{t} \}
\end{equation}
that coincides with (\ref{formula_cal_T_j_def2}) whenever $\inv{u} = \my_span{\sup z_j}$.

\begin{MS_Proposition}
\label{prop_cal_T_x}
Given ${\cal T}_x$ (\ref{formula_cal_T_x_def}) with \inv{u} made only of $k$ generators $\mygen_{i}$ and corresponding fictitious clause $z_x$ then
$$
{\cal T}_x = \{ t \in \OO{n} : M(z_x) \subseteq (\Identity, t) \} \dotinformula
$$
\end{MS_Proposition}
\begin{proof}
Given $t \in {\cal T}_x$ (\ref{formula_cal_T_x_def}) for any $\mygen_{i} \in \inv{u}$ is also in \inv{t} and $t \mygen_{i} = \pm \mygen_{i}$ accordingly with fictitious clause $z_x$ and thus $M(z_x) \subseteq (\Identity, t)$. Conversely given $t$ such that $M(z_x) \subseteq (\Identity, t)$ it follows that for any $\mygen_{i} \in \sup z_x$ then $t \mygen_{i} = \pm \mygen_{i}$ accordingly with fictitious clause $z_x$ and thus all $\mygen_{i} \in \inv{u}$ are also in \inv{t} and thus $\inv{u} \subseteq \inv{t}$ and $t \in {\cal T}_x$ (\ref{formula_cal_T_x_def}).
\end{proof}

\myseparation

Old Lemma~$5$

\begin{MS_lemma}
\label{lemma_eit_in_Tj_old}
Given ${\cal T}_x$ (\ref{formula_cal_T_x_def}) with \inv{u} made only of $k$ generators $\mygen_{i}$ and corresponding fictitious clause $z_x$ then $\mygen_{i} \mygen_{j} t \in {\cal T}_x$ if and only if $\{ \mygen_{i}, \mygen_{j} \} \cap \inv{u} = \emptyset$.
%
%Given any $t \in {\cal T}_j$ (\ref{formula_cal_T_j_def2}) and any generator $\mygen_{i}$ then $\mygen_{i} t \in {\cal T}_j$ if and only if $\mygen_{i} \notin \sup z_j$.
%
\end{MS_lemma}
\begin{proof}
For any $t \in {\cal T}_x$ by Proposition~\ref{prop_cal_T_x} $M(z_x) \subseteq (\Identity, t)$ and if $\mygen_{i}, \mygen_{j} \notin \inv{u}$ they do not act in $\my_span{\sup z_x}$ and thus $M(z_x) \subseteq (\Identity, \mygen_{i} \mygen_{j} t)$ and $\mygen_{i} \mygen_{j} t \in {\cal T}_x$. Conversely given $\mygen_{i} \mygen_{j} t \in {\cal T}_x$ by same proposition $M(z_x) \subseteq (\Identity, \mygen_{i} \mygen_{j} t)$ and since we know that also $M(z_x) \subseteq (\Identity, t)$ it follows that $\mygen_{i}, \mygen_{j} \notin \inv{u}$.
%
% old proof for less general case of $t \in {\cal T}_j$ (\ref{formula_cal_T_j_def2})
%
%For any $t \in {\cal T}_j$ $M(z_j) \subseteq (\Identity, t)$ and if $\mygen_{i} \notin \sup z_j$ it does not act in $\my_span{\sup z_j}$ and thus $M(z_j) \subseteq (\Identity, \mygen_{i} t)$ and thus $\mygen_{i} t \in {\cal T}_j$. Conversely given $\mygen_{i} t \in {\cal T}_j$ necessarily $M(z_j) \subseteq (\Identity, \mygen_{i} t)$ and since we know that also $M(z_j) \subseteq (\Identity, t)$ it follows that $\mygen_{i}$ does not act in $\my_span{\sup z_j}$ and thus $\mygen_{i} \notin \sup z_j$.
\end{proof}

\myseparation

\begin{MS_Proposition}
\label{prop_time_test_generic}
Given $t \in \OO{n}$ and $m$ sets ${\cal T}_j$ checking wether
$
t \stackrel{?}{\in} \sum_{j = 1}^m {\cal T}_j
$
can be done in polynomial time \eg \bigO{n^s}.
\end{MS_Proposition}
\begin{proof}
We provide a procedure to verify whether $t \in \sum_{j = 1}^m {\cal T}_j$ that starts building Givens expansion of $t$ (\ref{formula_Givens_expansion}) that we assume requires polynomial time \cite{Hoffman_Raffenetti_Ruedenberg_1972}. This expansion gives $t$ as the composition of $g \le {\bino{n}{2}}$ $t_{il}$ operating in $\my_span{\mygen_{i}, \mygen_{l}}$ and implicitly defines a succession $t_0, t_1, \ldots , t_k, \ldots t_g$ in which $t_0 = \Identity$, $t_k$ contains the partial product of the first $k$ of the $g$ terms of (\ref{formula_Givens_expansion}) and $t_g = t$.

By Lemma~\ref{lemma_Givens_all_in_set} $t \in \sum_{j = 1}^m {\cal T}_j$ if and only if for all terms of the succession $t_k \in \sum_{j = 1}^m {\cal T}_j$. We can thus start verifying whether $t_0 \in \sum_{j = 1}^m {\cal T}_j$ and continue incrementally up to $t_g$: only if all $g+1$ successive $t_k$ are in $\sum_{j = 1}^m {\cal T}_j$ we can conclude that $t \in \sum_{j = 1}^m {\cal T}_j$, if any $t_k$ fails the test this provides a certificate that $t \notin \sum_{j = 1}^m {\cal T}_j$.

Moreover by Proposition~\ref{prop_subspace_SSpinors} $\sup \psi_0 \subseteq \sup \psi_1 \subseteq \cdots \subseteq \sup \psi_g$.
\opt{margin_notes}{\mynote{mbh.note: we need to convert $t \to \psi_t$ here}}%
Let $J_k \subseteq \{ 1, \ldots, m\}$ be the subset of clauses such that $t_k \in \sum_{j \in J_k} {\cal T}_j$, then by Lemma~\ref{lemma_eiejt} $J_0 \subseteq J_1 \subseteq \cdots \subseteq J_g$.

We prove the proposition by induction on $k$; for $k = 0$, $t_0 = \Identity$ and $J_0$ can contain any clause having only affirmative Boolean variables since for these clauses $M(z_j) \subseteq (\Identity, \Identity)$ (obviously there may be more than one; should there be none then $t \notin \sum_{j = 1}^m {\cal T}_j$ by Lemma~\ref{lemma_Givens_all_in_set}). Clearly scanning for these clauses requires a time of \bigO{m}. For the induction step we need to prove that given $t_k \in \sum_{j \in J_k} {\cal T}_j$, verifying whether there exists or not a superset $J_k \subseteq J_{k+1}$ such that $t_{k+1} = t_{il} t_k \in \sum_{j \in J_{k+1}} {\cal T}_j$ requires polynomial time, \eg \bigO{n^s}. If we prove this than, being $g = \bigO{n^2}$, the complete test will take \bigO{n^{s+2}}.

We thus assume we are given a certain $t_k$ together with a subset of clauses $J_k$ such that $t_k \in \sum_{j \in J_k} {\cal T}_j$. By Lemma~\ref{lemma_eiejt} $t_{il} t_k \in \sum_{j \in J_{k+1}} {\cal T}_j$ if and only if $\mygen_{i} \mygen_{l} t_k \in \sum_{j \in J_{k+1}} {\cal T}_j$ and there are only two possibilities to proceed: either $J_k = J_{k+1}$ or $J_k \subset J_{k+1}$ and $t_{k+1} = t_{il} t_k \in \sum_{j \in J_{k+1}} {\cal T}_j$; if neither case is true we have a certificate that $t_{k+1} \notin \sum_{j = 1}^m {\cal T}_j$ and thus that $t \notin \sum_{j = 1}^m {\cal T}_j$.

To examine in detail the different cases we write
$$
t_k = \sum_{j \in J_{k}} t_{(j)} \qquad t_{(j)} \in {\cal T}_j
$$
namely for each term $t_{(j)}$ of the sum making up $t_k$ we have $M(z_j) \subseteq (\Identity, t_{(j)})$. We remark that knowing the explicit form of these $t_{(j)}$ could be prohibitively expensive since some of them could contain \bigO{2^n} Fock basis elements but we will not need explicit forms. To study $\mygen_{i} \mygen_{l} t_k$ we write
$$
\mygen_{i} \mygen_{l} t_k = \sum_{j \in J_{k}} \mygen_{i} \mygen_{l} t_{(j)}
$$
and we can easily check, also ignoring the explicit form of $t_{(j)}$, whether $\mygen_{i} \mygen_{l} t_{(j)}$ belongs or not to ${\cal T}_j$ applying Lemma~\ref{lemma_eit_in_Tj}: $\mygen_{i} \mygen_{l} t_{(j)} \in {\cal T}_j$ if and only if $\{ \mygen_{i}, \mygen_{l} \} \cap \sup z_j = \emptyset$.

If for all $j \in J_k$ we have $\mygen_{i} \mygen_{l} t_{(j)} \in {\cal T}_j$ this gives a sufficient condition to have $J_k =J_{k+1}$. This conditions is not necessary since it may also happen that for $j, j' \in J_k$ we have $\mygen_{i} \mygen_{l} t_{(j)} \in {\cal T}_{j'}$ so that we can conclude that: $J_k =J_{k+1}$ if and only if for any $j \in J_k$ then $\mygen_{i} \mygen_{l} t_{(j)} \in {\cal T}_{j'}$ for at least one $j' \in J_k$.

We can subdivide the various $\mygen_{i} \mygen_{l} t_{(j)}$ into two subsets: those $j \in J_k$ such that $\mygen_{i} \mygen_{l} t_{(j)} \in {\cal T}_j$ and for them we need to do nothing and the complementary subset of $j \in J_k$ such that $\mygen_{i} \mygen_{l} t_{(j)} \notin {\cal T}_j$. For each of these latter cases there are only three possibilities:
\begin{itemize}
\item we can prove that in $J_k$ there exists a clause $z_{j'}$ (or a sum of these clauses) such that $\mygen_{i} \mygen_{l} t_{(j)} \in {\cal T}_{j'}$ and thus that $t_{k+1} = t_{il} t_k \in \sum_{j \in J_k} {\cal T}_j$ and $J_k =J_{k+1}$;
\item we can prove that in the set of all clauses $\{ 1, \ldots, m\}$ there exists a clause $z_{j'}$ (or a sum of clauses) such that $\mygen_{i} \mygen_{l} t_{(j)} \in {\cal T}_{j'}$ and thus that $t_{k+1} = t_{il} t_k \in \sum_{j \in J_{k+1}} {\cal T}_j$ with $J_k \subset J_{k+1}$;
\item we can prove that a clause $z_{j'}$ (or a sum of clauses) such that $\mygen_{i} \mygen_{l} t_{(j)} \in {\cal T}_{j'}$ does not exist and thus that $t_{k+1} \notin \sum_{j = 1}^m {\cal T}_j$ and $t \notin \sum_{j = 1}^m {\cal T}_j$.
\end{itemize}

In all cases given $\mygen_{i} \mygen_{l} t_{(j)} \notin {\cal T}_j$ we need a procedure that searches for a clause $z_{j'}$ (or a sum of clauses) such that $\mygen_{i} \mygen_{l} t_{(j)} \in {\cal T}_{j'}$.

By Lemma~\ref{lemma_eit_in_Tj} $\mygen_{i} \mygen_{l} t_{(j)} \notin {\cal T}_{j}$ implies $\{ \mygen_{i}, \mygen_{l} \} \cap \sup z_j \ne \emptyset$ and we can define a modified clause $z_j' := \mygen_{i} \mygen_{l} z_j$ and from $M(z_j) \subseteq (\Identity, t_{(j)})$ (\ref{formula_cal_T_j_def2}) easily follows $M(\mygen_{i} \mygen_{l} z_j) = M(z_j') \subseteq (\Identity, \mygen_{i} \mygen_{l} t_{(j)})$.
\opt{margin_notes}{\mynote{mbh.note: easy proof}}%
%

%To complete this part we need to show how we can check wether $\mygen_{i} \mygen_{l} t_{(j)} \in {\cal T}_{j'}$ with $j \ne j'$ (the case $j = j'$ being dealt with by Lemma~\ref{lemma_eit_in_Tj}). Since we suppose $\mygen_{i} \mygen_{l} t_{(j)} \notin {\cal T}_{j}$ this implies $\{ \mygen_{i}, \mygen_{l} \} \cap \sup z_j \ne \emptyset$ and we define $z_j' := \mygen_{i} \mygen_{l} z_j$ and from $M(z_j) \subseteq (\Identity, t_{(j)})$ (\ref{formula_cal_T_j_def2}) easily follows $M(\mygen_{i} \mygen_{l} z_j) = M(z_j') \subseteq (\Identity, \mygen_{i} \mygen_{l} t_{(j)})$ and thus for any clause $z_{j'}$ for $j' \in J_k$ such that $M(z_{j'}) \subseteq M(z_j')$.
If we are lucky we can find among clauses the clause $z_{j'}$, or a clause $z_{j''}$ such that $M(z_{j''}) \subseteq M(z_{j'})$, and in the two cases we will have respectively $M(\mygen_{i} \mygen_{l} z_j) \subseteq (\Identity, \mygen_{i} \mygen_{l} t_{(j')})$ or $M(\mygen_{i} \mygen_{l} z_j) \subseteq (\Identity, \mygen_{i} \mygen_{l} t_{(j'')})$. Also in this lucky case the algorithm needed to perform this test is polynomial being \bigO{m^2} since we need to examine at worst any of the clauses of $J_k$ against all other clauses checking for a clause compatible with $z_j'$.
\opt{margin_notes}{\mynote{mbh.note: crucial point, more details needed for: clauses of different sizes, 'frozen' part of $t$, orthogonal subspaces (\eg $1,2$ and $3,4$) etc.}}%
%
%we get $\mygen_{i} \mygen_{l} t_{(j)} \in {\cal T}_{j'}$ and if this holds for any $j \in J_k$ such that $\mygen_{i} \mygen_{l} t_{(j)} \notin {\cal T}_j$ we may conclude that $\mygen_{i} \mygen_{l} t_k \in \sum_{j \in J_{k}} {\cal T}_j$. Also in this case the algorithm needed to do this test is polynomial being \bigO{m^2} since we need to examine at worst any of the clauses of $J_k$ against all other clauses of the same set.
%
%If there exists at least one $j \in J_k$ such that $\mygen_{i} \mygen_{l} t_{(j)}$ is not in any other ${\cal T}_{j'}$ for $j' \in J_k$ we necessarily conclude that $\mygen_{i} \mygen_{l} t_k \notin \sum_{j \in J_{k}} {\cal T}_j$. The only remaining possibility is that there exists a subset $J_k \subset J_{k+1}$ such that for any of the quoted $j \in J_k$ there exists at least one $j' \in J_{k+1}/J_k$
%%
%\opt{margin_notes}{\mynote{mbh.note: crucial point \#2, here we could need $k, k', k" \ldots \in J_{k+1}/J_k$.}}%
%%
%such that $\mygen_{i} \mygen_{l} t_{(j)} \in {\cal T}_{j'}$ and also this test requires \bigO{m^2} comparisons and is thus polynomial.

We are left with the 'unlucky' (and harder) case in which there are no single clauses compatible with $z_j'$ and thus we must look for a sum of clauses that may contain $M(z_j')$.
\end{proof}

\myseparation

\begin{proof}
(final part of old proof)
We thus assume we are given a certain $t_k$ together with $\inv{t_k}$ and with a set of $r$ subsets of clauses $J_k$ such that for any of these subsets $t_k \in \sum_{j \in J_k} {\cal T}_j$ (this descends from the property that for any two compatible clauses $z_j$ and $z_k$ ${\cal T}_j \cap {\cal T}_k \ne \emptyset$ and thus that $t_k$ can be in different subsets of clauses ${\cal T}_k$ and/or sum of clauses).

By Lemma~\ref{lemma_eiejt} $t_{il} t_k \in \sum_{j \in J_{k+1}} {\cal T}_j$ if and only if $\mygen_{i} \mygen_{l} t_k \in \sum_{j \in J_{k+1}} {\cal T}_j$. By hypothesis $t_k \in \sum_{j \in J_k} {\cal T}_j$ and there are only two possible cases: in the first case $\mygen_{i}, \mygen_{l} \notin \inv{\sum_j {\cal T}_j}$: by Lemma~\ref{lemma_eit_in_Tj} $t_{k+1} = t_{il} t_k \in \sum_{j \in J_k} {\cal T}_j$ and we can move to the next term of the succession: this case clearly requires polynomial time.

\opt{margin_notes}{\mynote{mbh.note: here start the problems!!!}}%
It remains the case in which one or both $\mygen_{i}, \mygen_{l} \in \inv{\sum_j {\cal T}_j}$ and we concentrate on the case that both generators are in $\inv{\sum_{j \in J_k} {\cal T}_j}$; by Lemma~\ref{lemma_eit_in_Tj} $\mygen_{i} \mygen_{l} t_k \notin \sum_{j \in J_k} {\cal T}_j$. Let us suppose for a moment that there exists another subset of clauses $J_k \subseteq J_{k+1}$ such that $t_{il} t_k \in \sum_{j \in J_{k+1}} {\cal T}_j$ and by Lemma~\ref{lemma_Givens_all_in_set} necessarily also $t_k \in \sum_{j \in J_{k+1}} {\cal T}_j$ while by hypothesis $t_k \in \sum_{j \in J_k} {\cal T}_j$ and $\mygen_{i} \mygen_{l} t_k \notin \sum_{j \in J_k} {\cal T}_j$. Having supposed that there exists $J_{k+1}$ it descends that $\mygen_{i} \mygen_{l} t_k \in \sum_{j \in J_{k+1}/J_k} {\cal T}_j$ and, by Lemma~\ref{lemma_Givens_all_in_set}, that $t_k \in \sum_{j \in J_{k+1}/J_k} {\cal T}_j$.
\opt{margin_notes}{\mynote{mbh.note: here we probably need a stronger version of Lemma~\ref{lemma_Givens_all_in_set}}}%
It follows that the subset $J_{k+1}$ is to be searched in the subsets $J_k$ that guarantee that $t_k \in \sum_{j \in J_k} {\cal T}_j$ and that contains also $\mygen_{i} \mygen_{l} t_k$.
\opt{margin_notes}{\mynote{mbh.note: here $\mygen_{i}, \mygen_{l} \notin \inv{t_k}$ ? but \inv{t_k} is given and fixed so ??}}%

We showed that if the required subset $J_{k+1}$ exists it can come only from the subsets of clauses $J_k$ that contained $t_k$ and that can be easily done examining these subsets.
\opt{margin_notes}{\mynote{mbh.note: we need to calculate the cost of maintaining the list of viable subsets $K$}}%
If, on the contrary, in all subsets $J_k$ there is not the needed subset $J_{k+1}$ we have a certificate that the set generated by the clauses ${\cal T}_j$ do not cover $\OO{n}$. In both cases only a polynomial time is required to give and answer and the proposition is thus proved.
\end{proof}

\myseparation

In other words any ${\cal T}_j$ induces naturally three sets: the set of generators $\mygen_{i}$ corresponding to Boolean variables in affirmative forms, the set of generators $\mygen_{j}$ corresponding to Boolean variables in negative forms and the remaining set of $n - k$ generators to which we associate the $\bino{n-k}{2}$ bivectors $\mygen_{i} \mygen_{j}$ and we can use these sets to build a graph with $n$ vertices corresponding to the $n$ generators and the edges of this graph are given by the bivectors.

Given ${\cal T}_j, {\cal T}_k$ and their respective graphs it is interesting to evaluate the graph associated to ${\cal T}_j + {\cal T}_k$ (\ref{formula_cal_T_j+k_def}).

Given $t \in \OO{n}$ we can define a graph with $n$ vertices and given the Givens expansion of $t$ it defines a graph with $n$ vertices corresponding to coordinates and links corresponding to the couples of coordinates $\mygen_{i} \mygen_{j}$ appearing in the Givens expansion of $t$.

\begin{MS_lemma}
\label{lemma_graph_t}
For any $t \in \OO{n}$ its induced graph is a subgraph of the complete graph of $n$ vertices.
\end{MS_lemma}
\begin{proof}
Given that any $t \in \OO{n}$ has a Givens expansion it induces a subgraph of the complete graph; that the complete graph correspond to a possible Givens expansion follows immediately from the fact that any composition of $\bino{n}{2}$ $\SO{2}$ isometries acting on $\mygen_{i} \mygen_{j}$ is in $\OO{n}$.
\end{proof}

We define $t \in \OO{n}$ to be \emph{full fledged} if it induces a graph with unique connected component connecting all vertices.

\myseparation

\begin{MS_theorem}
\label{theorem_SAT_in_O(n)_6}
A given \SAT{} problem is unsatisfiable if and only if the sets induced by its $m$ clauses ${\cal T}_j$ are such that the set of bivectors induced by $\sum_{j = 1}^m {\cal T}_j$ form a unique connected component of size $n$.
\opt{margin_notes}{\mynote{mbh.note: see log p. 808}}%
\end{MS_theorem}
\begin{proof}
Let the problem be unsatisfiable then by Theorem~\ref{theorem_SAT_in_O(n)} $\sum_{j = 1}^m {\cal T}_j = \OO{n}$ that induces a unique connected component of size $n$. Conversely if the induced bivectors form a unique connected component of size $n$ than we can build any $t \in \OO{n}$ and Theorem~\ref{theorem_SAT_in_O(n)} applies again.
\end{proof}

\newpage
\section{An actual algorithm testing \SAT{} unsatisfiability}
\label{sec_SAT_algorithm}
We wrap up previous results in an actual algorithm that tests \SAT{} unsatisfiability.
\begin{algorithm}
\caption{\SAT-unsatisfiability algorithm}
\label{SAT_unsat_algorithm}
\begin{algorithmic}[5]
\Require A $k$\SAT{} problem $\myBooleanS$ in $n$ Boolean variables
\Ensure $\myBooleanT$ if $\myBooleanS$ is satisfiable, $\myBooleanF$ otherwise
\State {}

\State Start: from $m$ clauses of $\myBooleanS$ build $m$ sets ${\cal T}_{j}$
%
%\If{$\myBooleanS$ is empty}
%\State return $\mathrm{F}$ \Comment{$0$ is fully symmetric}
%\EndIf
%
%\If{$\myBooleanS$ contains any satisfiable problem}
%\State return $\mathrm{T}$ \Comment{a satisfiable problem is not symmetric}
%\EndIf
%
%\State choose next literal $\mylitrl_i$ \Comment{this is a delicate point}

\For{$det = -1$ to $+1$}
\For{$i = 1$ to $k$}
\State newT = 0
\For{$j1 = 1$ to $m$}
\For{$j2 = j1 + 1$ to $m$}
\If{incidence of ${\cal T}_{j_1}$ and ${\cal T}_{j_1} == n - 2$} %\Comment
\State generate ${\cal T}_{j_1 + j_2}$ with current det
\State ++newT
\EndIf
\EndFor
\EndFor

\If{newT == 0}
\State return $\myBooleanT$ \Comment{if no new $n$ dimensional null subspace the problem is satisfiabile}
\EndIf

m += newT

\EndFor
\EndFor

\State return $\myBooleanF$ \Comment{Fock basis fully covered: the problem is unsatisfiabile}

\end{algorithmic}
\end{algorithm}

We remark that the algorithm is the same for $k=2$ or $k > 2$ \SAT{} problems and that the addition of two clauses with one common opposite Boolean variable is well known in standard \SAT{} algorithms where is named 'resolution'.

%\newpage
\section*{Appendix}
\label{Appendix_old}
We begin recalling the classical Givens expansion of an orthogonal isometry $t \in \OO{n}$ \cite{Hoffman_Raffenetti_Ruedenberg_1972}: given a real space $\R^n$ with its standard basis $\mygen_{i}, i = 1, \ldots, n$ any $t \in \OO{n}$ may be decomposed in a succession of up to $\bino{n}{2}$ $\SO{2}$ rotations acting in subspaces $\my_span{\mygen_{i}, \mygen_{j}}$ and moreover the expansion is not unique and the order of $\my_span{\mygen_{i}, \mygen_{j}}$ can be chosen at will (that is not to say that they commute, this being in general false).

We remark that the two dimensional subspaces $\my_span{\mygen_{i}, \mygen_{j}}$ are in one to one correspondence with the bivectors $\mygen_{i} \mygen_{j}$ of $\myClg{}{}{\R^{n,n}}$ and that altogether $\mygen_{i} \mygen_{j}$ form the standard basis for the grade $2$ elements of $\myClg{}{}{\R^{n,n}}$ (bivectors).

\begin{MS_lemma}
\label{lemma_SO2_expansion}
Any $t \in \SO{2} \myisom \R(2)$ acting in $\my_span{\mygen_{i}, \mygen_{j}}$ can be decomposed in the linear combination of two $\SO{2}$ isometries.
\end{MS_lemma}
\begin{proof}
For any $t \in \SO{2}$ we may write
$$
t = \left(\begin{array}{r r} \cos \theta_{i,j} & - \sin \theta_{i,j} \\ \sin \theta_{i,j} & \cos \theta_{i,j} \end{array}\right) = \cos \theta_{i,j} \left(\begin{array}{r r} 1 & 0 \\ 0 & 1 \end{array}\right) + \sin \theta_{i,j} \left(\begin{array}{r r} 0 & -1 \\ 1 & 0 \end{array}\right) \dotinformula
$$
\end{proof}
In this expansion we recognize the vectorial representation of an isometry acting in $\my_span{\mygen_{i}, \mygen_{j}}$ that in the spinorial representation would be
$$
\psi' = v_1 v_2 \psi = (v_1 \cdot v_2 + v_1 \wedge v_2) \psi = (\cos \frac{\theta_{i,j}}{2} + \sin \frac{\theta_{i,j}}{2} \mygen_{i} \mygen_{j}) \psi \qquad v_1, v_2 \in \my_span{\mygen_{i}, \mygen_{j}}
$$
where supposing $v_1^2 = v_2^2 = 1$ we easily obtain $v_1 \cdot v_2 = \cos \frac{\theta_{i,j}}{2}$ and $v_1 \wedge v_2 = \sin \frac{\theta_{i,j}}{2} \mygen_{i} \wedge \mygen_{j} = \sin \frac{\theta_{i,j}}{2} \mygen_{i} \mygen_{j}$ and clearly expansion (\ref{formula_Givens_expansion2}) give rise to a sum that can contain up to $2^{\bino{n}{2}}$ terms and that contains all possible combinations of $\left(\begin{array}{r r} 1 & 0 \\ 0 & 1 \end{array}\right)_{i,j}$ and $\left(\begin{array}{r r} 0 & -1 \\ 1 & 0 \end{array}\right)_{i,j}$ so that it is easy to prove
\begin{MS_Proposition}
\label{prop_Fock_Givens_expansion}
Any $t \in \OO{n}$ can be expressed as a linear composition of%
\opt{margin_notes}{\mynote{mbh.note: WRONG!! any $\sum_\lambda \alpha_\lambda \lambda$ is necessarily \emph{diagonal}! Probably the matrices are those of the Hyperoctahedral group.}}%
\begin{equation}
\label{formula_Fock_Givens_expansion}
t = \sum_\lambda \alpha_\lambda \lambda \qquad \alpha_\lambda \in \R, \quad \lambda \in \O1{n} \dotinformula
\end{equation}
\end{MS_Proposition}
\begin{proof}
Substituting in Givens expansion the expression of $t$ we easily get
$$
\cos \theta_{i,j} \Identity_n + \sin \theta_{i,j}
\left(\begin{array}{r r r r r}
1 & 0 & & ... & 0 \\
0 & 1 & & ... & 0 \\
& & ... & & \\
0 & 0 & ... & -1 & 0 \\
& & ... & & \\
0 & 1 & 0 & ... & 0 \\
0 & 0 & 0 & ... & 1 \\
\end{array}\right)
$$
where the effect of the second matrix is that of swapping $ii$ and $jj$ terms of $\Identity_n$ substituting the last one by $-1$ and it is simple to prove by induction on $n$ that all elements of $\lambda \in \O1{n}$ can be generated in this way.
\end{proof}
This is the equivalent formulation of Fock basis spinor expansion in $\OO{n}$ and we leave as an exercise to verify that the basis vectors $\lambda \in \O1{n}$ form a proper basis of $\OO{n}$.

} % note finali: stampate solo se all'inizio c'è l'opzione final_notes

\end{document}